\documentclass[structabstract]{aa}
\pdfoutput=1
\usepackage{natbib}
\bibpunct{(}{)}{;}{a}{}{,}
\usepackage{graphicx}
\usepackage{lscape}
\usepackage{aalongtable}
\usepackage{subfigure}
\usepackage{txfonts}
\usepackage[update,prepend]{epstopdf}
\usepackage[pdfborder={0 0 0.5},colorlinks=false]{hyperref}

\newcommand{\Ha}{\mbox{H$\alpha$}}
\newcommand{\Hb}{\mbox{H$\beta$}}
\newcommand{\Hd}{\mbox{H$\delta$}}
\newcommand{\Hg}{\mbox{H$\gamma$}}
\newcommand{\CHbeta}{\mbox{$C$(H$\beta$)}}

\newcommand{\AV}{\mbox{$A_V$}}
\newcommand{\kms}{\mbox{km~s$^{-1}$}}

\begin{document}
\title{Mapping Luminous Blue Compact Galaxies with VIRUS-P:}

\subtitle{morphology, line ratios and kinematics}

\titlerunning{Mapping Luminous Blue Compact Galaxies with VIRUS-P}

\author{L.~M. Cair{\'o}s \inst{1}
        \and
        N. Caon \inst{2,3}
	\and  
	B. Garc{\'\i}a Lorenzo \inst{2,3}
	\and
	A. Kelz \inst{1}
	\and
	M. Roth \inst{1}
	\and
	P. Papaderos \inst{4}
	\and
	O. Streicher \inst{1}
}

\institute{Leibniz-Institut f{\"u}r Astrophysik, 
           An der Sternwarte 16, D-14482 Potsdam, Germany\\
           \email{luzma; akelz; mmroth; ole@aip.de}
           \and
           Instituto de Astrof{\'\i}sica de Canarias, 
           E-38200 La Laguna, Tenerife, Spain 
           \and 
           Departamento de Astrof{\'\i}sica, Universidad de la Laguna, 
           E-38206 La Laguna, Tenerife, Spain\\ 
           \email{nicola.caon; bgarcia@iac.es}
           \and
           Centro de Astrof{\'\i}sica and Faculdade de Ci{\^e}ncias, 
	   Universidade do Porto, Rua das Estrelas, 4150-762 Porto, Portugal \\ 
           \email{papaderos@astro.up.pt}}
\date{Received May 15, 2012; accepted August 29, 2012}

 
\abstract
{Blue Compact Galaxies (BCG) are narrow emission-line systems, undergoing a
violent burst of star formation. They are compact, low-luminosity galaxies,
with blue colors and low chemical abundances, that offer us a unique
opportunity to investigate collective star formation and its effects on
galaxy evolution in a relatively simple, dynamically unperturbed
environment. Spatially resolved spectrophotometric studies of BCGs are
essential for a better understanding of the role of starburst-driven feedback
processes on the kinematical and chemical evolution of low-mass galaxies near
and far.}
{We carry out an integral field spectroscopy (IFS) study of a sample of 
luminous BCGs, with the aim to probe the morphology, kinematics, dust 
extinction and excitation mechanisms of their warm interstellar medium (ISM).}
{IFS data for five luminous BCGs were obtained using VIRUS-P, the prototype
instrument for the Visible Integral Field Replicable Unit Spectrograph,
attached to the 2.7~m Harlan J. Smith Telescope at the McDonald Observatory.
VIRUS-P consists of a square array of 247 optical fibers, which covers a 
$109\arcsec \times 109\arcsec$ field of view, with a spatial sampling of
$4\farcs2$  and a 0.3 filling factor. We observed in the 3550--5850 \AA\
spectral range, with a resolution of 5 \AA\ FWHM. From these data we built
two-dimensional maps of the continuum and the most prominent emission-lines
([\ion{O}{ii}]~$\lambda3727$, \Hg, \Hb\ and [\ion{O}{iii}]~$\lambda5007$), and
investigate the morphology of diagnostic emission-line ratios and the
extinction patterns in the ISM as well as stellar and gas kinematics.
Additionally, from integrated spectra we infer total line fluxes and
luminosity-weighted extinction coefficients and gas-phase metallicities.}
{All galaxies exhibit an overall regular morphology in the stellar continuum,
while their warm ISM morphology is more complex: in II~Zw~33 and Mrk~314, the
star-forming regions are aligned along a chain-structure; Haro~1,
NGC~4670 and III~Zw~102 display several salient features, such as extended
gaseous filaments and bubbles.  A significant intrinsic absorption by dust is
present in all galaxies, the most extreme case being III~Zw~102. Our data
reveal a manifold of kinematical patterns, from overall regular gas and
stellar rotation to complex velocity fields produced by structurally and
kinematically distinct components.}
{}

\keywords{galaxies: starburst - galaxies: dwarf - galaxies: abundances - 
galaxies: kinematics and dynamics}
   
\maketitle
%

\section{Introduction}
\label{Section:Introduction}

Blue Compact Galaxies (BCGs) are low-luminosity ($M_{B}$ ranges from $-12$ to
$-21$) and low-metallicity ($7.0 \leq 12+\log(\mathrm{O/H}) \leq 8.4$)
systems, which have optical spectra similar to those of \ion{H}{ii} regions
in spiral galaxies  \citep{ThuanMartin1981, Kunth2000}. With star formation
rates between $0.1$ and $1\,M_{\sun}$ yr$^{-1}$
\citep{Fanelli1988,HunterElmegreen2004}, they represent the most extreme
manifestation of star-forming (SF) activity in late-type dwarf galaxies in the
local Universe. 

BCGs are of paramount importance in extragalactic astronomy and observational
cosmology research, as they hold key clues to understand fundamental topics such 
as galaxy formation and evolution, and the ignition and self-regulation of SF
activity in relatively isolated late-type galaxies.  

They represent an unparalleled link to the early Universe, since in the well
accepted framework of a Cold Dark Matter (CDM) Universe, structure formed
hierarchically, with small scale objects (low mass halos) collapsing first
at relatively large redshifts. About two percent of the BCGs have very low
metallicities, ($12+\log\mathrm{(O/H)} \leq 7.65$); these objects, often
referred to as eXtremely Metal Deficient (XBCD) galaxies \citep{Papaderos2008}, 
constitute the best local analogs to the distant subgalactic units from which 
larger galaxies are formed \citep{Papaderos1998,Kniazev2004}. Also, the more luminous BCGs (or
different subsamples of them) have been regarded as the local counterparts of
different high redshift SF galaxy populations  \citep{MasHesse2003,
Grimes2009}. Detailed analysis of these nearby objects are essential to
interpret the observations of the more distant SF galaxy population, as their
proximity allows for studies focused on their stellar content, kinematics and
abundances, with an accuracy and spatial resolution that cannot be achieved at
higher redshifts \citep{Cairos2009-mrk409,Cairos2009-mrk1418,Cairos2010}. 

Moreover, low-luminosity systems are the best laboratories to study the
star-formation process. Without spiral density waves to act as a triggering
mechanism for star formation, these galaxies enable us to investigate 
in an unbiased manner all of the other factors that trigger and control 
SF activities. In addition, BCGs offer a unique opportunity to test 
star-formation models at very low metallicities, similar to those 
of the early Universe. 

In spite of the strong interest gained for BCGs in the last
decades, several key issues in the field, namely, the evolutionary status of
BCGs, their Star Forming History (SFH) or the mechanisms that trigger their
star-forming activity, are still open.

One of the reasons is probably that most of the BCG work published so far deal
with analysis of samples by means of surface photometry
\citep{Papaderos1996a,Cairos2001-II,Cairos2001-I,Doublier1997,Doublier1999,GildePaz2003,GildePaz2005},
and spectroscopic studies typically only cover the brightest SF regions of
these systems. Studies combining surface photometry with spatially resolved
spectroscopy and spectral  synthesis, and aiming at a quantitative study of
the formation history of  stellar populations in BCGs, are scarce.

That very few spectrophotometric analyses can be found in the literature, and
virtually all of them focused on one single object
\citep{Guseva2003_SBS1129,Guseva2003_HS1442,Guseva2003_SBS1415,Cairos2002,Cairos2007},
is essentially due to the large amount of observing time that conventional
observational techniques require: acquiring images in several broad-band and
narrow-band filters, plus a sequence of long-slit spectra sweeping the region
of interest, translate  into observing times of a few nights per galaxy,
making analysis of statistically meaningful samples not feasible in terms of
observing time. Moreover, such observations usually suffer from varying
instrumental and atmospheric conditions, which make combining all these data
together complicated. 

Integral Field Spectroscopy (IFS) offers a way to approach spectrophotometric
BCG studies in a highly effective manner \citep{Izotov2006,GarciaLorenzo2008,
Vanzi2008,James2009,Cairos2009-mrk409,Cairos2009-mrk1418,Cairos2010}. IFS
provides simultaneous spectra of each spatial resolution element, under
identical instrumental and atmospheric conditions, which is not only a more
efficient way of observing, but also guarantees the homogeneity of the
dataset. From only one Integral Field Unit (IFU) frame we can produce a series
of  broad-band images in a large number of filters, narrow-band images in a
large set of bands, as well as derive a large collection of observables
from the spectrum (e.g. line indices). In terms of observing time, IFS
observations of BCGs are one order of magnitude more efficient than
traditional observing techniques, implying that now 
spectrophotometric studies of large samples of BCG galaxies have become
viable.

The work presented here is part of the IFS-BCG survey, a challenging long-term
project that aims to perform an exhaustive spectrophotometric survey of a
large sample of BCGs by means of IFS. The main scientific goals of this
projects include:  i)~to disentangle and characterize the different stellar
populations in BCGs (e.g. constrain their ages, Initial Mass Function and 
metallicity); ii)~to investigate the evolutionary status of the galaxies and
constrain their SFH; iii)~to probe the mechanism responsible for the actual
burst of star formation; iv)~to analyze the feedback effects between massive
stars and the Interstellar Medium (ISM) in dwarf galaxies; v)~to investigate
the recent suggestion that some BCGs harbor Active Galactic Nuclei (AGN)
associated with intermediate mass black holes \cite{Izotov2010};  
vi)~to provide an accurate 
dataset of photometric and spectroscopic parameters for a large sample of
nearby SF dwarf galaxies, the essential template to understand the results of
the investigations at intermediate and high-z.  

This is the fourth in a series of papers presenting the findings of the
project, in which  we report on results derived for five luminous BCGs
observed with VIRUS-P. In the first and second papers of the series
\citep{Cairos2009-mrk409,Cairos2009-mrk1418} we illustrated the full potential
of this study by showing results on two representative BCGs, Mrk~1418 and
Mrk~409, both observed with the Potsdam multi-aperture spectrophotometer
(PMAS), attached at the 3.5-m telescope at Calar Alto Observatory. A more
detailed description of the project objectives, IFU observation and reduction
techniques and results for another eight objects observed with PMAS, was
presented in \cite{Cairos2010}. 

This paper is structured as follows: In
Sect.~\ref{Section:ObservationsandData} we describe the observations, the data
reduction process and the method employed to derive the maps. In
Sect.~\ref{Section:Results} we present the main outcomes of the work, that is,
the flux, line ratio and velocity maps, as well as the results derived
from the analysis of the integrated spectra. Finally, the main findings for
each galaxy are discussed in Sect.~\ref{Section:Discussion}, and summarized in
Sect.~\ref{Section:Summary}.

\section{Observations and Data Processing}
\label{Section:ObservationsandData}

\subsection{Galaxy sample and observations} 
\label{SubSection:Observations} 

The whole survey includes the mapping of about 40 galaxies, chosen as to cover
the wide range of luminosities ($-12\geq M_{B}\geq-21$) and metallicities
($7.0 \leq 12+\log(\mathrm{O/H}) \leq 8.4$) found among BCDs. We took also
care that the different morphological classes observed among BCGs be well
represented in the sample \citep{Cairos2001-II}. 

In this paper we present results for five luminous BCGs
\citep{BergvallOstlin2002,Cairos2002}. These objects are the most luminous ($M_{B}\leq
-17$), and typically also the more irregular BCGs; they tend to have an
important contribution of older stars, significant amounts of dust and very
distorted starburst morphologies \citep{Cairos2003,Bergvall2012}. As we have
already stated in the introduction, these luminous BCGs have a major relevance
in cosmological studies, as they (or different subsamples of them) have been
regarded as the local counterparts of different galaxies populations at higher
redshifts ($z \approx$ 2-6), as Lyman-break galaxies \citep{Grimes2009} or
Lyman-$\alpha$ emitters \citep{MasHesse2003}.

\begin{figure*}[htb]
\mbox{
\centerline{
\hspace*{0.0cm}\subfigure{\includegraphics[width=0.27\textwidth]{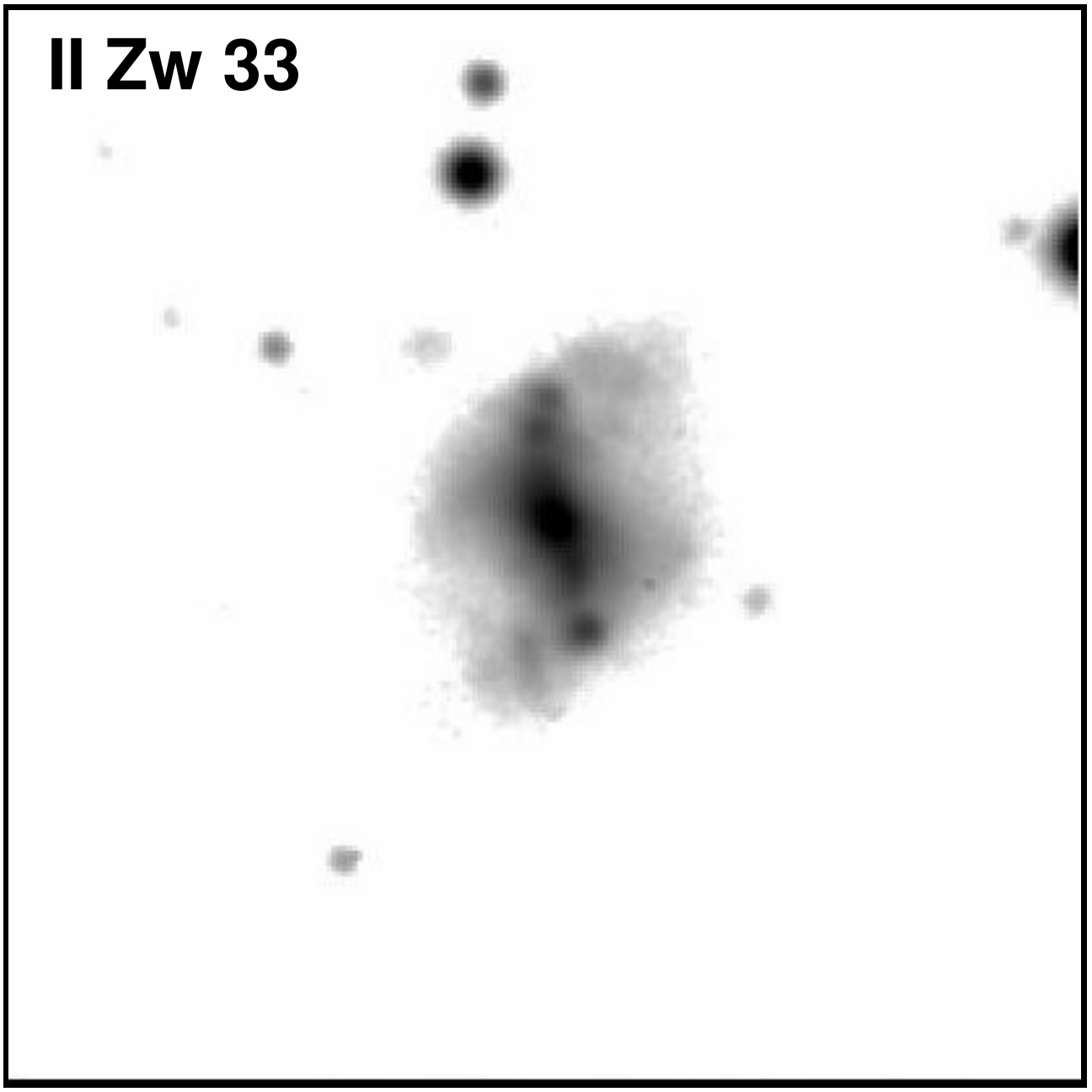}}
\hspace*{0.0cm}\subfigure{\includegraphics[width=0.27\textwidth]{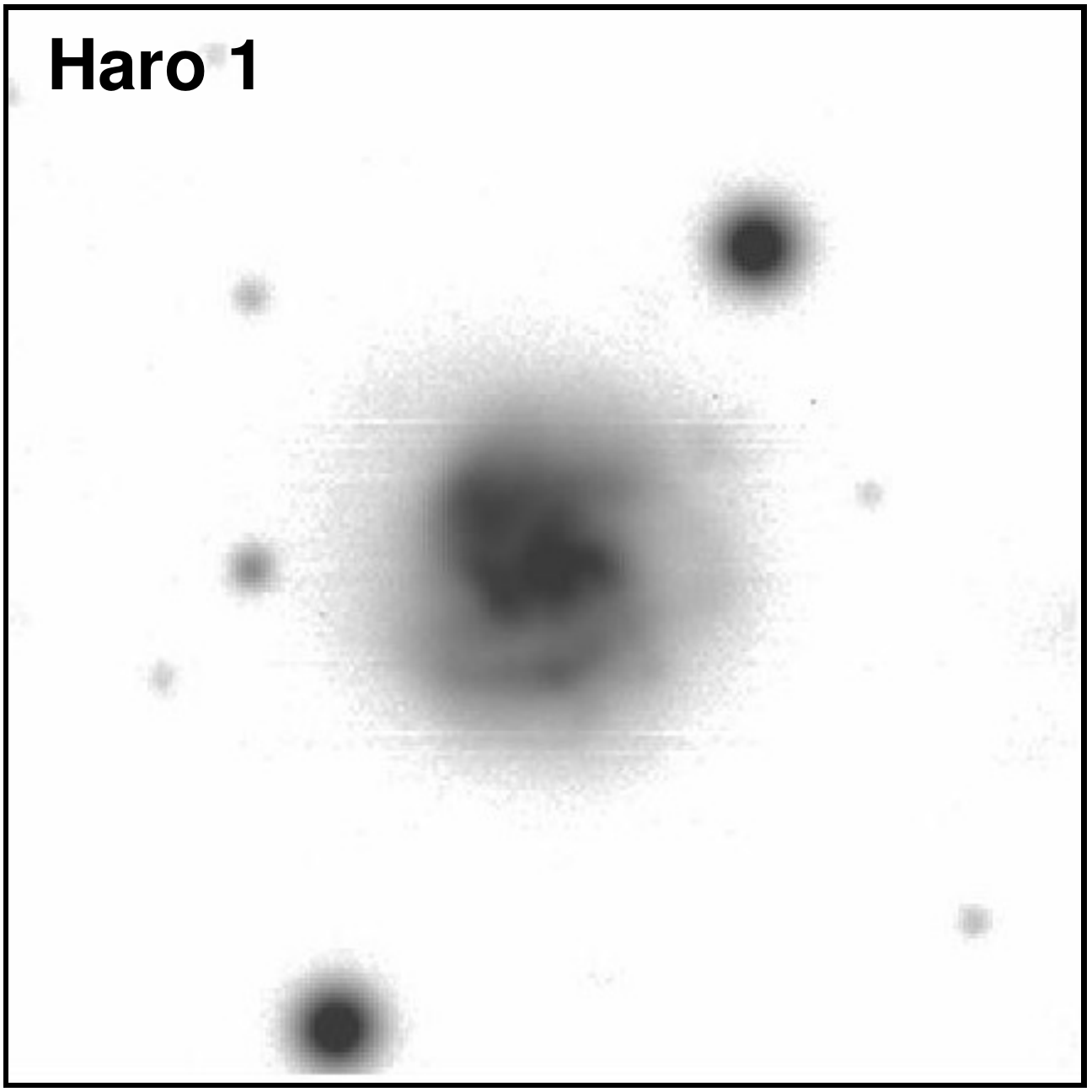}}
\hspace*{0.0cm}\subfigure{\includegraphics[width=0.27\textwidth]{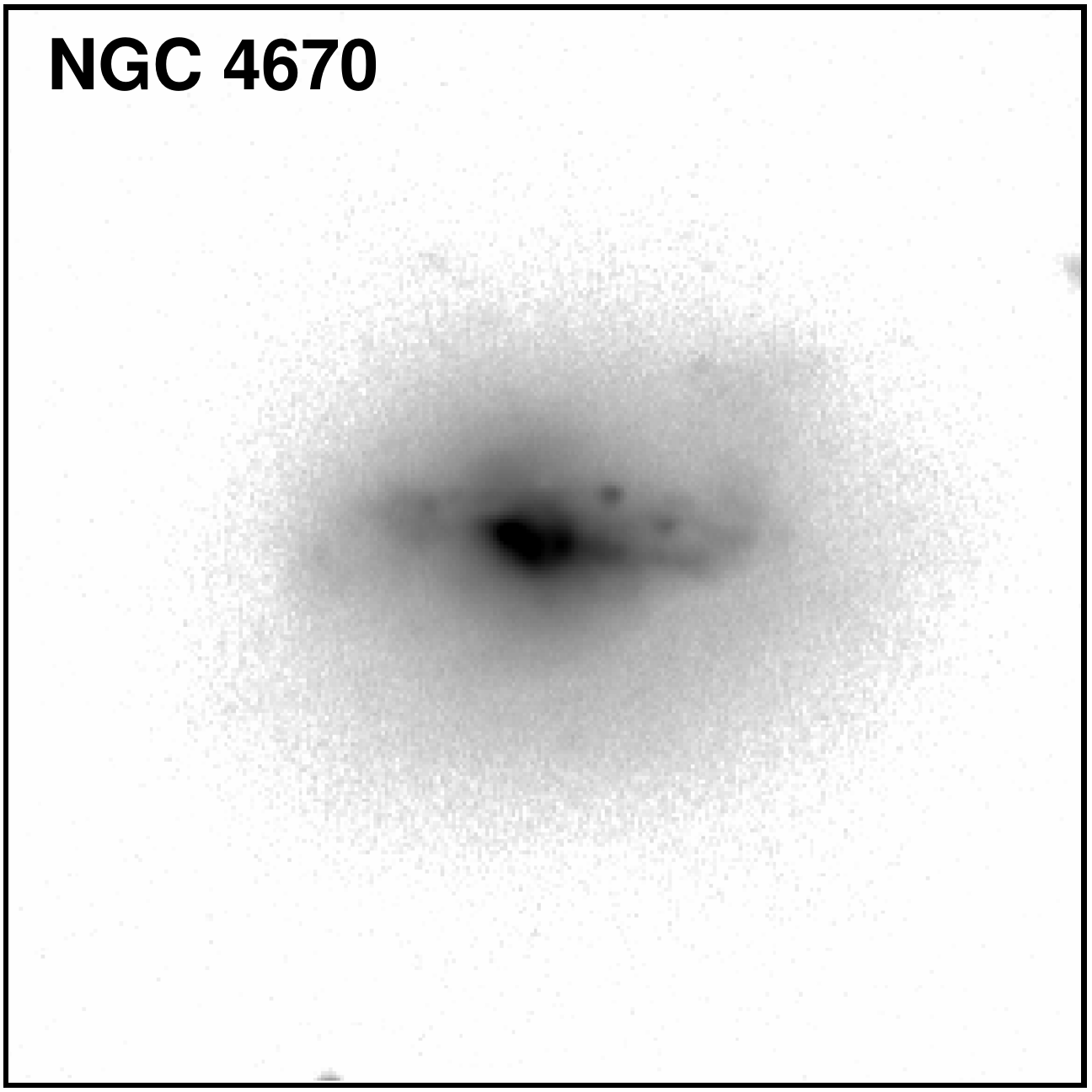}}
}}   
\mbox{
\centerline{
\hspace*{0.0cm}\subfigure{\includegraphics[width=0.27\textwidth]{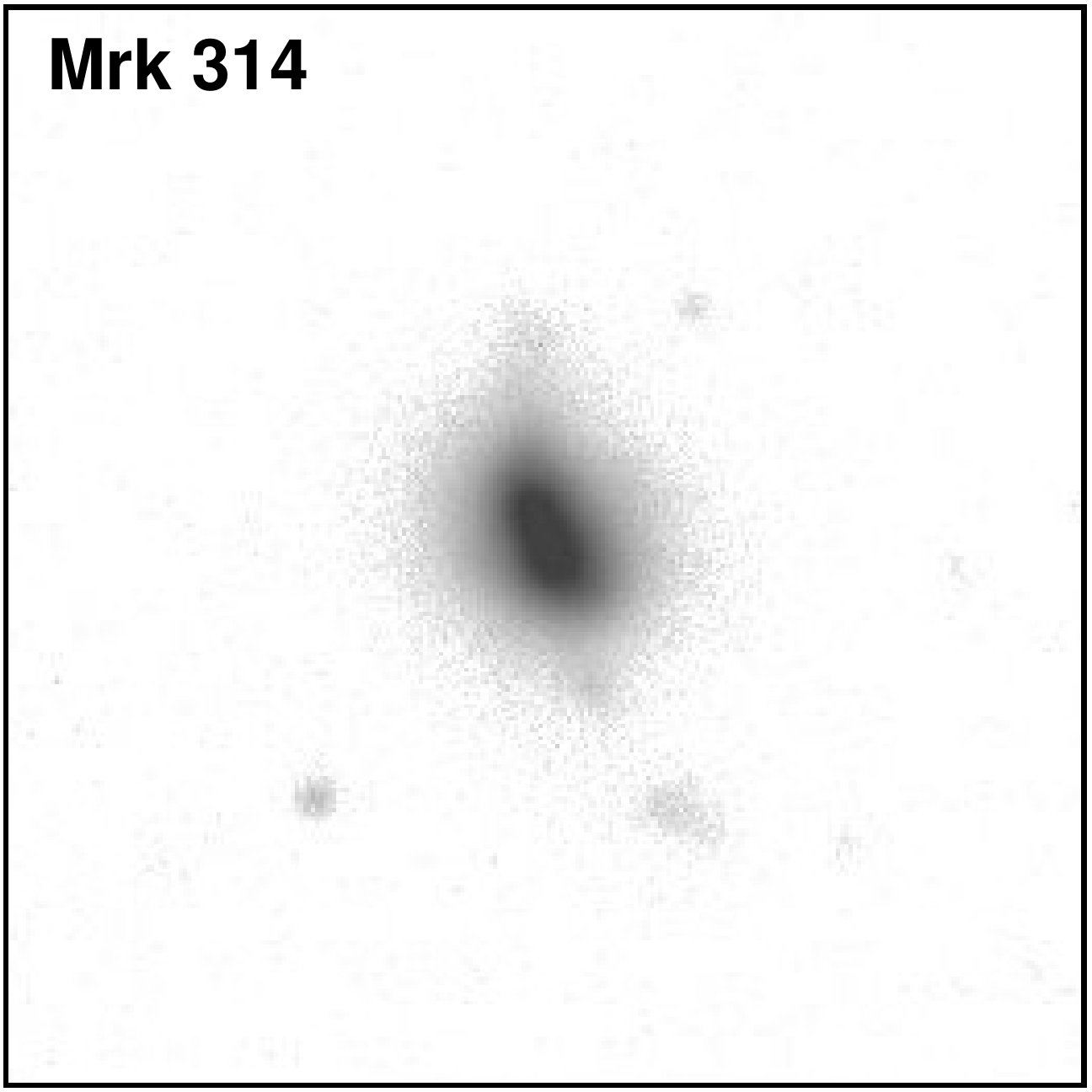}}
\hspace*{0.0cm}\subfigure{\includegraphics[width=0.27\textwidth]{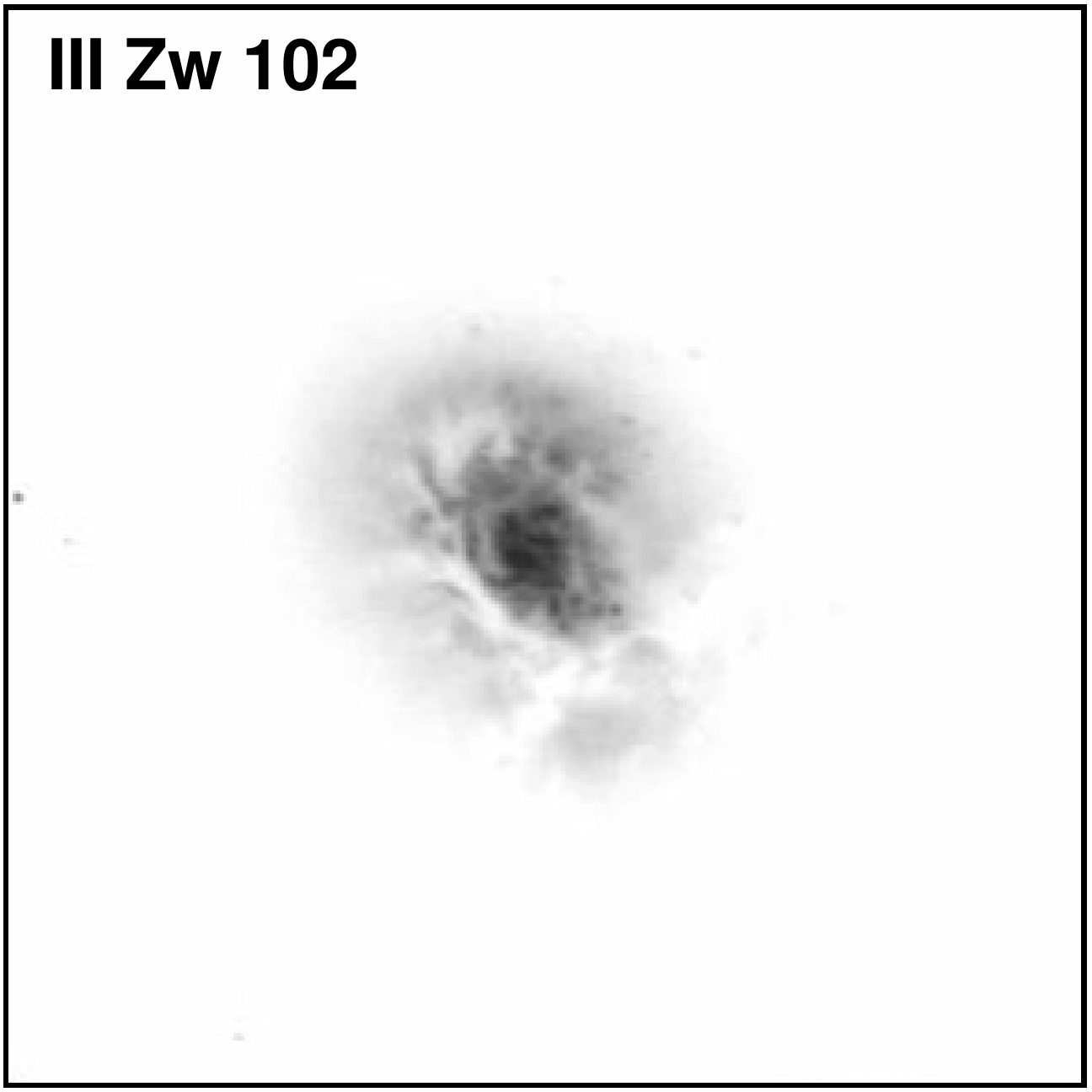}}
}} 
\caption{$B$-band images of the galaxies (in parenthesis the
instrument/telescope at which the images were obtained): II~Zw~33 (CAFOS at
the 2.2~m CAHA, \citealt{Cairos2001-I}); Haro~1 (Prime Focus at 3.5-m CAHA, 
\citealt{Cairos2001-I}); NGC~4670 (60 inch telescope on Mt. Palomar, 
\citealt{GildePaz2003}); Mrk~314 (prime Focus at the 3.5-m CAHA, 
\citealt{Cairos2001-I}) and III~Zw~102 (ALFOSC/NOT, unpublished). 
North is up, east to the left. 
The FOV is $109\arcsec\times109\arcsec$ (the FOV of the VIRUS-P instrument). 
All images are shown in logarithmic  scale. 
}
\label{Figure:Bbandimages}
\end{figure*}

The basic data for the galaxies and the complete log of the observations are
listed in Table~\ref{Table:sample} and Table~\ref{Table:log} respectively.
$B$-band images of the galaxies are displayed in
Figure~\ref{Figure:Bbandimages}. IFS of III~Zw~102 and Mrk~314 has been
previously published by \cite{GarciaLorenzo2008}, who studied the central
$33\farcs6\times29\farcs4$ of  III~Zw~102 and $16\arcsec\times12\arcsec$ of
Mrk~314 using INTEGRAL at the William Herschel Telescope. In both cases, the
much smaller field of view (FOV) of INTEGRAL allows only for a partial mapping of the
starburst regions.

Spectral data for the five galaxies listed in Table~\ref{Table:sample} were collected
using the VIRUS-P IFU spectrograph \citep{Hill2008}, working at the 2.7m
Harlan J. Smith Telescope at the McDonald Observatory (TX, USA). VIRUS-P is
the prototype instrument for VIRUS (= Visible Integral-field Replicable Unit
Spectrograph), a massively replicated IFU made of 150 units like VIRUS-P,
which will be installed on the 9.2~m Hobby-Eberly Telescope
\citep{Hill2006,Hill2007}.


\begin{table*}
\caption{Galaxy Sample\label{Table:sample}}
\begin{center}
\centering
\begin{tabular}{lcccccccc} 
\hline\hline 
Galaxy                & Other designations        & RA         & DEC          & $m_{B}$               & $D$   & $M_{B}$  & Morph & $A_{V}$  \\
                      &                           &  (2000)    &   (2000)     &                       & (Mpc) & (mag)    &       &  (mag)   \\
   (1)                &    (2)                    &  (3)       &   (4)        & (5)                   & (6)   & (7)      & (8)   &  (9)     \\
\hline
\object{II Zw 33}   & UGCA~102, Mrk~1094             & 05 10 48.1 & $-$02 40 54  & 14.57\tablefootmark{a}  & 37.4  & $-18.29$ & Ch    & 0.288    \\
\object{Haro 1}     & NGC~2415, UGC~3930             & 07 36 56.7 &    35 14 31  & 12.61\tablefootmark{a}  & 52.9  & $-21.00$ & Ext   & 0.117    \\ 
\object{NGC 4670}   & UGC~7930, Arp~163, Haro~9      & 12 45 17.2 &    27 07 32  & 13.02\tablefootmark{b}  & 22.4  & $-18.73$ & Nuc   & 0.041    \\ 
\object{Mrk 314}    & NGC~7468, UGC~12329            & 23 02 59.2 &    16 36 19  & 13.78\tablefootmark{a}  & 29.0  & $-18.53$ & Ch    & 0.234    \\ 
\object{III Zw 102} & NGC~7625, UGC~12529, Arp~212   & 23 20 30.1 &    17 13 32  & 12.50\tablefootmark{a}  & 23.1  & $-19.31$ & Ext   & 0.069    \\ 
\hline
\end{tabular}
\end{center}
\tablefoot{
Cols. (3) and (4): Units of right ascension are hours, 
minutes, and seconds, and units of declination are degrees, arcminutes, and
arcseconds.
Col.~(5): $B$-band magnitudes: \tablefoottext{a} asymptotic photometry obtained by 
extrapolating the growth curves, and corrected for galactic extinction 
\citep{Cairos2001-II}. 
The asymptotic magnitudes listed in \cite{Cairos2001-II} were
corrected for Galactic extinction following \cite{Burstein1982};
here they have been recomputed using the \cite{Schlegel1998} 
extinction values; \tablefoottext{b} integrated magnitudes from \cite{GildePaz2003}, 
corrected for Galactic extinction.
Col.~(6): Distances, computed assuming 
a Hubble flow, with a Hubble
constant $H_0=73$ km  s$^{-1}$ Mpc$^{-1}$, and taking into account 
the influence of the Virgo cluster, the Great Attractor and the Shapley 
supercluster, are taken from NED (http://nedwww.ipac.caltech.edu).
Col.~(7): Absolute magnitude in the $B$ band, computed from the 
tabulated $B$-band magnitude and distance.
Col.~(8): morphological classification following \cite{Cairos2001-II}.
Col.~(9): galactic extinction, from the \cite{SchlaflyFinkbeiner2011} 
recalibration, as listed in NED.
}
\end{table*}

VIRUS-P (with the VP-1 IFU bundle used in these observations) consists of a
square array of 247 optical fibers (of which three were dead in the used
configuration), which samples a $109\arcsec \times 109\arcsec$ FOV 
with a 0.3 filling factor.
Each fiber is circular and has a diameter of $4\farcs2$ on the sky. 
We observed in
the spectral range 3550--5850 \AA, with a spectral resolution (FWHM) of
$\approx 5$ \AA\ ($340$ km s$^{-1}$ at \Hb).

Observations were carried out during two observing runs, in 2007 October and
in 2008 March. During the latter run, due to camera alignment issues, the
spectra of five fibers fell off the chip, and therefore only 239 fibers were
usable. 

We followed the same strategy in the two observing campaigns. Calibration
frames consisted of bias frames, sky-flats and comparison lamps taken at the
beginning and at the end of the night. Sky-flats are required to define and
trace the fibers (apertures) on the detector as well as to perform the
throughput correction. Spectra of emission-line lamps (Hg-Cd lamp in our case)
are needed to carry out the wavelength calibration. Because VIRUS-P is an
instrument with practically no flexures, comparison lamp spectra taken once
during the night will suffice, and no arc exposures interspersed with the
science observations are required.

As the filling factor of the IFU is 30\%, three dithered exposures were taken
in order to ensure a coverage of about 90\% of the FOV. The standard observing
sequence consists of a set of three exposures (of 20 to 30 minutes each) at 
each of the three positions (the telescope offsets are first $3\farcs9$ 
west, $2\farcs1$ south and then $3\farcs75$ east, $2\farcs1$ 
south).

Flux spectrophotometric standards were also observed, albeit using a six
position dither pattern, which guarantees the collection of (almost) their 
total flux. The spectrophotometric standard stars BD+25d4655,
G~191-B2B  and BD+40d4032 were observed during the October
run, and Feige 34 and BD+33d2642 in the March run.

\begin{table*}
\caption{Log of the Observations\label{Table:log}}
\begin{center}
\begin{tabular}{lccccc} 
\hline\hline 
Galaxy          & Date        & Exp. time    & Airmass range & Seeing      & Spatial scale         \\
                &             & (s)          &               & (arcsec)    & (pc arcsec$^{-1}$)   \\
\hline
II~Zw~33        & Oct 2007    & 3600         & 1.20--1.40    & 2.2--2.4    & 760                   \\
Haro~1          & Mar 2008    & 3600         & 1.01--1.10    & 2.5--3.2    & 1080                  \\
NGC~4670        & Mar 2008    & 4800         & 1.00--1.13    & 2.0--3.0    & 460                   \\
Mrk~314         & Oct 2007    & 5400         & 1.03--1.20    & 1.8--2.0    & 590                   \\
III~Zw~102      & Oct 2007    & 4200         & 1.03--1.25    & 2.0--2.8    & 470                   \\
\hline
\end{tabular}
\end{center}
\end{table*}

\subsection{Data reduction} 
\label{SubSection:dataprocess} 

The data have been processed using standard IRAF\footnote{IRAF is distributed
by the National Optical Astronomy Observatories, which are operated by the
Association of Universities for Research in Astronomy, Inc., under cooperative
agreement with the National Science Foundation.} tasks.

The first step in the data reduction was to remove the bias pedestal level,
whose value was computed using the overscan regions in the science
frames, and subtracted out. As the bias frames showed some smooth large-scale
structure, we averaged all of them to obtain a master bias, to which we
subtracted its  mean value. This master bias was then subtracted from all the
overscan-corrected frames.

Next, apertures were defined and traced on the detector. The apertures need to
be traced and defined in well exposed frames: sky-flats taken at the beginning
or at the end of the night were used for that. We defined the apertures using
the IRAF task \emph{apall}; this task first finds the centers of each fiber
(the emission peak) along the spatial axis at some specified position, and
then the user is prompted for the size of the extraction window, 
which we set to 7 pixels, a width that, in our dataset, maximized the
overall signal-to-noise ratio (cross-talk is negligible, see \citealt{Murphy2011}). 
The apertures were then traced by fitting a
polynomial to the centroid along the dispersion axis. A fifth degree Legendre
polynomial was found to provide good fits, with a typical RMS of about 0.02
pixels.

Once the apertures were defined and traced in the continuum frames, we used
again \emph{apall} to extract them in all the science frames. The extraction
consists of summing, for each fiber, the pixels along the spatial direction
into a one-dimensional spectrum.  The output is thus a ``row-stacked
spectrum'' (RSS) of the 244 spaxels (239 for the 2008 March run).

We then performed the wavelength calibration. First, in the comparison spectra
we identified several emission features of known wavelength in a reference
fiber. Second, using \emph{identify}, a polynomial was fitted along the 
dispersion direction; the standard deviation (RMS) of the polynomial fit gives
an estimate of the uncertainty in the wavelength calibration. We  obtained a
typical RMS of about 0.01 \AA\ by fitting a fifth degree Legendre polynomial.
A total of 13 lines were fitted. Finally, with \emph{reidentify} we
identified the emission lines in all the remaining fibers of the arc frame,
using the selected one as a reference.  

Subsequently we corrected for throughput (spaxel-to-spaxel variations) by
using the sky flat exposures. This step was carried out with the task
\emph{msresp1}.

In each spectrum, the sky background was computed by taking the median of 
$\sim 40-50$ spaxels in the periphery of the FOV where the galaxy line and 
continuum emission is negligible, and subtracted out.
The spectrum was then corrected for atmospheric transparency 
variations, gauged by means of the flux of the guide star in the guider saved
frames, as explained in \cite{Blanc2009}, and for atmospheric extinction
(the guide star exposures were also used to estimate the seeing during the
observations, see Table~\ref{Table:log}).

The three spectra for each dithering position were averaged together
using the IRAF task \emph{imcombine}, and then stacked 
to form a final spectrum of 732 rows (717 for the 2008 march observations).

\subsection{Flux calibration} 
\label{SubSection:FluxCalibration} 

Spectra of several spectrophotometric standard stars were obtained and reduced
in the same way as the galaxy spectra.

\begin{figure}
\includegraphics[width=0.50\textwidth]{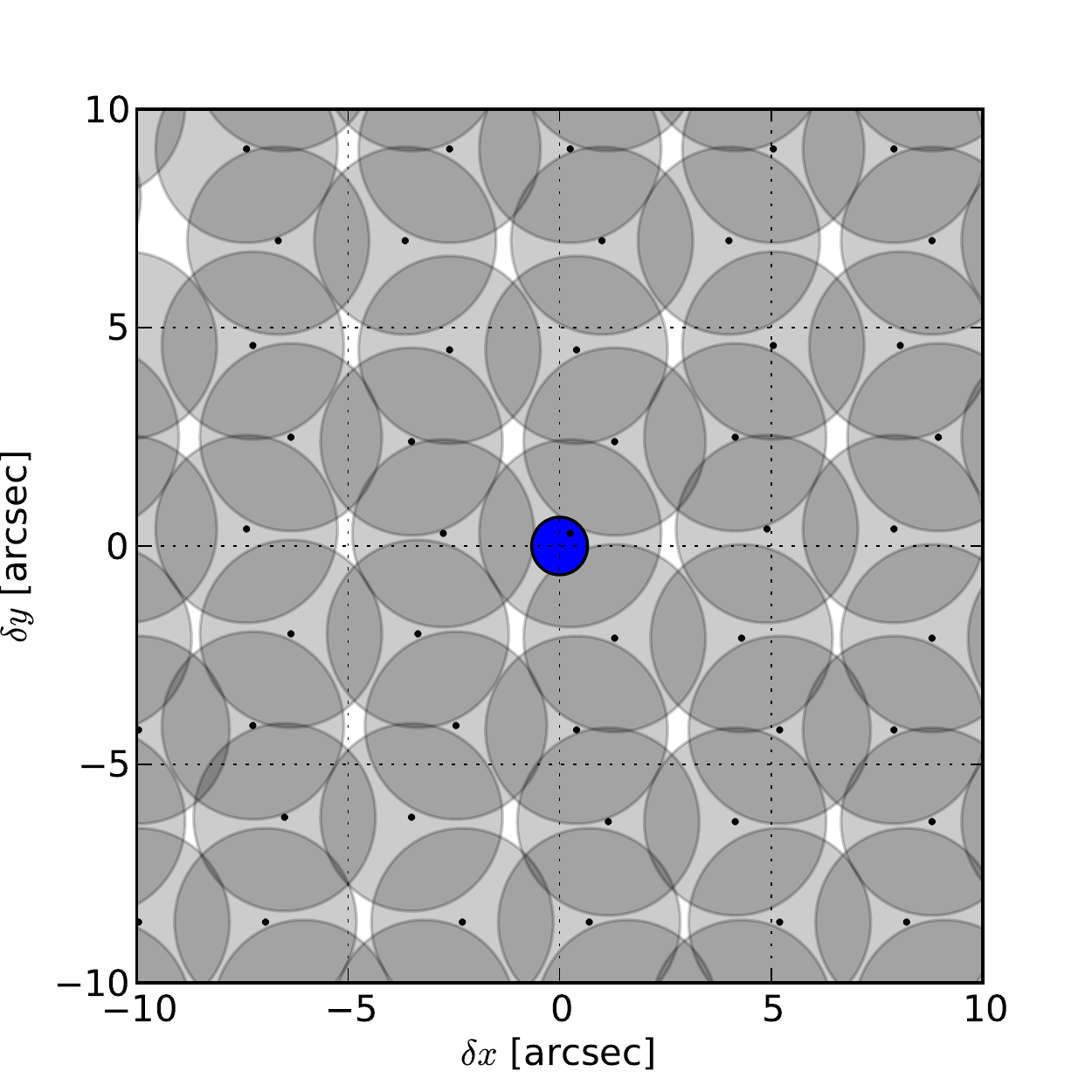}
\caption{
Size and position of the fibers, centered on the
position of the star (G 191-B2B). The fiber diameter is $4\farcs2$, the
distance between adjacent fibers in the same exposure is $7\farcs5$. The
blue circle shows the computed FWHM of this star ($\simeq1\farcs4$).
}
\label{Figure:fiberpos_ole}
\end{figure}

The only remarkable difference is that the standard stars were observed 
with a six-position dithering pattern, which ensures that only about 3\% of 
FOV is ``lost'' in inter-fibers gaps, while about two thirds of the area is
covered by two overlapping fibers, as shown in
\ref{Figure:fiberpos_ole}.
Combined with the PSF undersampling, this dithering pattern does 
not allow an accurate direct flux estimation by simple summation of all the 
relevant spaxels.

If we summed together all the spaxels with sufficiently strong stellar
continuum (typically about 30 spaxels  in the row-stacked final spectrum), we
would overestimate the actual stellar spectrum, because of the dithering
geometry. Moreover, the overestimation factor is in principle
wavelength-dependent, because of both the increase of the stellar PSF width
towards bluer wavelengths, and the shift of the light centroid due to
differential atmospheric refraction.

To better understand it, we can think of a hypothetical star with an extremely
narrow PSF, whose blue centroid happens to fall on one of the inter-fibers
gaps while the red centroid falls on a spatial location covered by two
overlapping fibers: we would measure very little or zero blue flux, while the
red flux would be overestimated by a factor of two.

\begin{figure}
\includegraphics[width=0.50\textwidth]{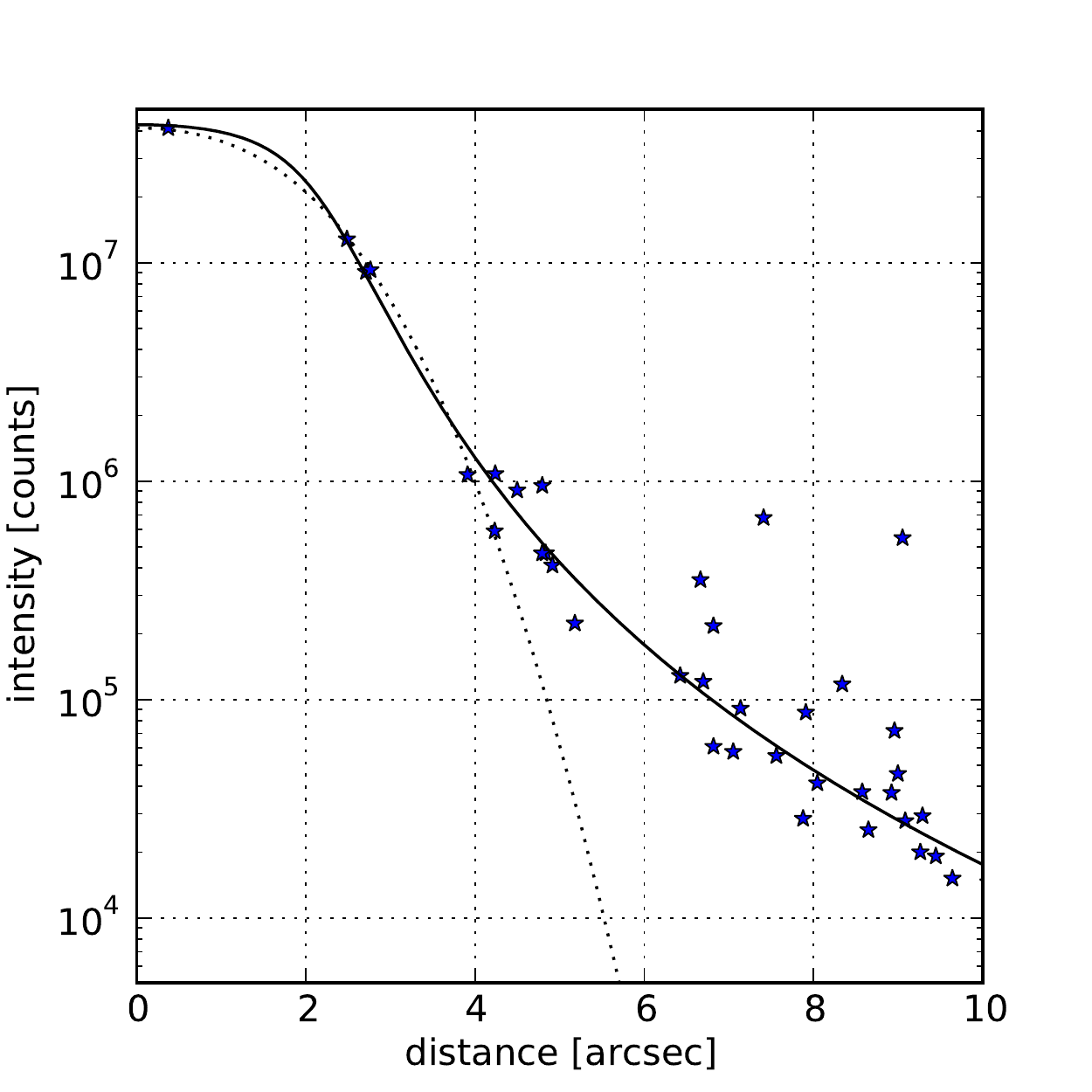}
\caption{
Measured and fitted intensity for the source G 191-B2B. The X axis
shows the distance to the fitted source center, Y the
intensity. Stars indicate the positions of the fiber centers. The curves are
the best Moffat (solid) and Gaussian (dotted) fits (both integrated over the
fiber area).
}
\label{Figure:moffat_ole}
\end{figure}

To assess the importance of this effect, we simulated the response of each
fiber and did a $\chi^2$ fit to determine the PSF parameters. The fiber
response is calculated by a Monte Carlo integration of the PSF over the circle
area of the fiber. For the PSF, a Moffat profile is chosen to follow the
instrument response to the outer wings, as shown in
Fig.~\ref{Figure:moffat_ole}. Fit parameters were the total flux, source
position and the PSF width $\alpha$ and exponent $\beta$. The Moffat fit was
done individually for each wavelength bin.

\begin{figure}[h]
\centering
\includegraphics[angle=0, width=0.50\textwidth, bb=50 50 570 740]{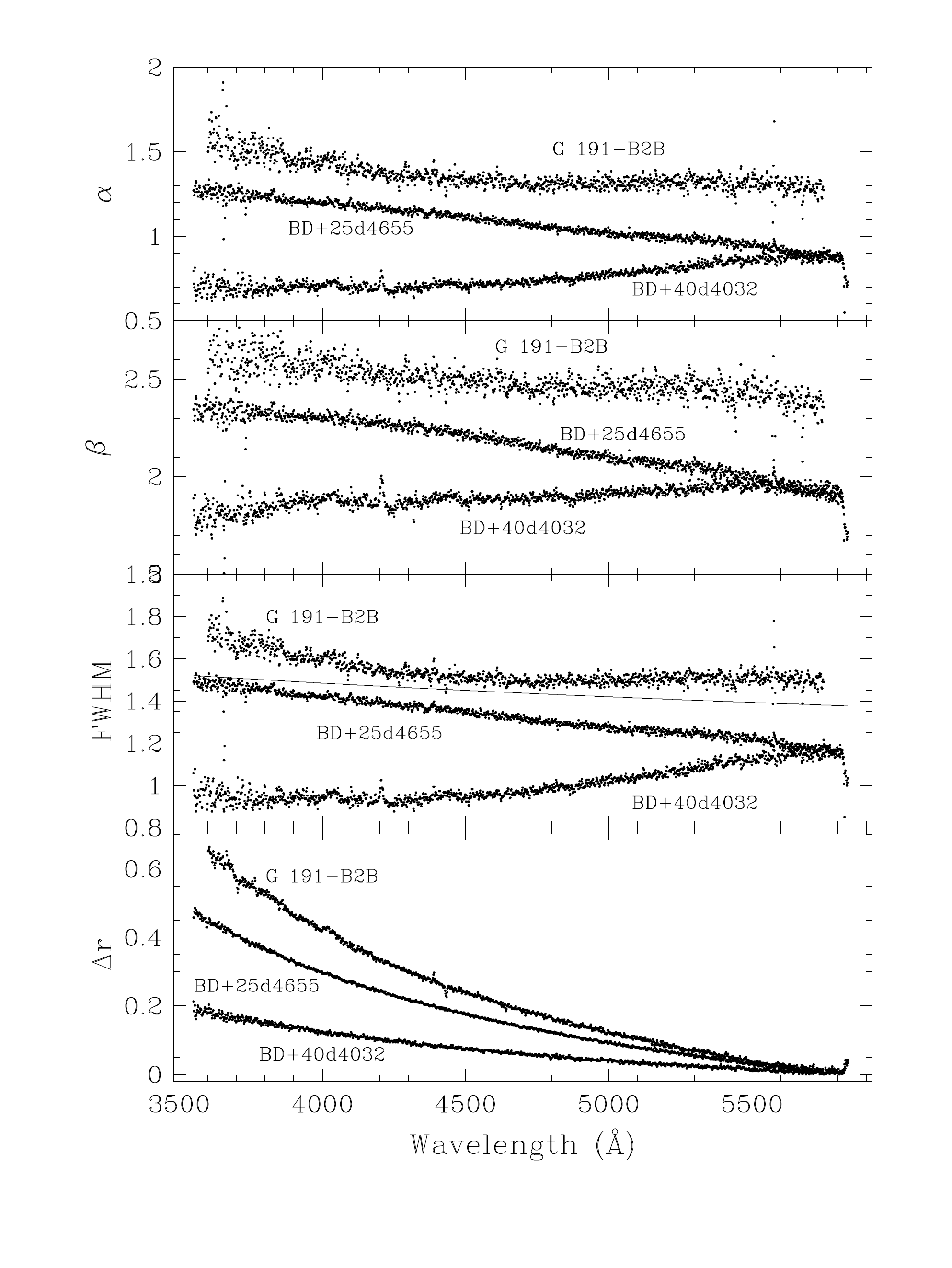}
\caption{The Moffat $\alpha$ and $\beta$ parameters are plotted as a function
of the wavelength for three spectrophotometric standards. The bottom panels
show the computed FWHM (the solid line represents the FWHM $\propto \lambda^{-0.2}$
relation) and the shift of the centroid with respect to a reference point at
$\lambda=5750$ \AA.}
\label{Figure:MoffatParameters}
\end{figure}

We found a tight correlation between the Moffat function parameters and
wavelength, as depicted in Fig.~\ref{Figure:MoffatParameters}; also, the
observed shift of the PSF centroid agrees well with the shift  computed
theoretically from the airmass and the parallactic angle of the
spectrophotometric standard stars spectra.

\begin{figure}[h]
\centering
\includegraphics[angle=270, bb=100 50 540 750, width=0.50\textwidth]{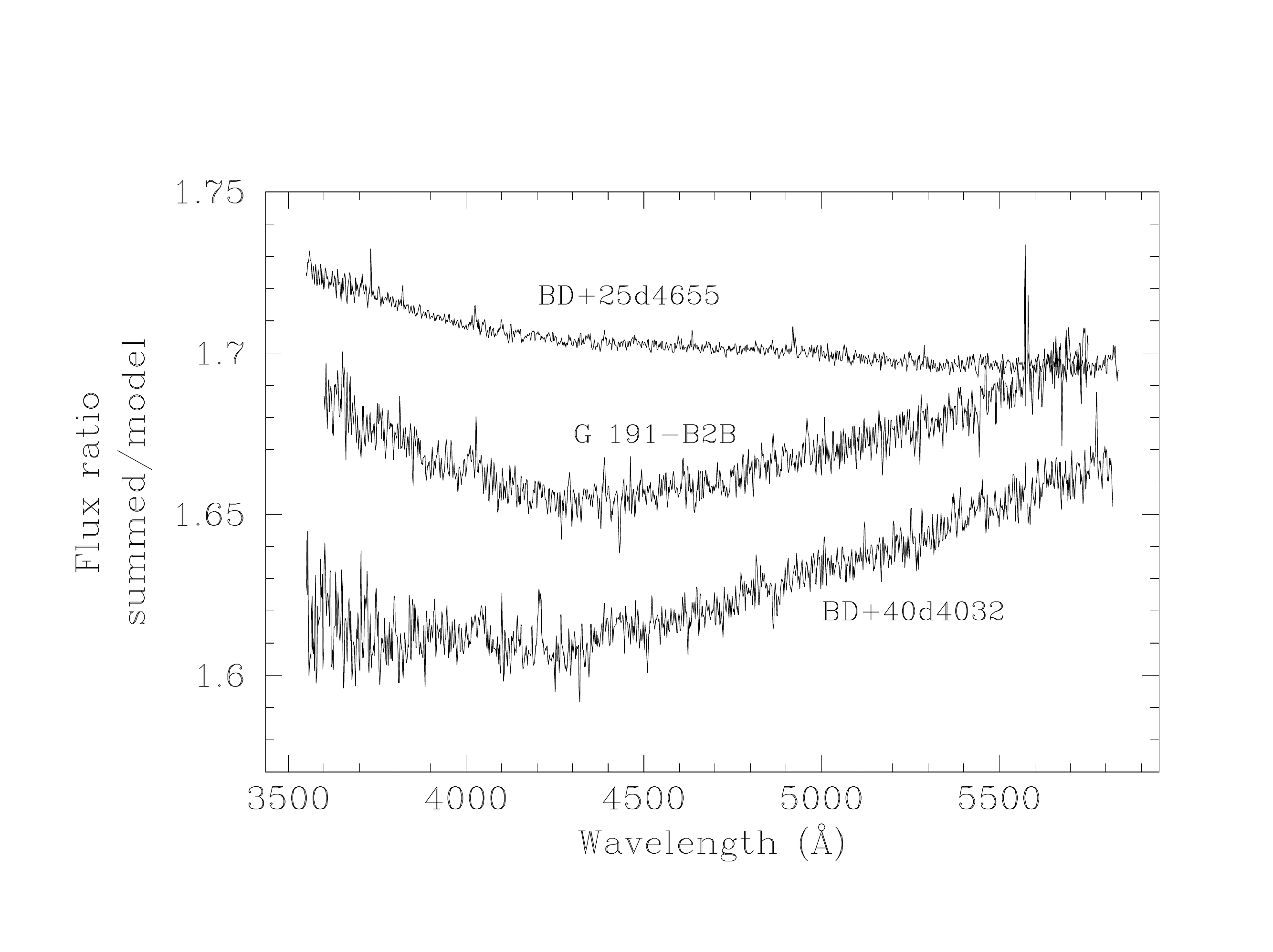}
\caption{Ratio between the flux computed by a direct sum of the spectra of all
spaxels with sufficiently good signal, and the flux derived by the Moffat fit
to the data, as a function of wavelength, for three spectrophotometric
standard stars. The direct integration overestimates the stellar flux
by a factor of about 5/3, which depends very little on wavelength.}
\label{Figure:Overestimate}
\end{figure}

However, the overestimation factor turns out to have only a very small 
($<0.03$ mag) dependence on wavelength (Fig.~\ref{Figure:Overestimate}),
because of the relatively poor seeing, which spreads out the star flux over a
large number of spaxels, and of having observed the standard stars at low
airmasses.

To derive the final flux calibration function, for each spectrophotometric 
standard star we summed together all spaxels with sufficient strong signal in
the stacked  spectrum, and  applied the overestimating factor correction. 
Then the IRAF tasks \emph{standard} and \emph{sensfunc} were used to compute
the  calibration function, after combining together the data for the
different  spectrophotometric stars.

By comparing the sensitivity curves for the different standard stars in the
two observing runs, we can estimate that the relative uncertainty on the
calibration factor is generally less than 2\%, except at the blue end of the
spectrum, where the uncertainty increases to about 7\%.

As the diameter of the fibers and the average seeing values are much larger
than differential atmospheric refraction (DAR) effects, no correction for DAR
was applied to the data.

\subsection{Emission-line fit}
\label{SubSection:LineFit}

\subsubsection{Recovering the {\em pure} emission-line spectrum}

The measurement of the emission-line properties (e.g. fluxes, equivalent
widths, or velocity dispersion) provides a great deal of information in regard
to the nebula physical conditions and the properties of the ionizing stellar
population. However, and in particular when dealing with SF galaxies, the 
derivation of accurate emission-line fluxes and equivalent widths is not free
from problems. At any given spatial position, the observed galaxy spectrum is
the sum of the nebular and stellar emission; underlying stellar absorption can
under certain circumstances dramatically decrease the measured  intensity and
equivalent width of Balmer emission lines. 

Therefore, in order to obtain precise measurements of the
emission-line parameters of the ionized gas, the flux emitted by the stellar
population must be modeled and subtracted. In the case of IFU data, this
process must be done for every spaxel, i.e., the technique chosen to model the
stellar population has to be applied to each individual spectrum of the IFS
mosaic.

In order to model the stellar population, we used the penalized pixel-fitting
(pPXF) method\footnote{URL: http://www-astro.physics.ox.ac.uk/~mxc/idl/}
developed by \cite{Cappellari2004}; the pPXF software allows to simultaneously
fit the optimal linear combination of stellar templates to the observed
spectrum and to derive the stellar kinematics, using a maximum likelihood
approach to suppress noise solutions.  We made use of the MILES stellar
library~\footnote{URL: http://miles.iac.es/} \citep{SanchezBlazquez2006}, which
contains a total of 985 stars spanning a large range in stellar age and
metallicity.

As a first step, we produced, for every galaxy, a high signal-to-noise
integrated spectrum, adding up all the spectra of the science spaxels. Using 
this spectrum we find the "optimal stellar template" for every object, i.e.,
the selected stars from the MILES library that produce the best fit. The fit
to the stellar continuum was done after masking out the gas emission-lines
expected in the spectral range under study, as well as residuals from skyline
subtraction. The number of stars combined to create the "optimal stellar
template" for each galaxy was as follows: 17 stars for II~Zw~33, 20 stars for
Haro~1, 22 stars for NGC~4670, 20 stars for Mrk~314, and 25 stars for
III~Zw102. 


After that, the pPXF method was run on the spectra of each individual
spaxel, using the "optimal stellar template" and allowing the weights of each
star to vary. The application of the "optimal stellar template" to each
individual spectrum considerably speed up the fitting process. As the
automatization of this process could lead to bad fittings, a careful
individual inspection by eye was done.  

Once the stellar continuum has been successfully reproduced by a sum of
templates for every spaxel, it was subtracted from the original spectrum to
get the \emph{pure} emission-line spectrum.  

We must remark here that the fit to the stellar continuum is not meant to 
provide a physical description of the stellar populations making up the
galaxy, a topic which is outside the scope of this paper, but that the purpose
of this modeling is to obtain a good representation of the underlying stellar
continuum in order to decouple it from the emission lines produced by the
ionized gas, and to derive more accurate values of emission-line fluxes. 

The stellar continuum fit only works out in the central regions of 
the galaxies, while it fails, producing meaningless results, in the outer regions 
where the signal-to-noise of the continuum is too small.

\subsubsection{Fitting the lines}

Emission-line fluxes for every spaxel were measured on the \emph{pure}
emission-line spectrum. However, the line emission extends well beyond
the galaxy region in which the \emph{pure} emission-line spectrum could
be computed.

To compute the fluxes in those spaxels outside such region, the original 
stacked spectra were used, fitting the continuum, typically 30--50
\AA\ on both sides, by a straight line.

In order to measure the relevant parameters of the emission lines (position,
flux and width), they were fitted by single Gaussians. The fit procedure was
carried out by using the the Levenberg-Marquardt non-linear fitting algorithm 
implemented by C.~B. Markwardt in the \emph{mpfitexpr} IDL library
\footnote{URL: http://purl.com/net/mpfit}. The
[\ion{O}{iii}]~$\lambda\lambda4959, 5007$  doublet was fitted imposing that
both lines have the same redshift and width. 
Uncertainties on the fit output parameters were computed adopting a simplified 
poissonian model on the non-flux-calibrated spectra, where the readout noise
was set equal to the rms of the nearby continuum, and the gain set to 1.

Criteria such as flux, error on flux, velocity and width were used to do a
first automatic assessment of whether to accept or reject a fit. For instance,
lines with too small (less than the instrumental width) or too large widths
were flagged as rejected, as well as lines with an error on the flux more than
about 10\% of the flux (the exact limits depending on the specific line and on
the overall quality of the spectrum). Such criteria were complemented by a
visual inspection of all fits, which led to override, in a few cases, the
automated criteria decision (by accepting a fit flagged as rejected, or vice
versa).

\subsection{Creating the 2D maps}
\label{SubSection:CreatingMaps}

The emission-line fit procedure gives, for each line and for each line
parameter (for instance flux), a table with the fiber ID number, its 
coordinates, the measured value and the acceptance/rejection flag. 

The gaussian fit parameters for the continuum-subtracted and the original
spectra were merged together. For the \Hg\ and the \Hd\ Balmer lines, due to
their strong absorption components, only the results from the
continuum-subtracted spectra were used.

Line ratios were computed by simply dividing the fluxes of the corresponding
lines.

To display all these maps, we explored several options. One is to interpolate
the data on a regularly spaced grid, by means of an algorithm such as the 
\cite{RenkaCline1984} two-dimensional interpolation method. However, our own
experience is that the interpolation can produce spurious oscillations,
especially over non-contiguous spaxels, and gives the false impression of a
spatial resolution higher than the actual value.

An alternative is to represent each spaxel with measured data with a circle
with the fiber's diameter at its spatial position. However, the irregular
inter-fiber gaps (due to the fact that the actual position of the fibers is
slightly and randomly offset from the nominal 
one) distract the eye of the viewer and interfere with a comfortable and 
effective visual assessment of the global characteristics of the map.

The solution we adopted was to depict each fiber by means of an hexagon
centered on the nominal (not the actual) position of each fiber  (the mean
difference between nominal and actual positions is about $0\farcs4$ with a
maximum of $1\farcs8$). This we found to be the best compromise between a
truthful representation of the data (as no rebinning or interpolation was
done: the only manipulation was to slightly  change the fibers positions and
shape) and a map which is both pleasant  to the eye and easy to appraise and
interpret.


\section{Results}
\label{Section:Results}

\subsection{Emission-line and continuum intensity maps}
\label{SubSection:IntensityMaps}

IFS provides a simultaneous mapping of the light emitted by a galaxy in a wide
spectral range. Accordingly, from one single exposure we can retrieve maps at
a given wavelength or maps equivalent to narrow- or broad-band images within
the whole observed range. 

\begin{figure*}[htb]
\mbox{
\centerline{
\hspace*{00pt}\subfigure{\begin{minipage}{0.20\textwidth}\qquad\qquad{\large II~Zw~33}\end{minipage}}
\hspace*{00pt}\subfigure{\begin{minipage}{0.20\textwidth}\qquad\qquad{\large Haro~1}\end{minipage}}
\hspace*{00pt}\subfigure{\begin{minipage}{0.20\textwidth}\qquad\quad{\large NGC~4670}\end{minipage}}
\hspace*{00pt}\subfigure{\begin{minipage}{0.20\textwidth}\qquad\qquad{\large Mrk~314}\end{minipage}}
\hspace*{00pt}\subfigure{\begin{minipage}{0.20\textwidth}\qquad\quad{\large III~Zw~102}\end{minipage}}
}} 
\mbox{
\centerline{
\hspace*{0.0cm}\subfigure{\includegraphics[width=0.20\textwidth]{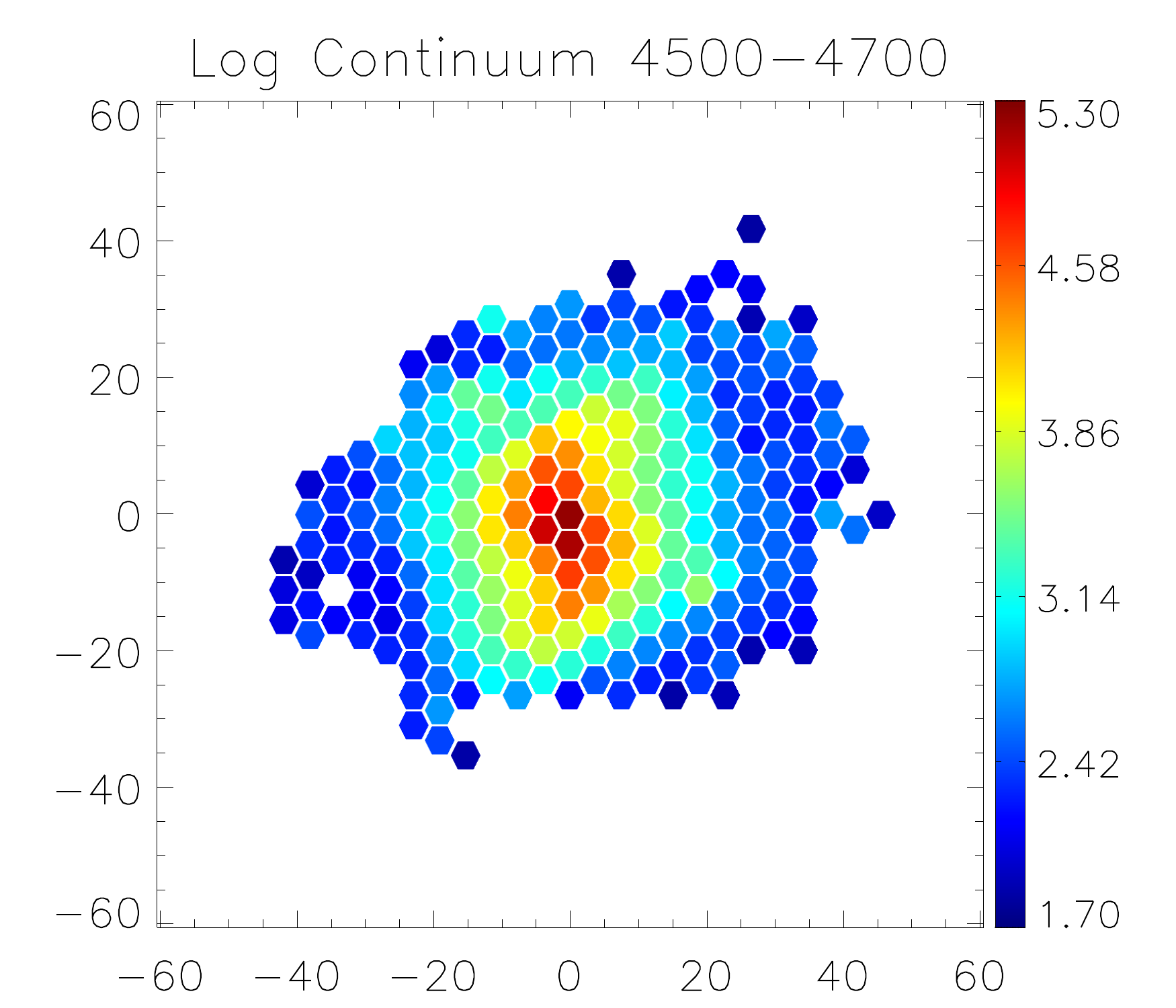}}
\hspace*{0.0cm}\subfigure{\includegraphics[width=0.20\textwidth]{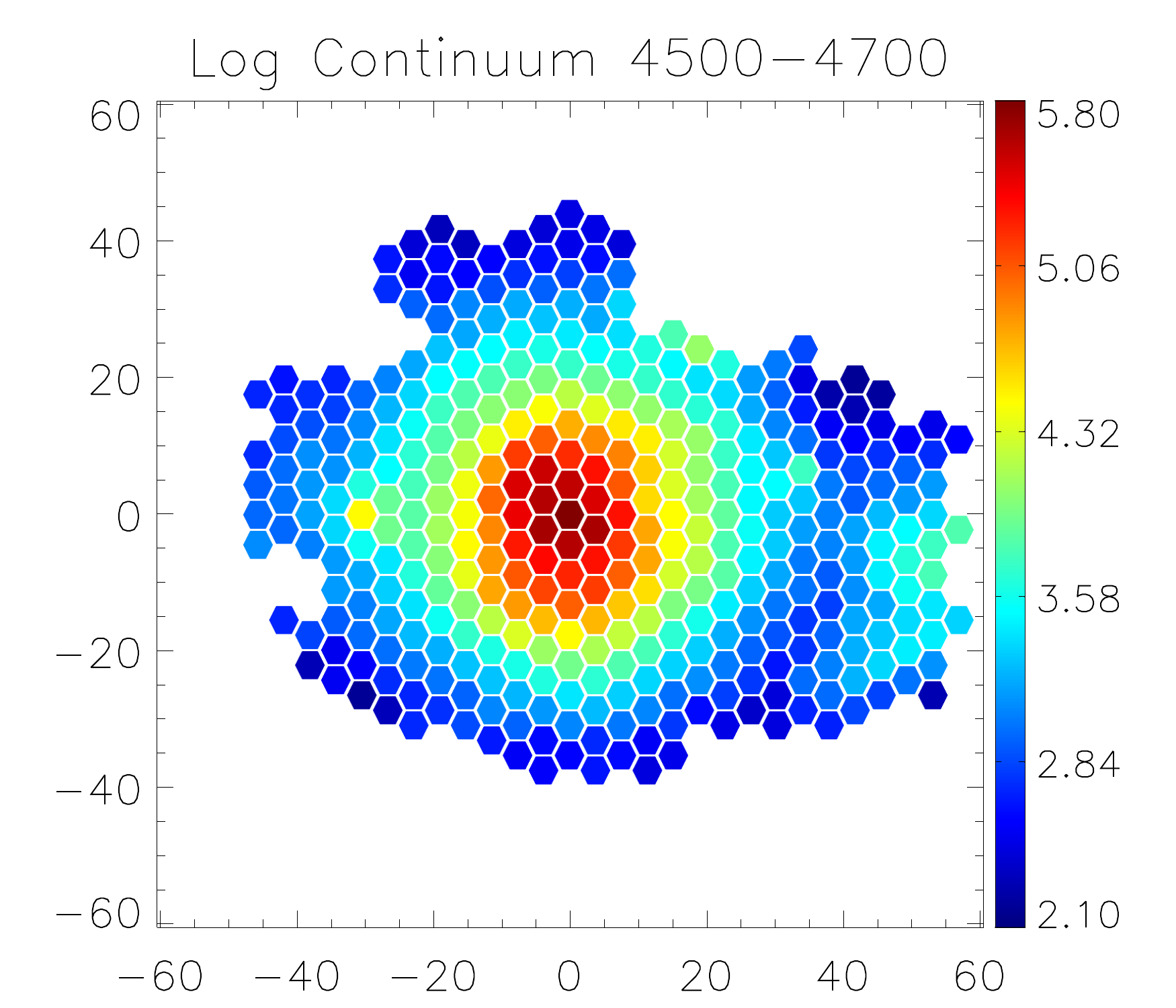}}
\hspace*{0.0cm}\subfigure{\includegraphics[width=0.20\textwidth]{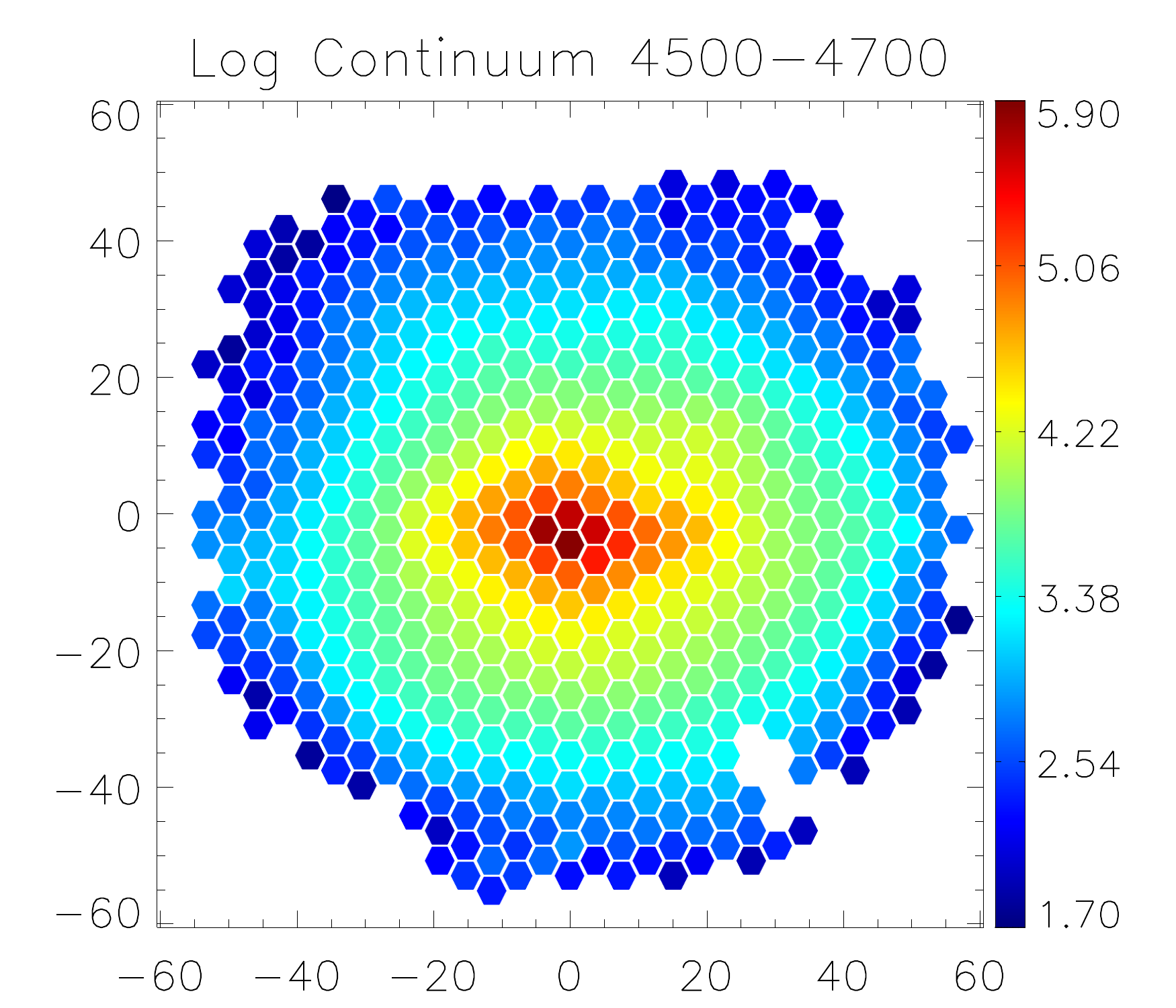}}
\hspace*{0.0cm}\subfigure{\includegraphics[width=0.20\textwidth]{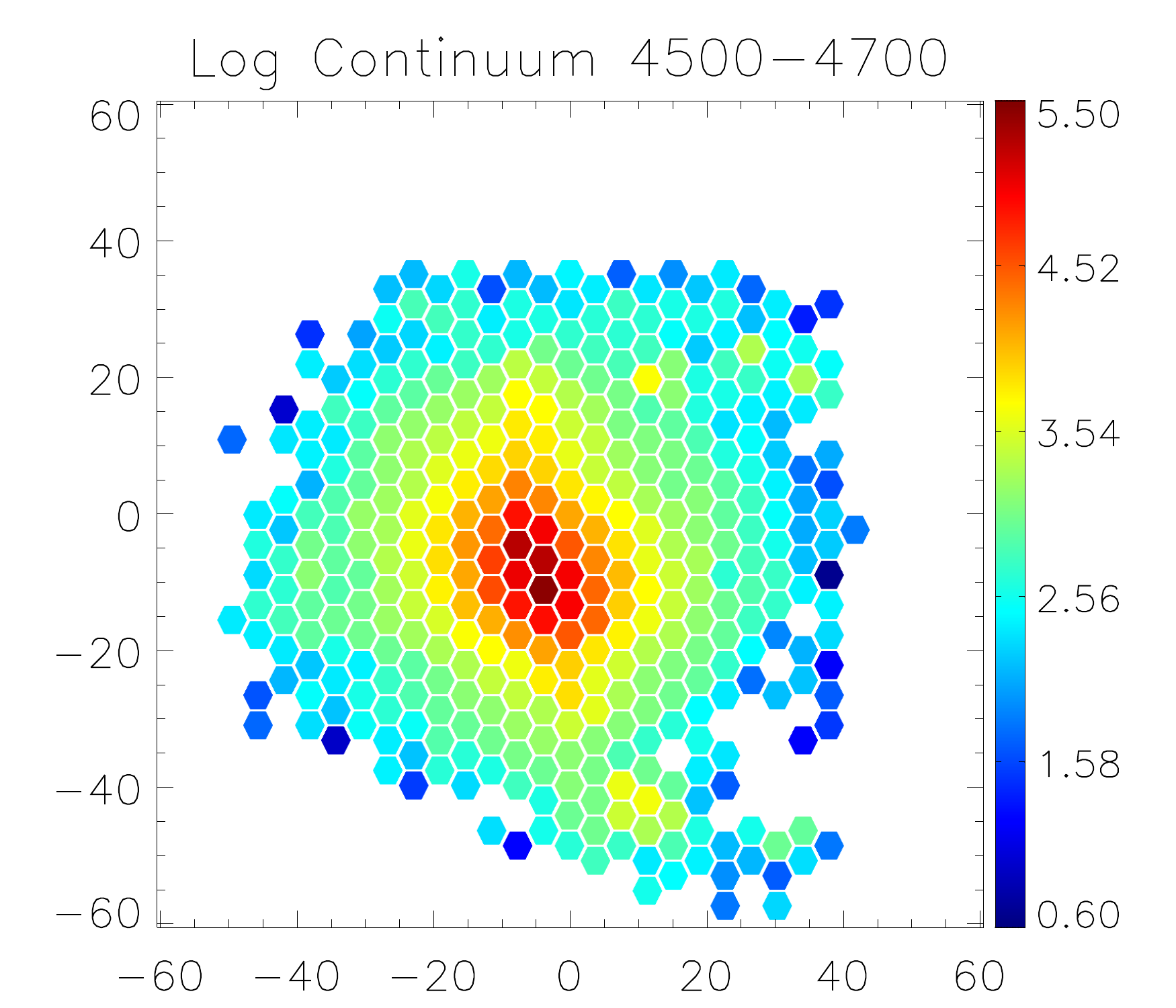}}
\hspace*{0.0cm}\subfigure{\includegraphics[width=0.20\textwidth]{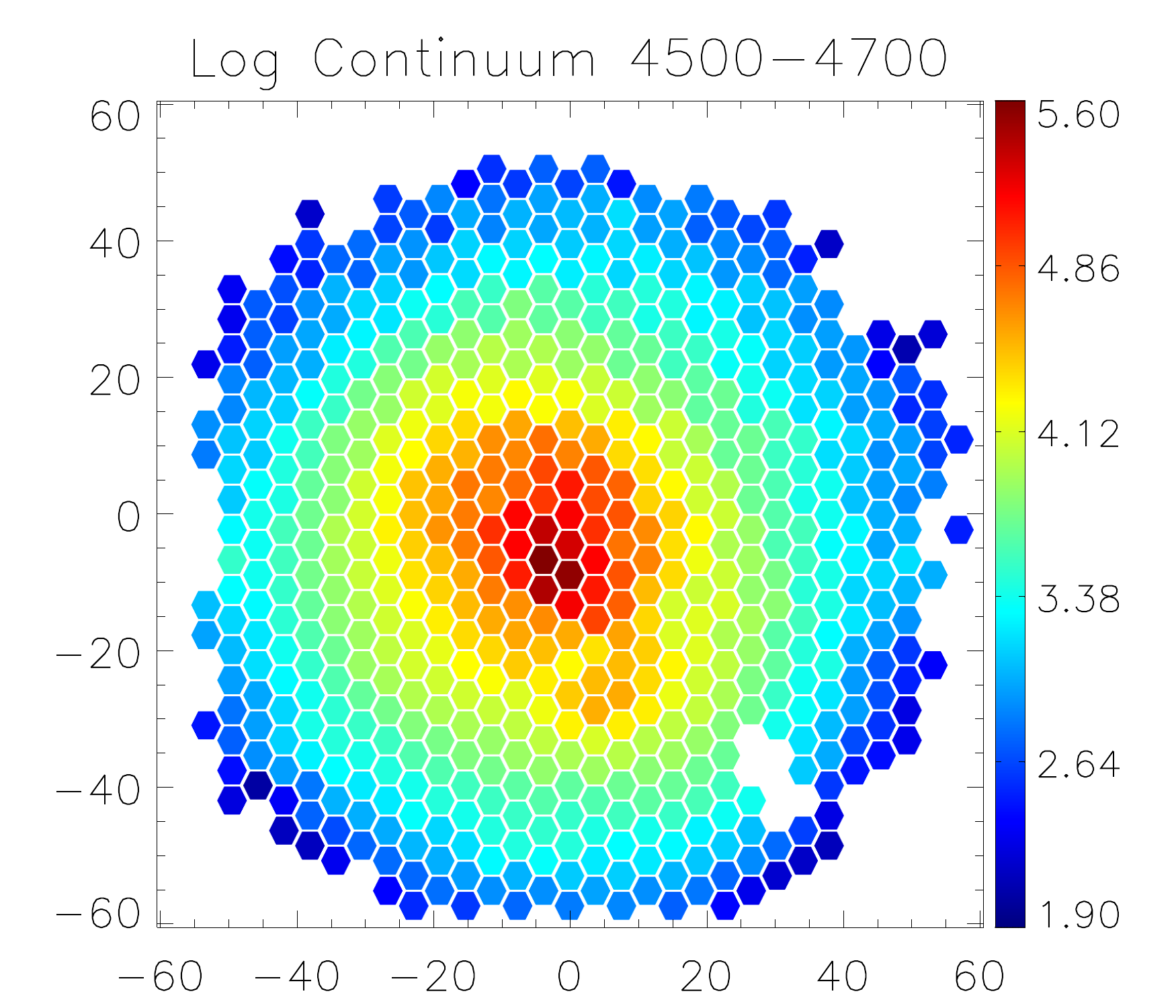}}
}} 
\mbox{
\centerline{
\hspace*{0.0cm}\subfigure{\includegraphics[width=0.20\textwidth]{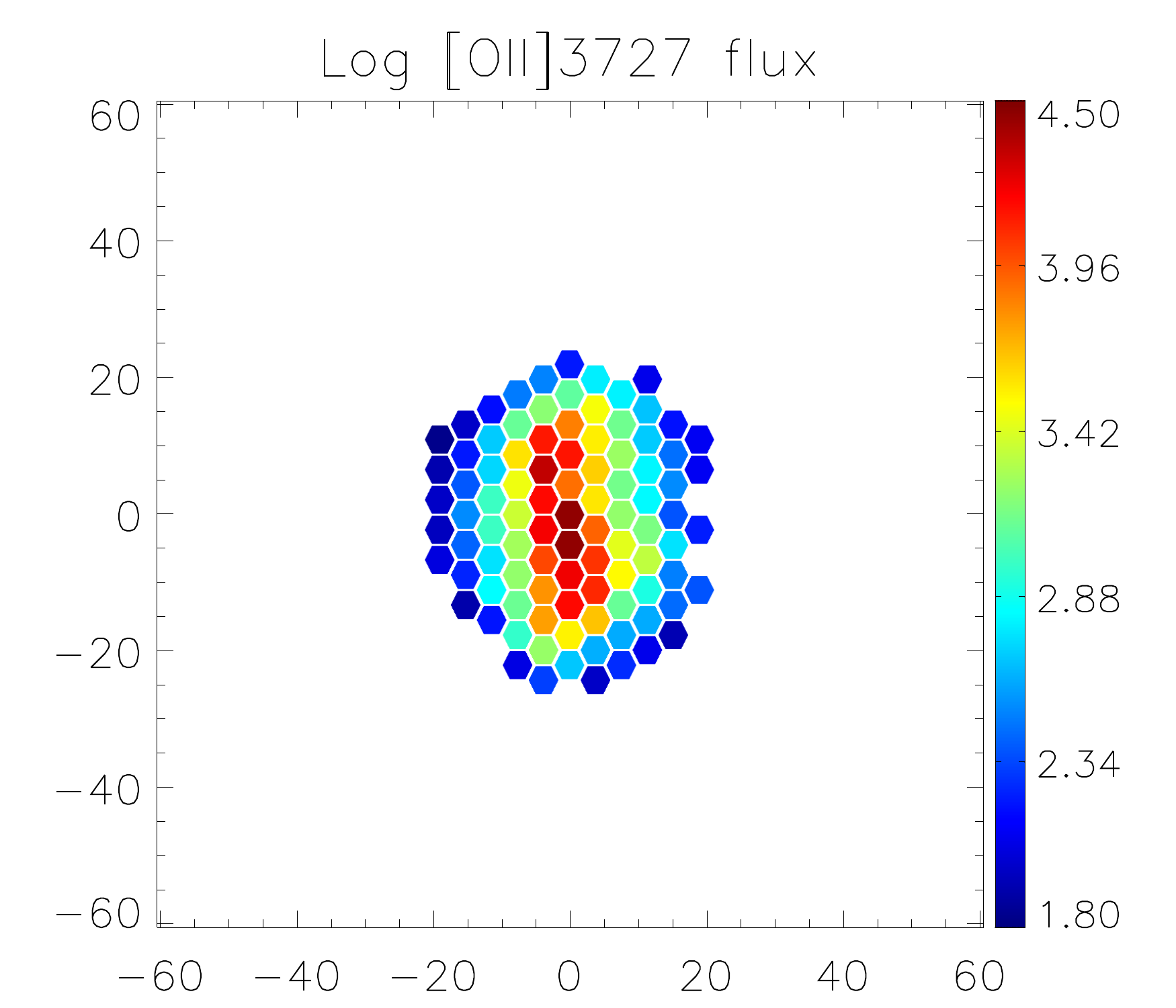}}
\hspace*{0.0cm}\subfigure{\includegraphics[width=0.20\textwidth]{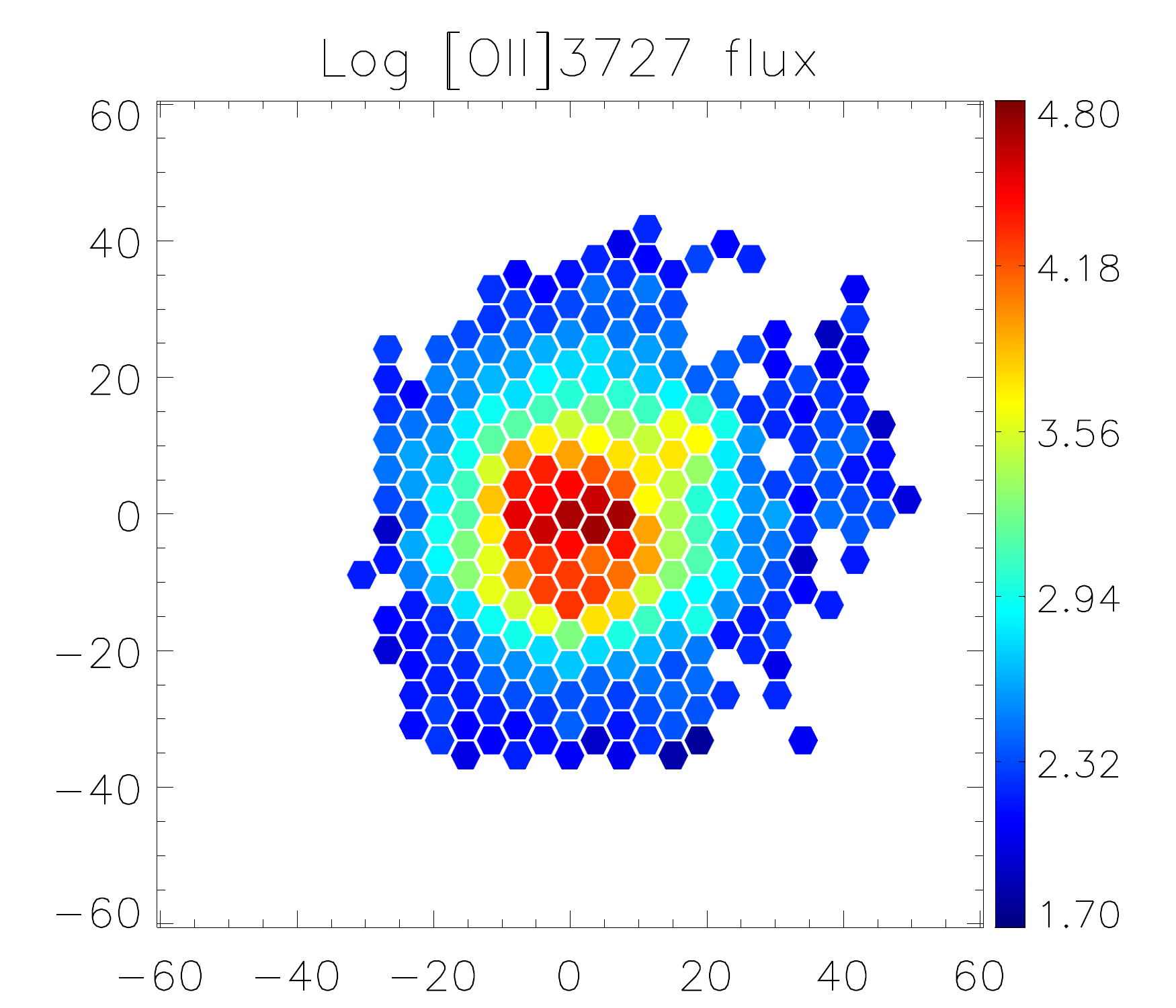}}
\hspace*{0.0cm}\subfigure{\includegraphics[width=0.20\textwidth]{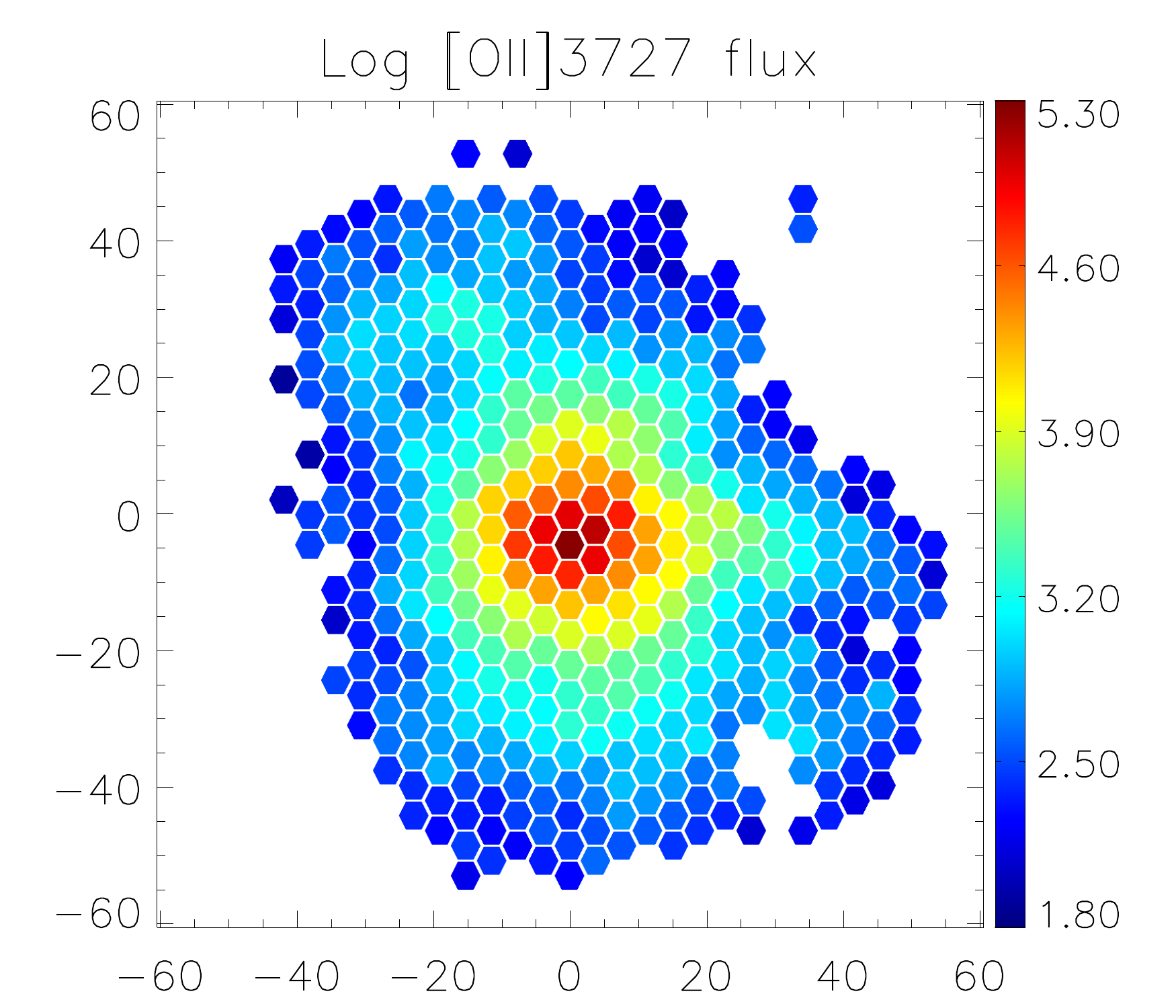}}
\hspace*{0.0cm}\subfigure{\includegraphics[width=0.20\textwidth]{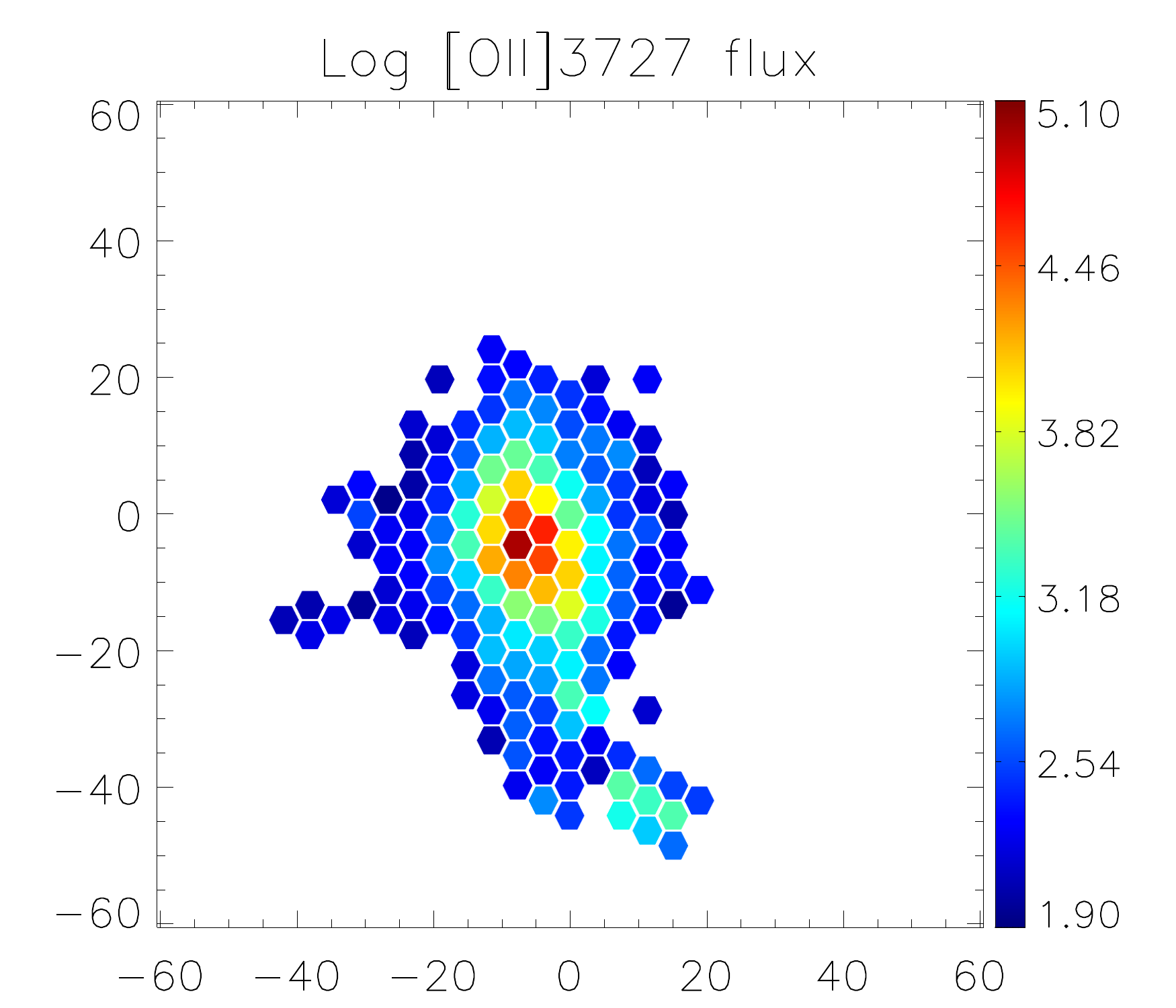}}
\hspace*{0.0cm}\subfigure{\includegraphics[width=0.20\textwidth]{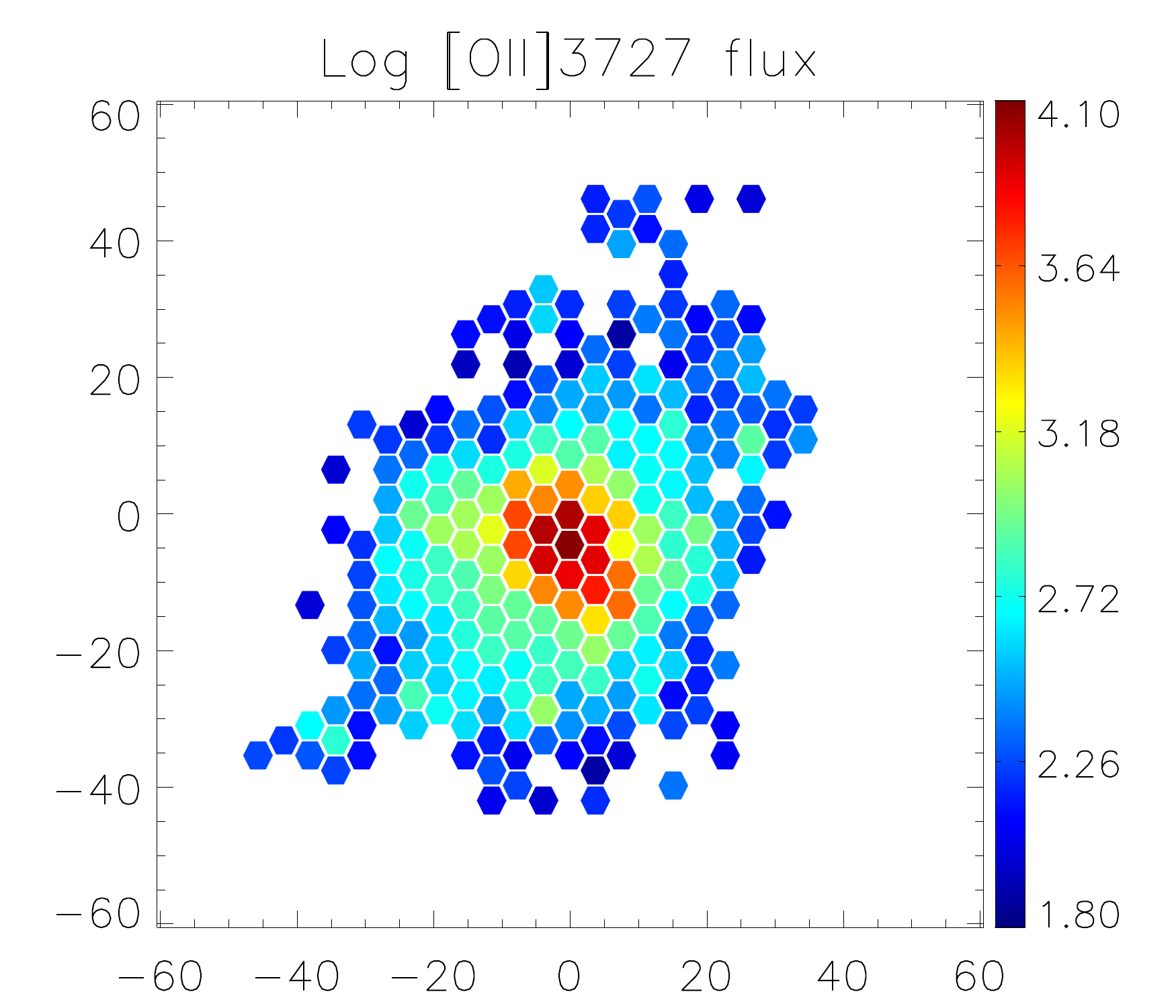}}
}} 
\mbox{
\centerline{
\hspace*{0.0cm}\subfigure{\includegraphics[width=0.20\textwidth]{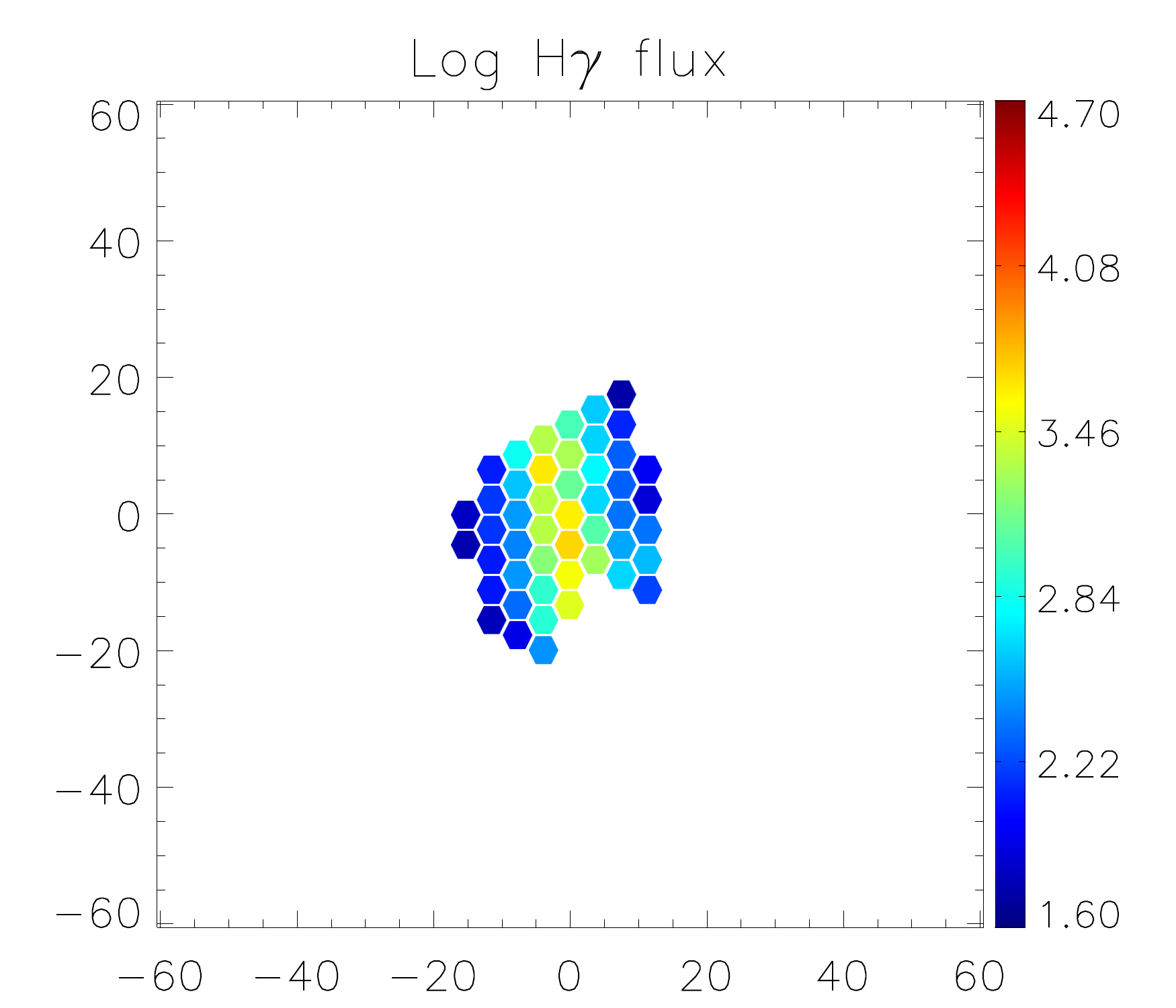}}
\hspace*{0.0cm}\subfigure{\includegraphics[width=0.20\textwidth]{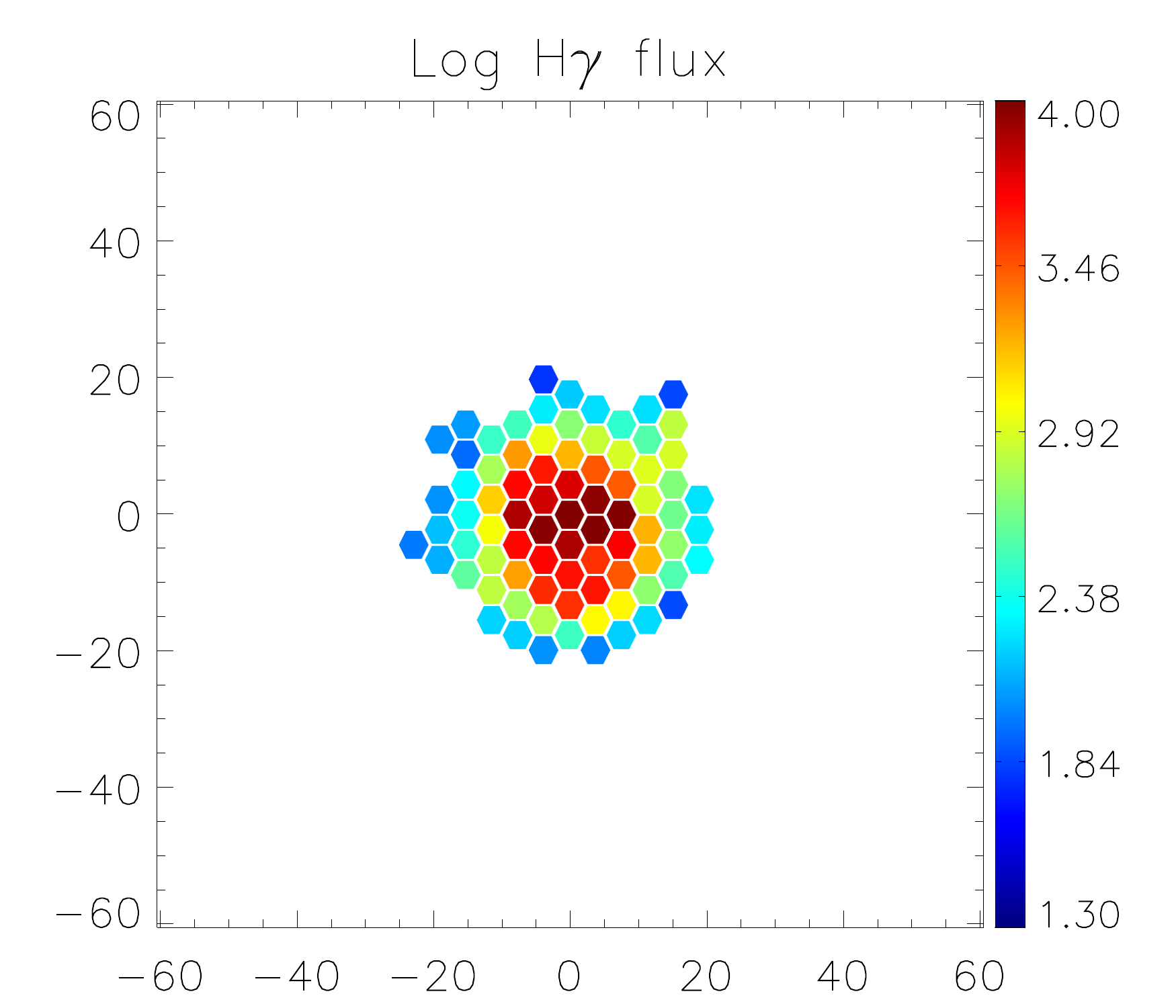}}
\hspace*{0.0cm}\subfigure{\includegraphics[width=0.20\textwidth]{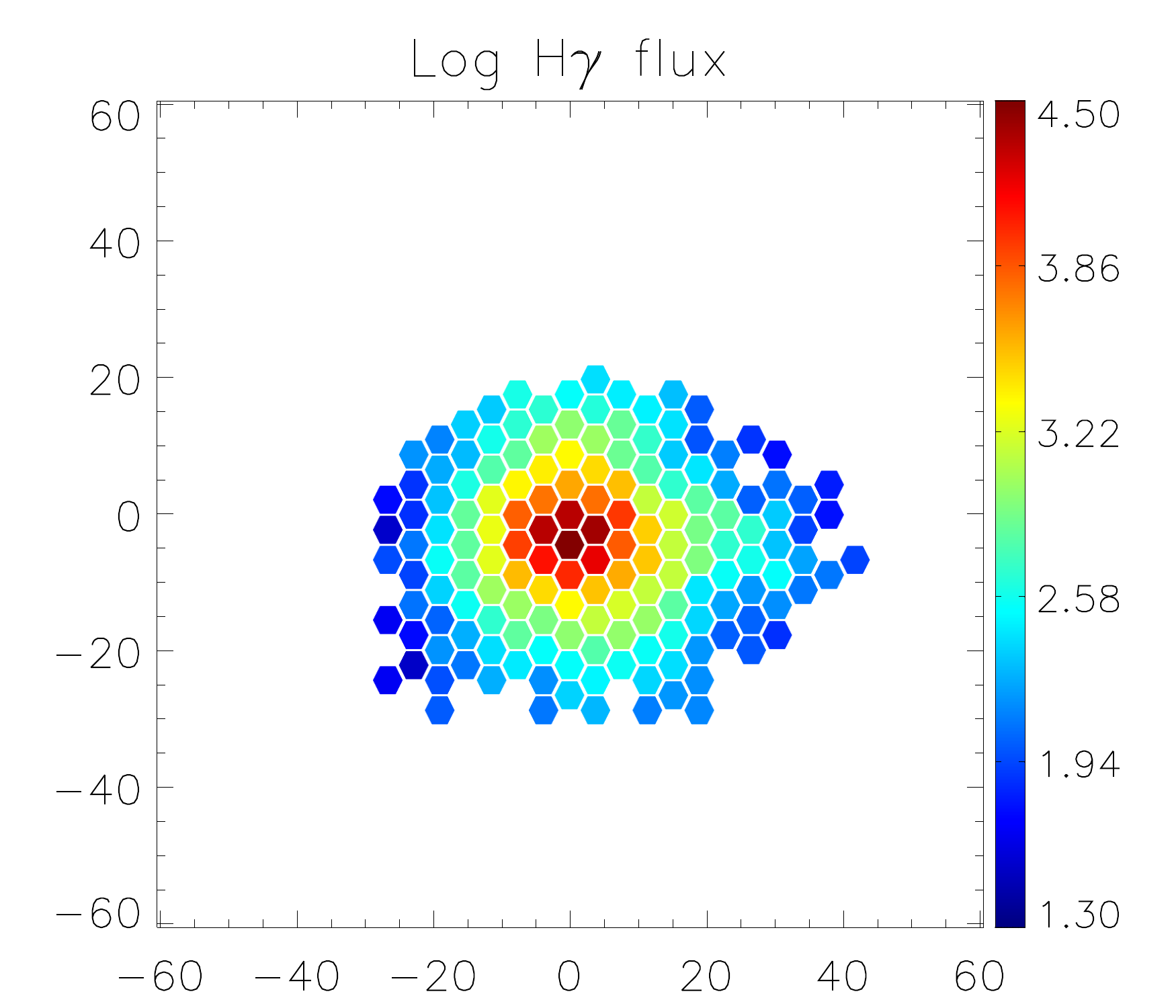}}
\hspace*{0.0cm}\subfigure{\includegraphics[width=0.20\textwidth]{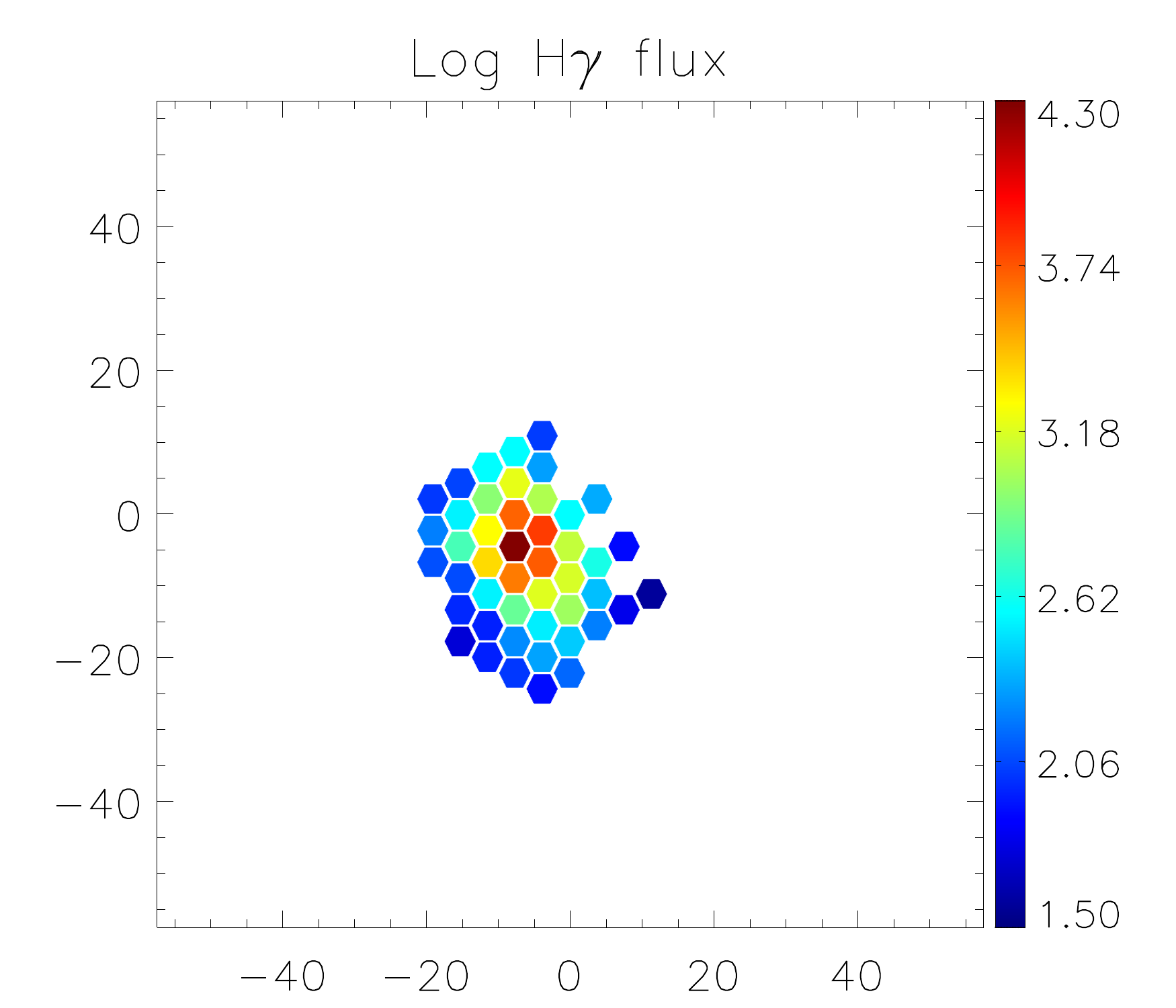}}
\hspace*{0.0cm}\subfigure{\includegraphics[width=0.20\textwidth]{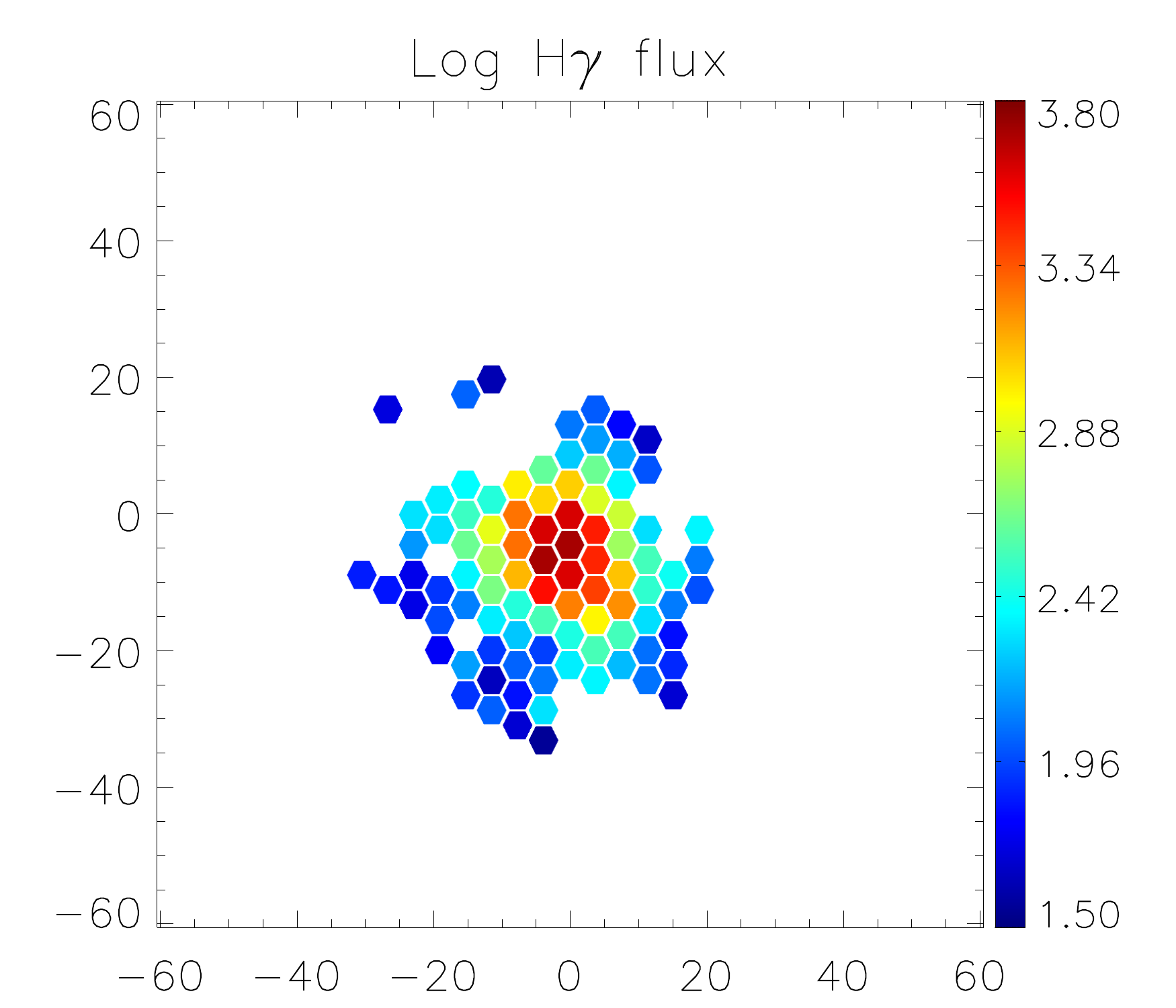}}
}} 
\mbox{
\centerline{
\hspace*{0.0cm}\subfigure{\includegraphics[width=0.20\textwidth]{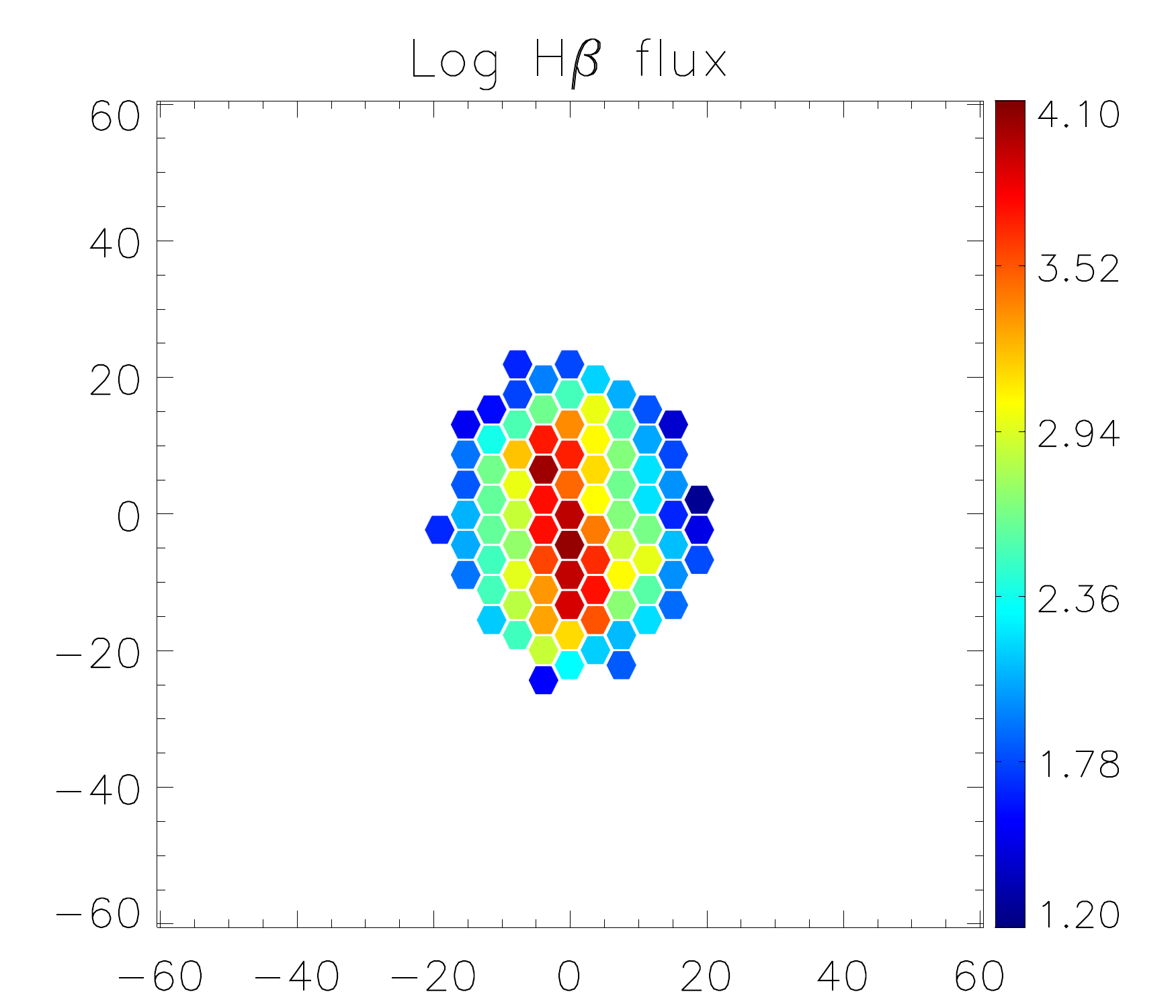}}
\hspace*{0.0cm}\subfigure{\includegraphics[width=0.20\textwidth]{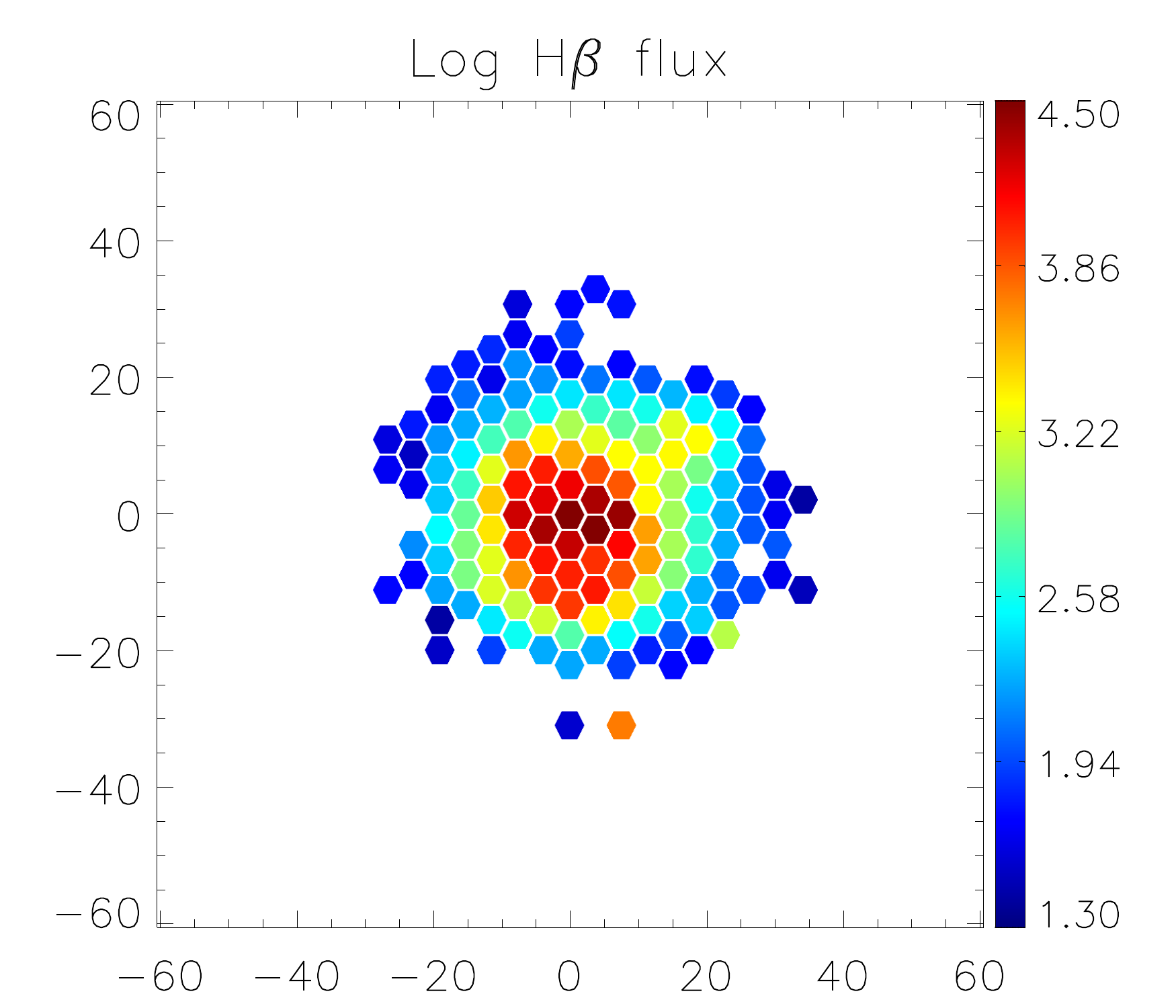}}
\hspace*{0.0cm}\subfigure{\includegraphics[width=0.20\textwidth]{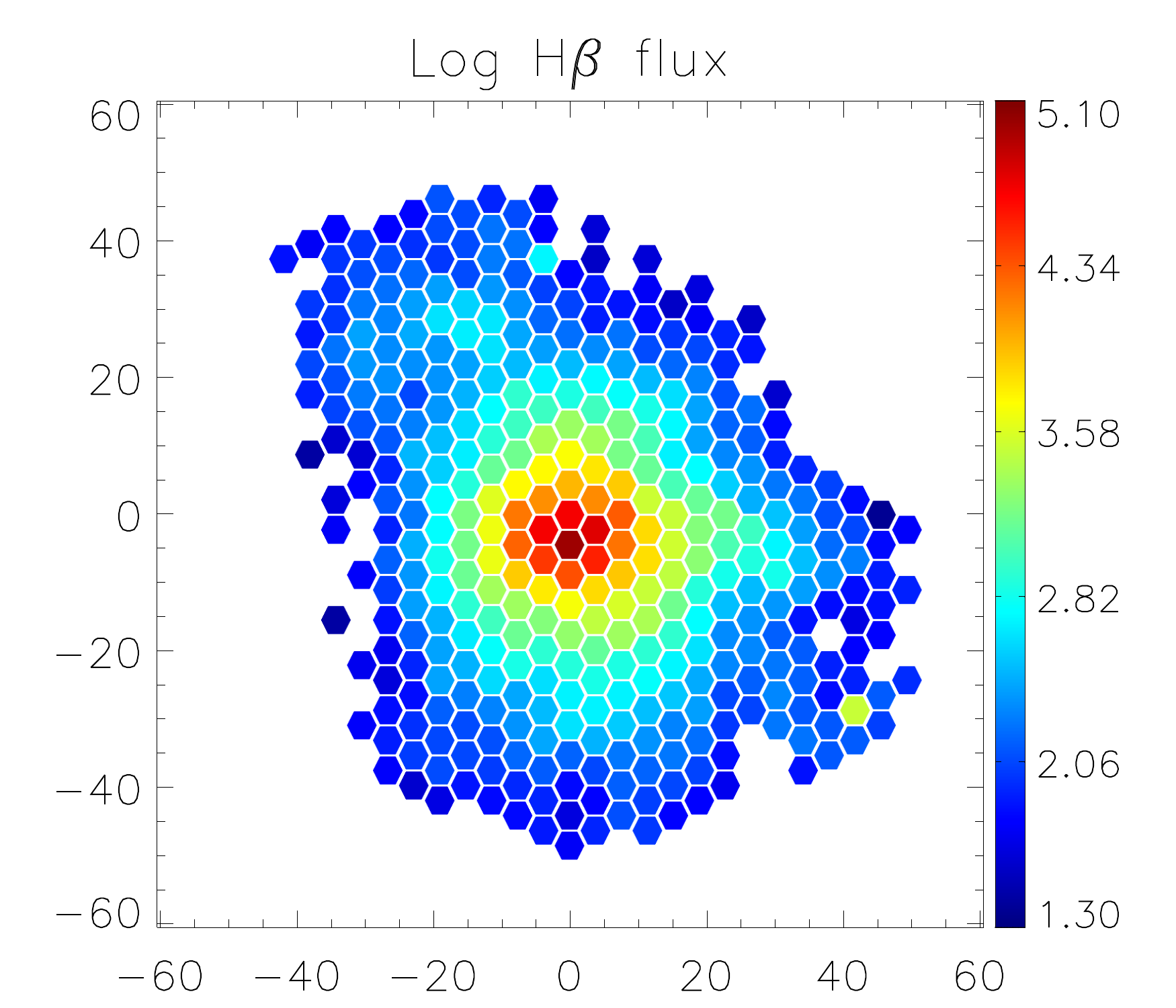}}
\hspace*{0.0cm}\subfigure{\includegraphics[width=0.20\textwidth]{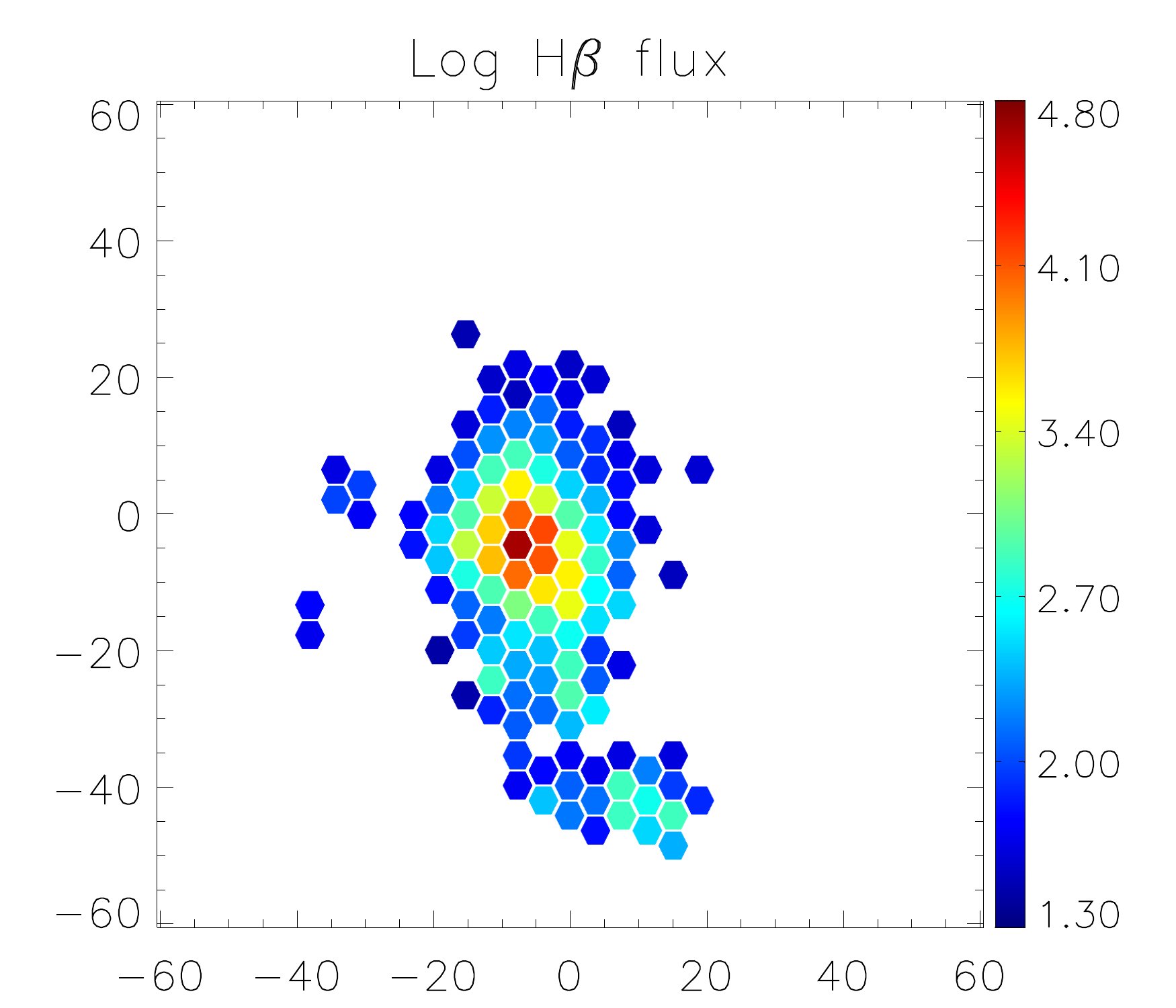}}
\hspace*{0.0cm}\subfigure{\includegraphics[width=0.20\textwidth]{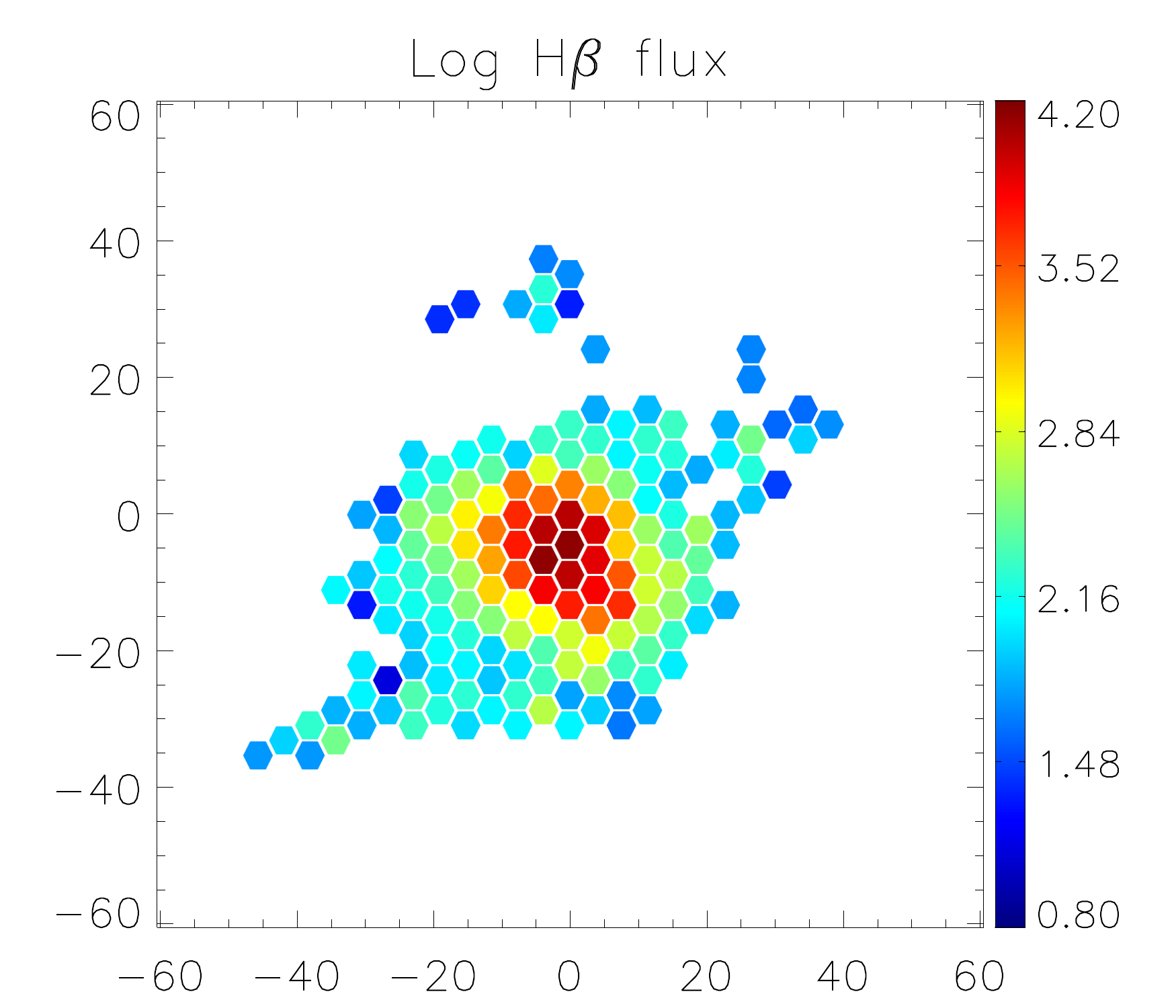}}
}} 
\mbox{
\centerline{
\hspace*{0.0cm}\subfigure{\includegraphics[width=0.20\textwidth]{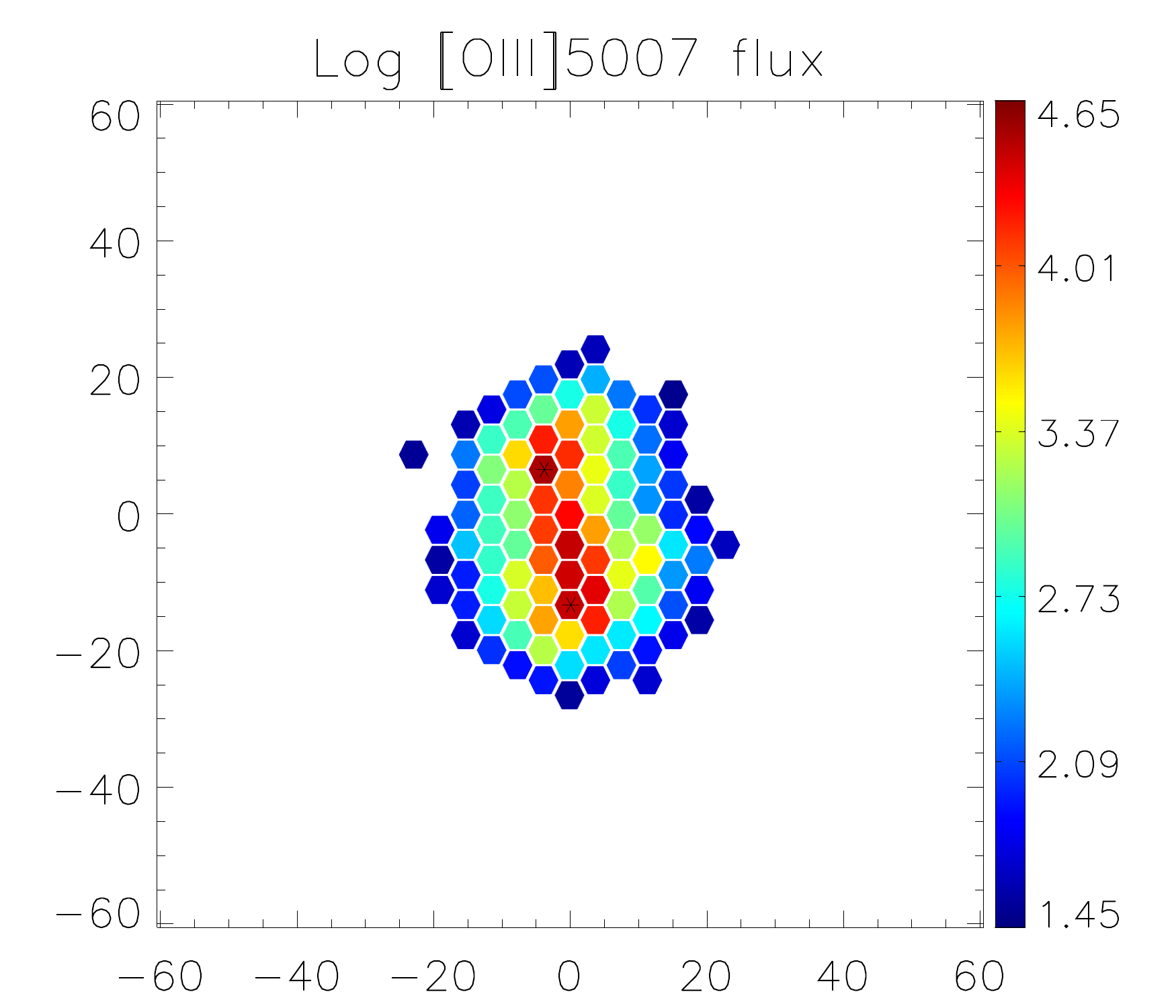}}
\hspace*{0.0cm}\subfigure{\includegraphics[width=0.20\textwidth]{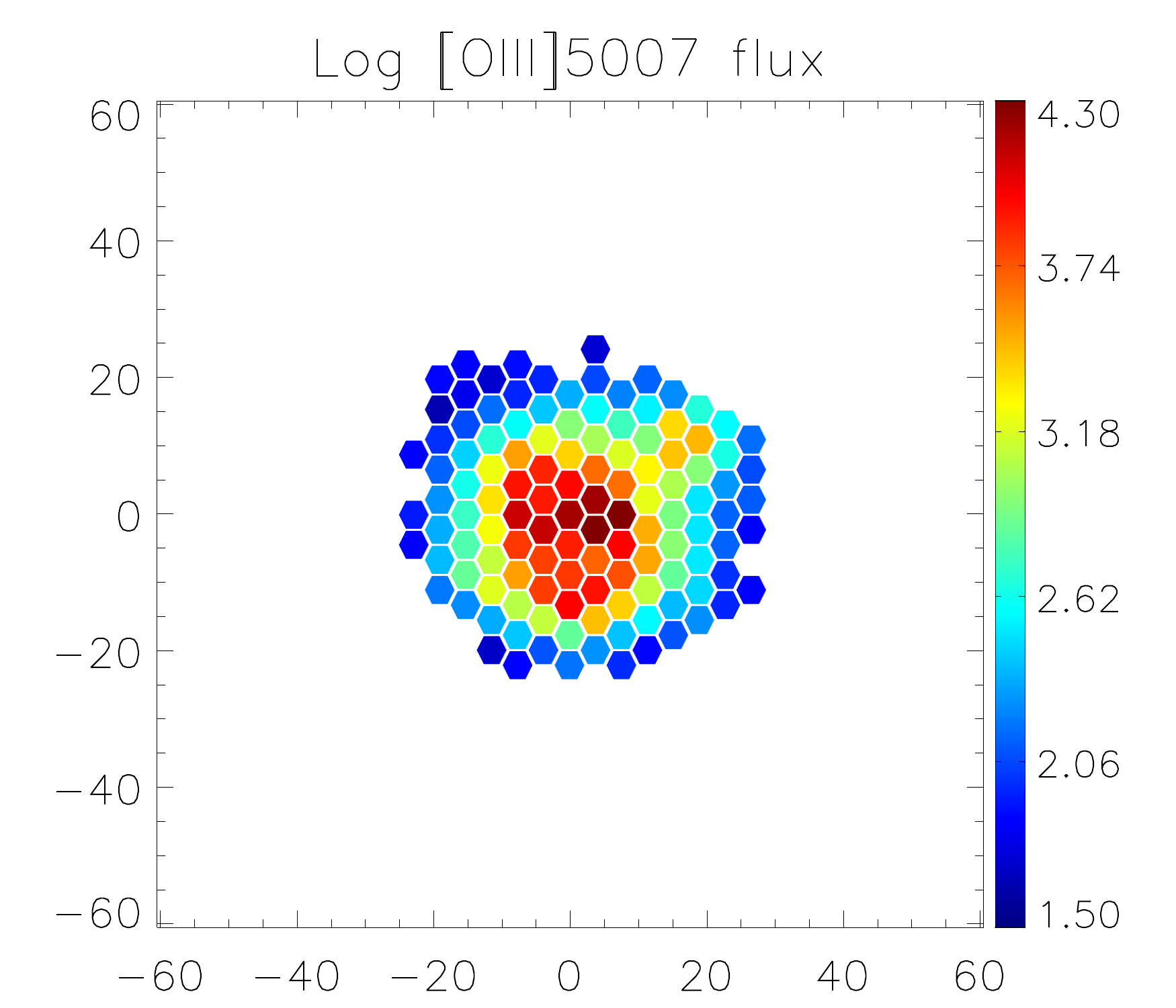}}
\hspace*{0.0cm}\subfigure{\includegraphics[width=0.20\textwidth]{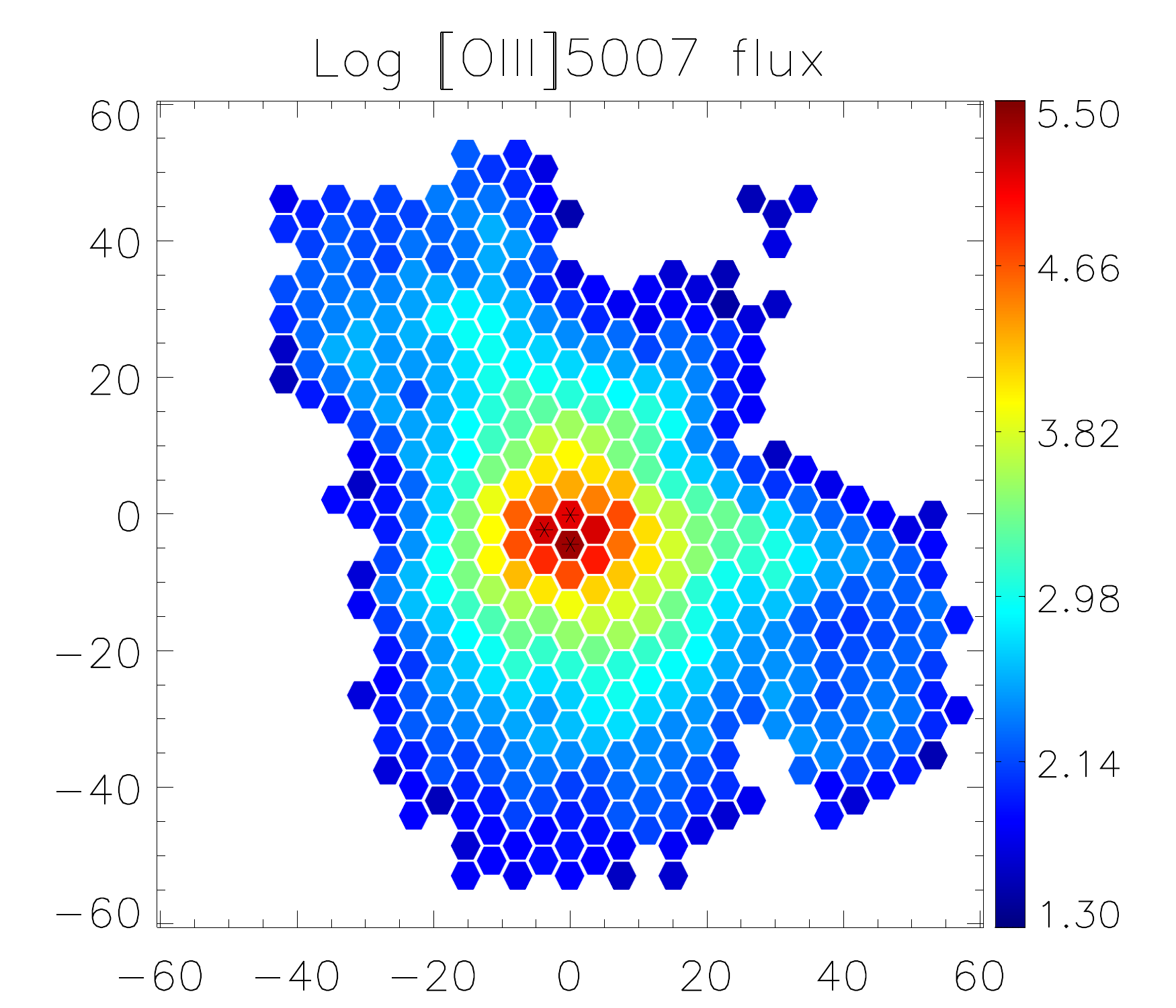}}
\hspace*{0.0cm}\subfigure{\includegraphics[width=0.20\textwidth]{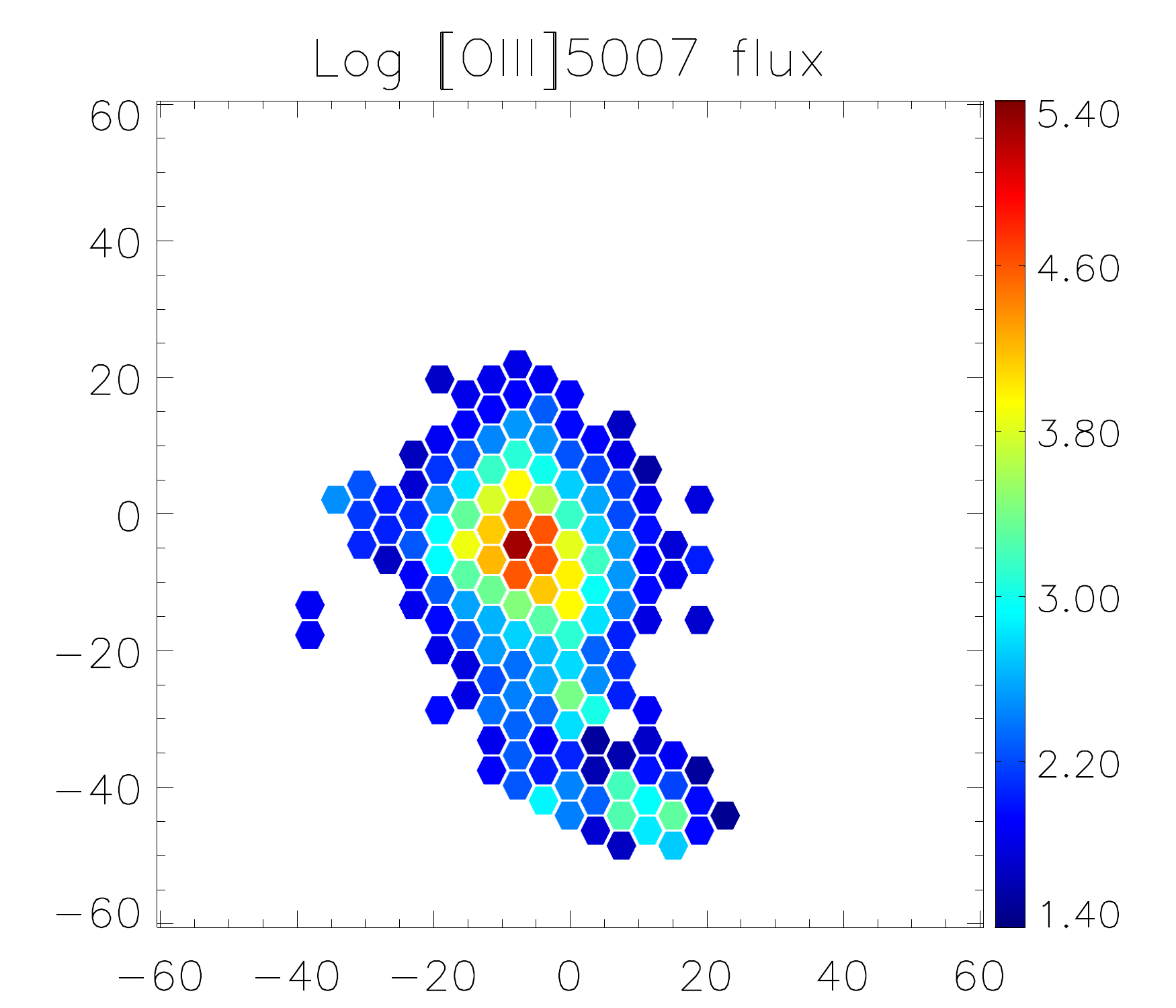}}
\hspace*{0.0cm}\subfigure{\includegraphics[width=0.20\textwidth]{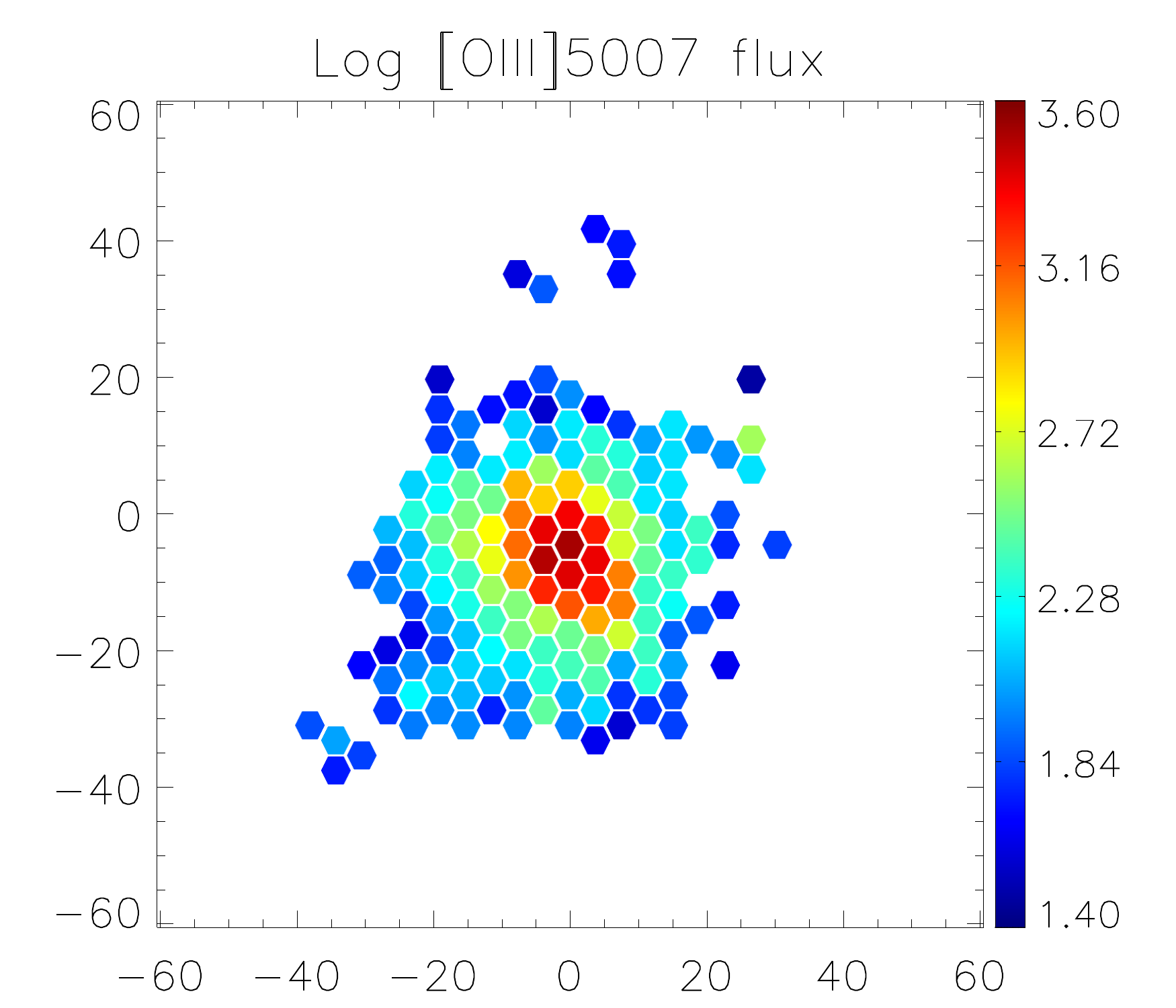}}
}}  
\caption{Observed intensity maps (not reddening-corrected).
First row: continuum emission between 4500 and
4700 \AA\ (a spectral region free from gas emission). Rows 2 to 5:
emission-line flux maps derived from Gaussian fits to each line and
spaxel (see the text for details). Axis units are arcseconds; north is up,
east to the left. All images are shown in logarithmic scale. 
Flux units are $10^{-18}$ ergs cm$^{-2}$ s$^{-1}$. 
Spaxels where the blue Wolf-Rayet bump is visible are marked by stars in the 
[\ion{O}{iii}] map.
}
\label{Figure:emissionmaps}
\end{figure*}

Following the steps described in Section~\ref{SubSection:CreatingMaps}  we
built emission-line flux maps for the brightest observed emission-lines,
namely [\ion{O}{ii}]~$\lambda3727$, \Hg, \Hb, and
[\ion{O}{iii}]~$\lambda5007$; these maps are displayed in
Figure~\ref{Figure:emissionmaps} (rows~2 to 5). 

Continuum maps were obtained by summing the flux within specific wavelength
intervals, selected so as to avoid emission lines or strong residuals from the
sky spectrum subtraction.
Continuum maps in the spectral interval 4500-4700 \AA\ are shown in 
Figure~\ref{Figure:emissionmaps} (row~1): this is the wavelength range
where the continuum signal is stronger and better portrays the galaxy outer regions
(Wolf-Rayet features are not an issue, as they are only visible in just a few 
central spaxels in two objects, see Sect.~\ref{Section:Discussion}).

The emission-line maps depict the interstellar gas ionized by massive stars,
while the continuum maps mainly trace the stellar light from the galaxy. The
five galaxies display an irregular morphology in emission-lines; in \object{II Zw 33}
and \object{Mrk 314}, the SF regions are aligned to form a chain, while in \object{Haro 1},
\object{NGC 4670} and \object{III Zw 102} the ionized gas departs from the inner regions, 
giving rise to a complex pattern, where the presence of several bubbles and
filaments is remarkable. 

On the other hand, the continuum morphology is regular in the five objects;
(the jagged appearance of the Haro~1 continuum map is due to the presence of a
few foreground stars, which have been masked out, and a foreground galaxy at
the west). In all the galaxies, the continuum peak is located roughly at the
center of the outer isophotes, and its position does not depend on which
specific wavelength range is used to build the continuum maps.


\subsection{Line ratio maps}
\label{SubSection:LineRatioMaps}

\begin{figure*}[htb]
\mbox{
\centerline{
\hspace*{00pt}\subfigure{\begin{minipage}{0.20\textwidth}\qquad\qquad{\large II~Zw~33}\end{minipage}}
\hspace*{00pt}\subfigure{\begin{minipage}{0.20\textwidth}\qquad\qquad{\large Haro~1}\end{minipage}}
\hspace*{00pt}\subfigure{\begin{minipage}{0.20\textwidth}\qquad\quad{\large NGC~4670}\end{minipage}}
\hspace*{00pt}\subfigure{\begin{minipage}{0.20\textwidth}\qquad\quad\enspace{\large Mrk~314}\end{minipage}}
\hspace*{00pt}\subfigure{\begin{minipage}{0.20\textwidth}\qquad\quad{\large III~Zw~102}\end{minipage}}
}} 
\mbox{
\centerline{
\hspace*{0.0cm}\subfigure{\includegraphics[width=0.20\textwidth]{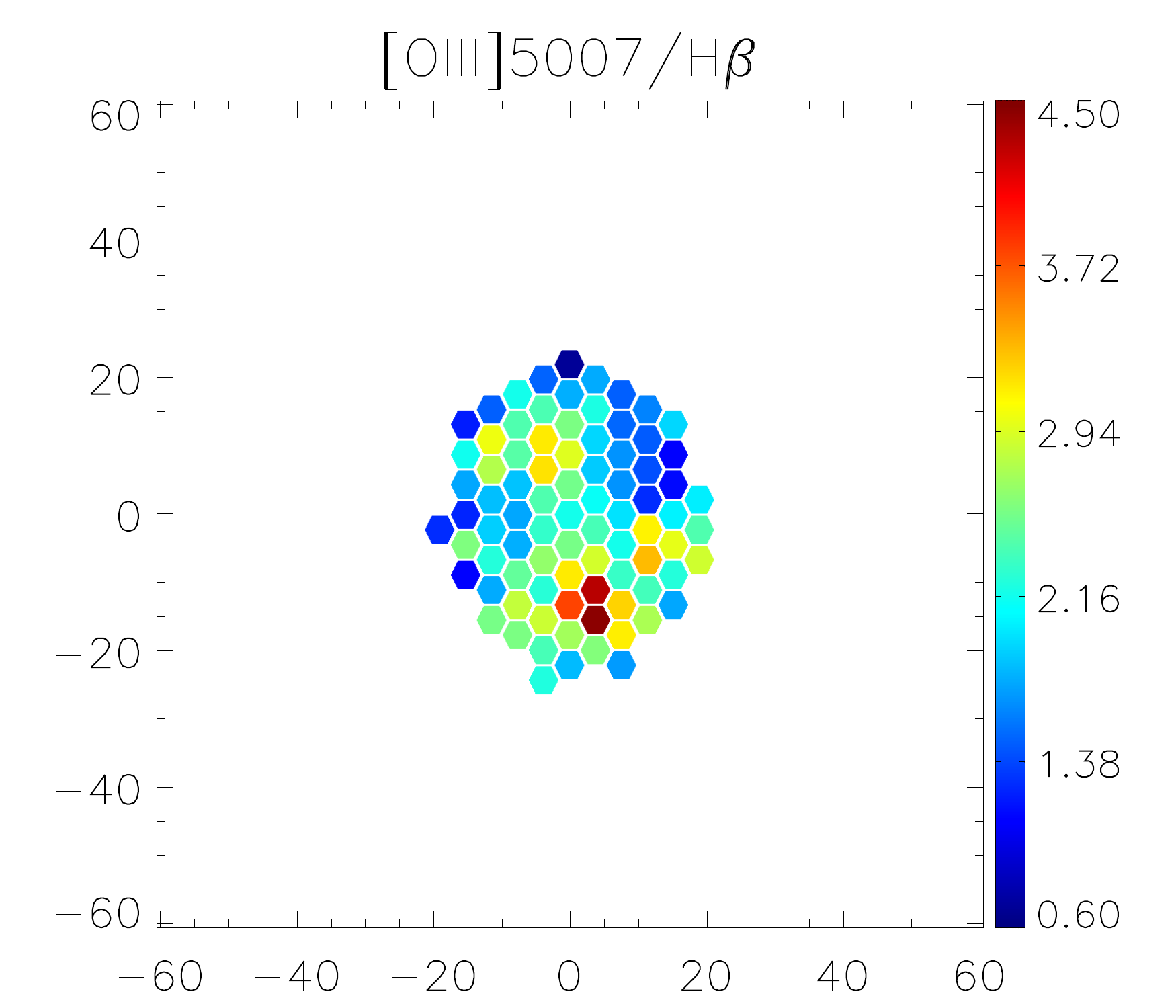}}
\hspace*{0.0cm}\subfigure{\includegraphics[width=0.20\textwidth]{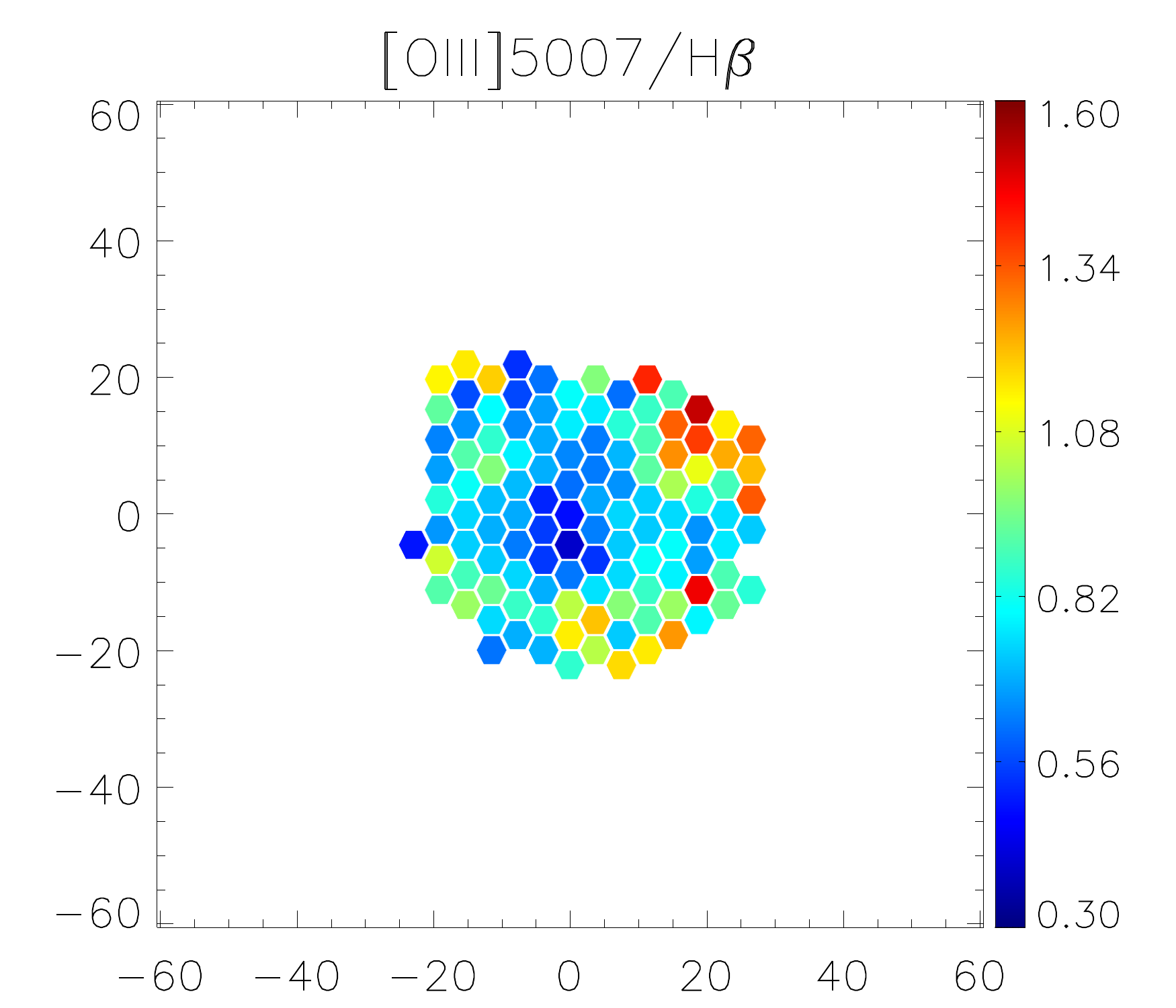}}
\hspace*{0.0cm}\subfigure{\includegraphics[width=0.20\textwidth]{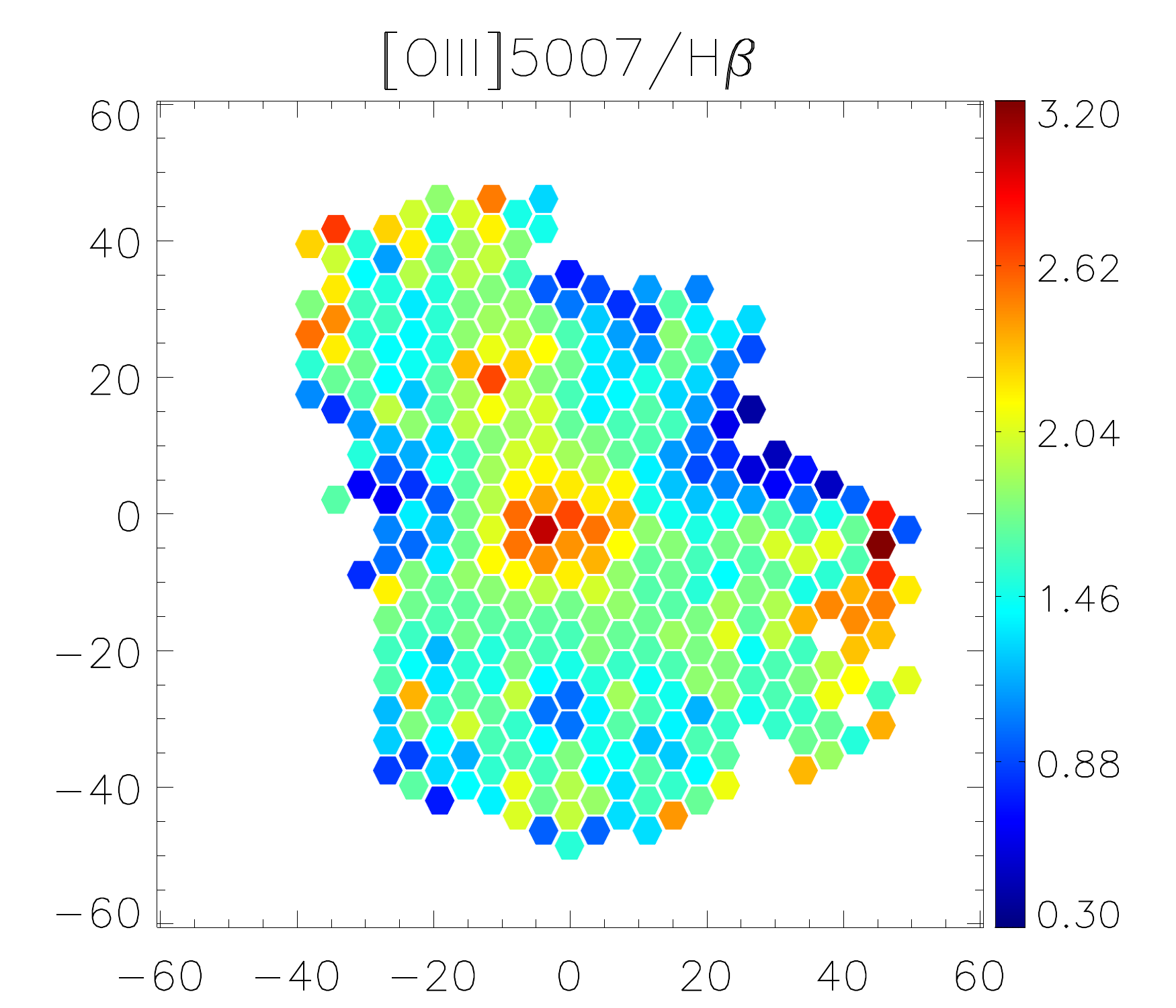}}
\hspace*{0.0cm}\subfigure{\includegraphics[width=0.20\textwidth]{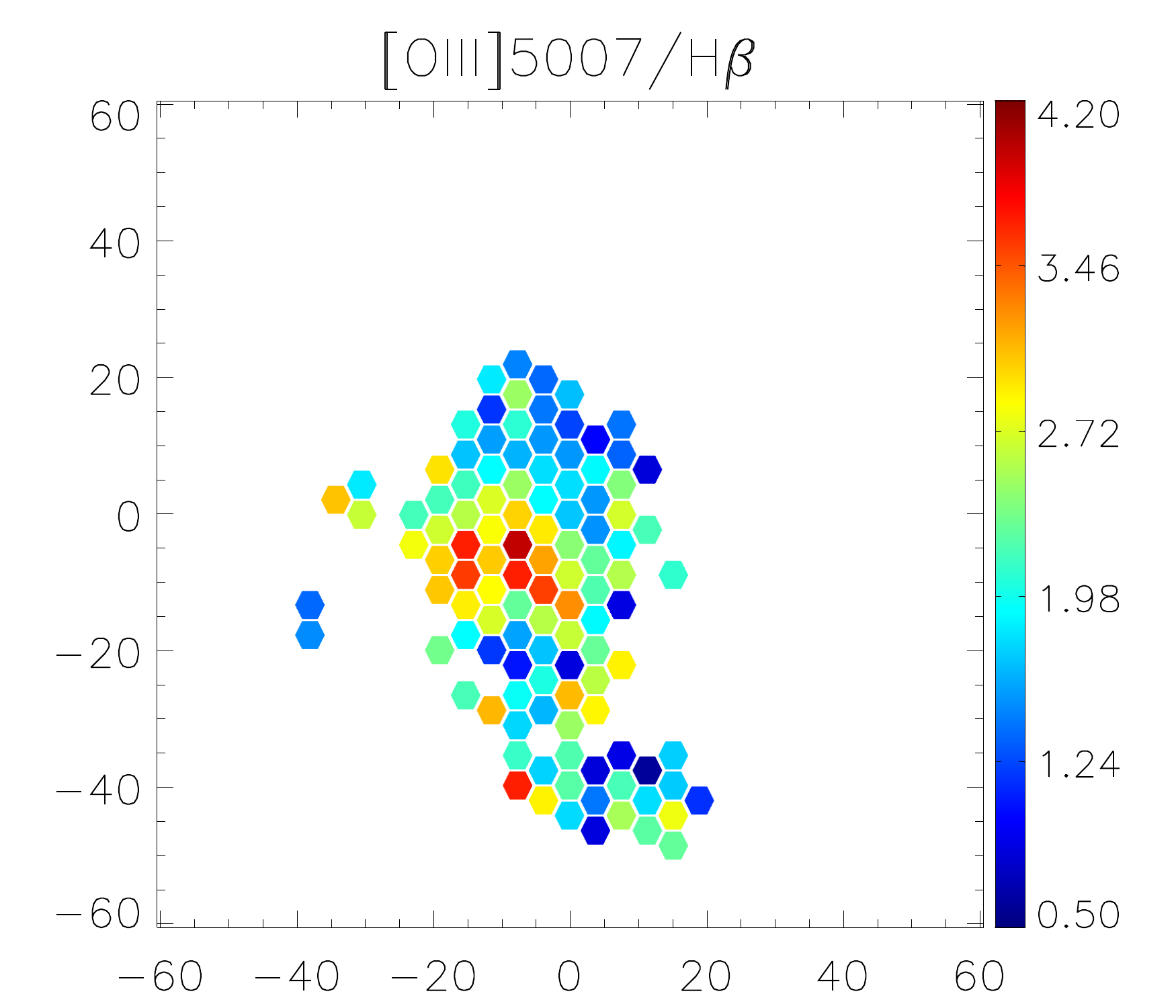}}
\hspace*{0.0cm}\subfigure{\includegraphics[width=0.20\textwidth]{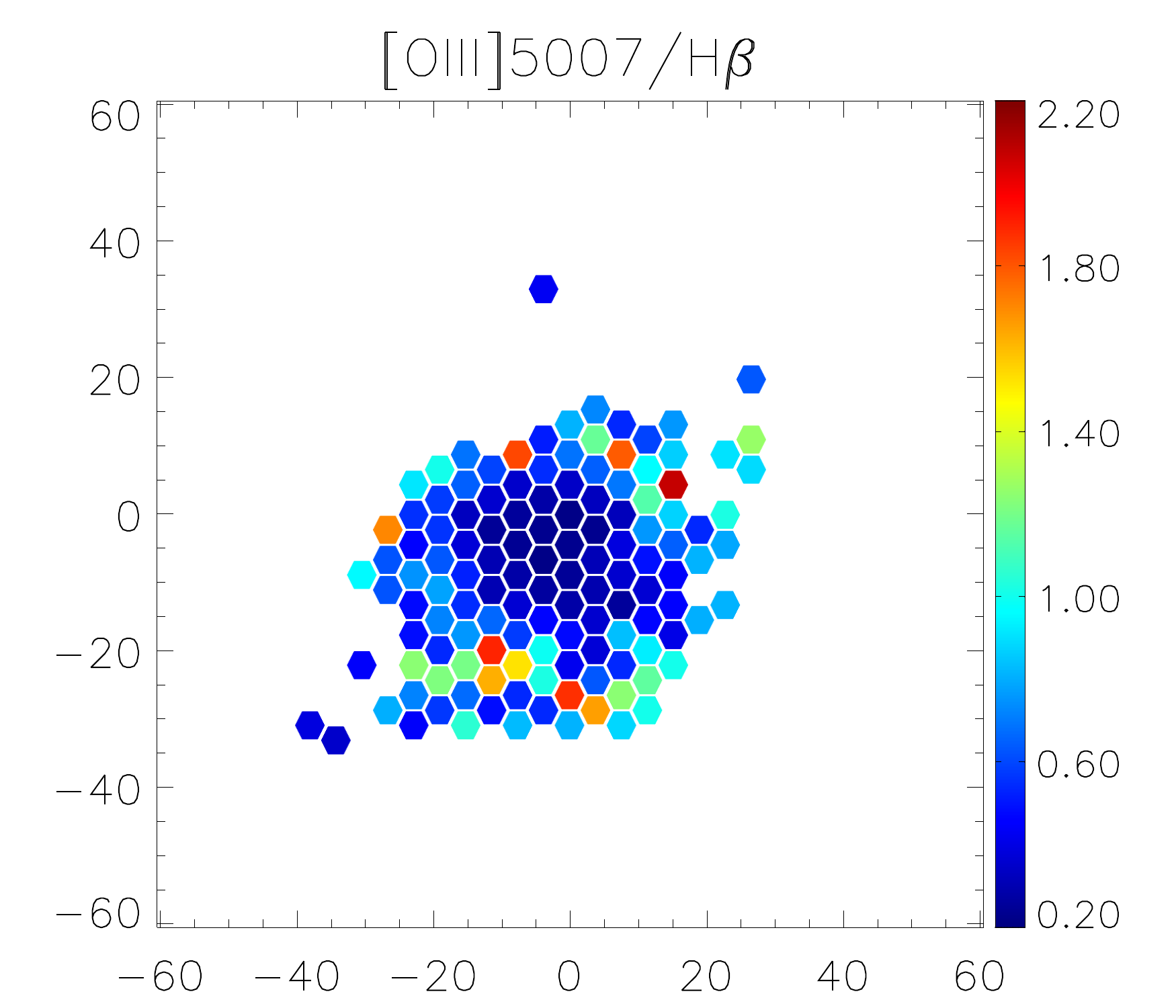}}
}}   
\mbox{
\centerline{
\hspace*{0.0cm}\subfigure{\includegraphics[width=0.20\textwidth]{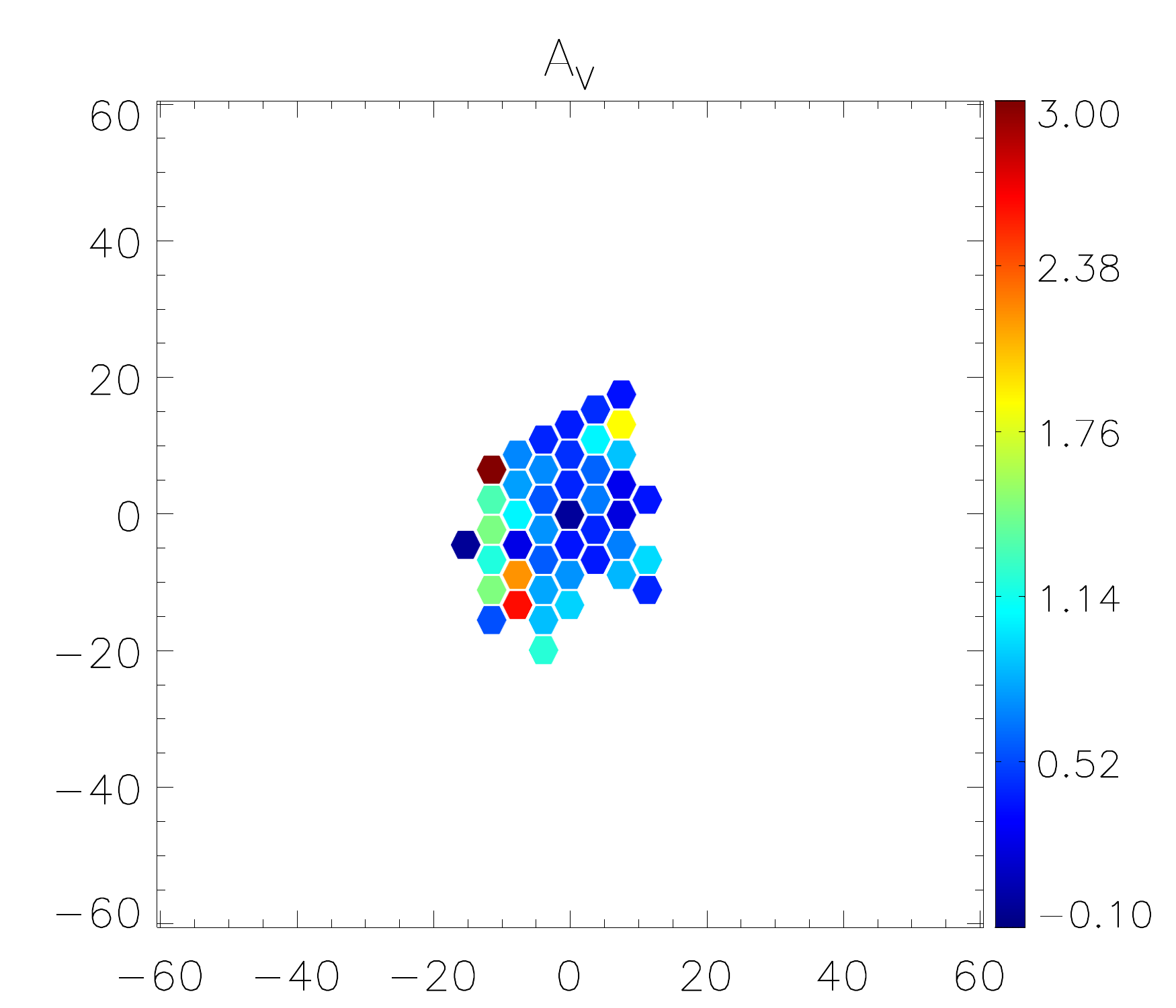}}
\hspace*{0.0cm}\subfigure{\includegraphics[width=0.20\textwidth]{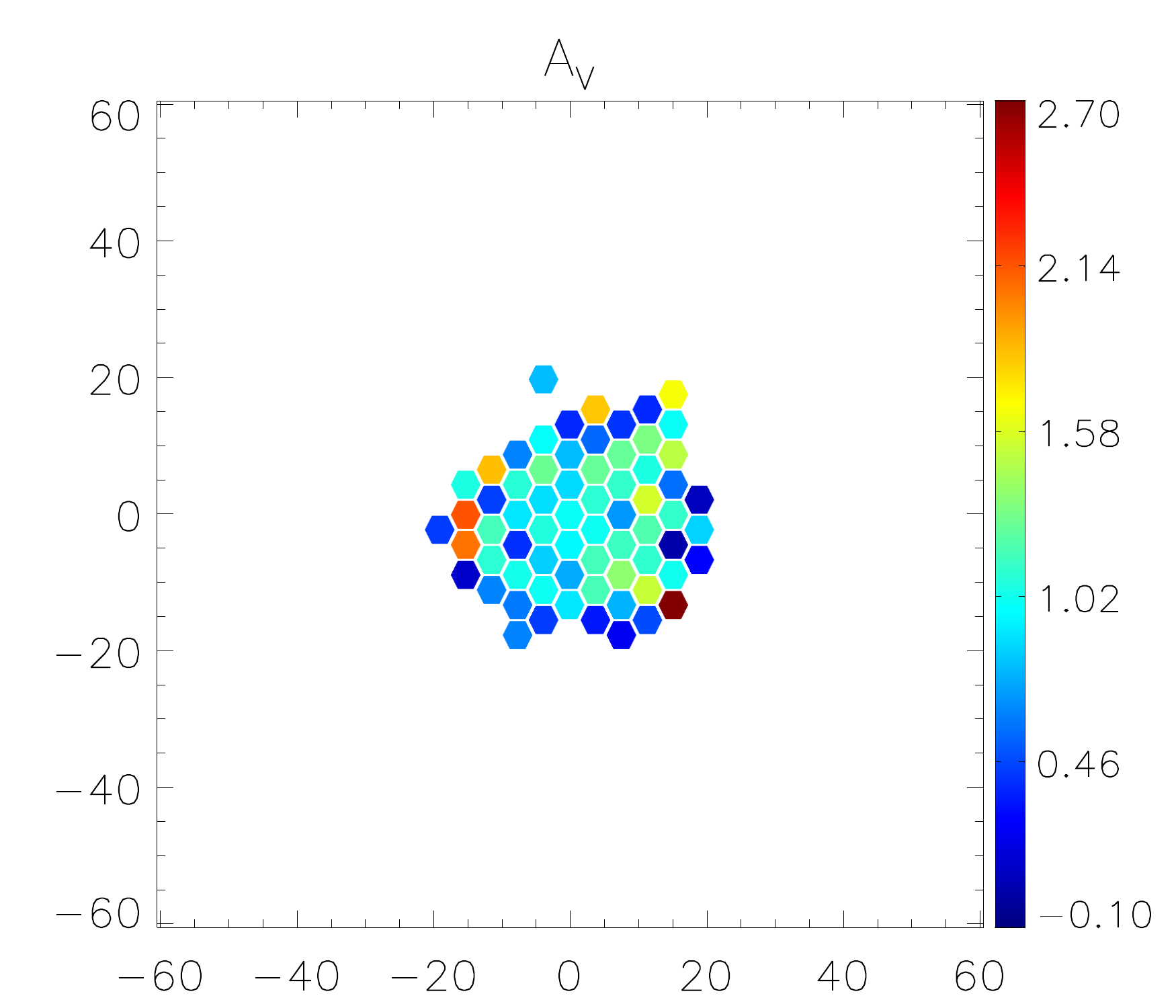}}
\hspace*{0.0cm}\subfigure{\includegraphics[width=0.20\textwidth]{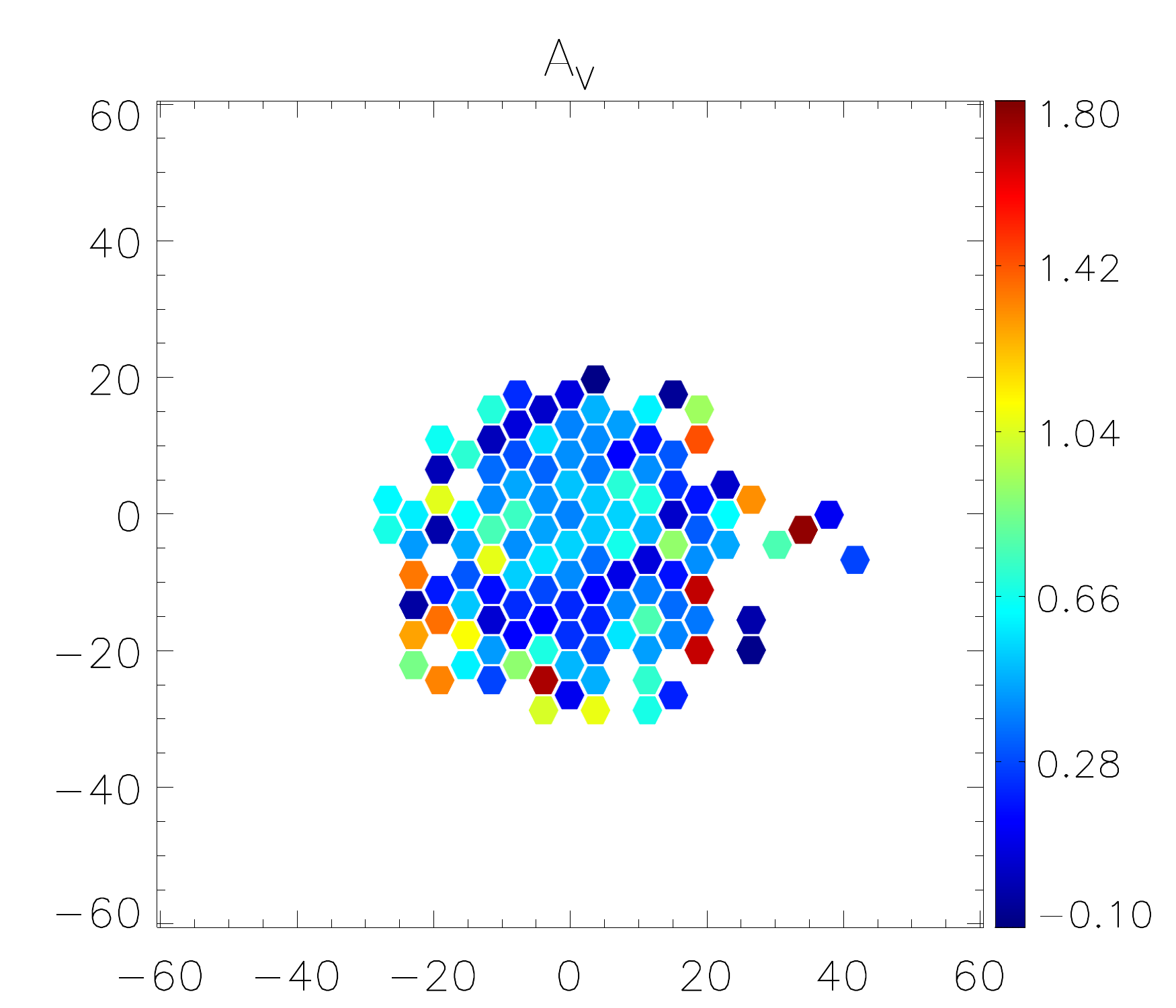}}
\hspace*{0.0cm}\subfigure{\includegraphics[width=0.20\textwidth]{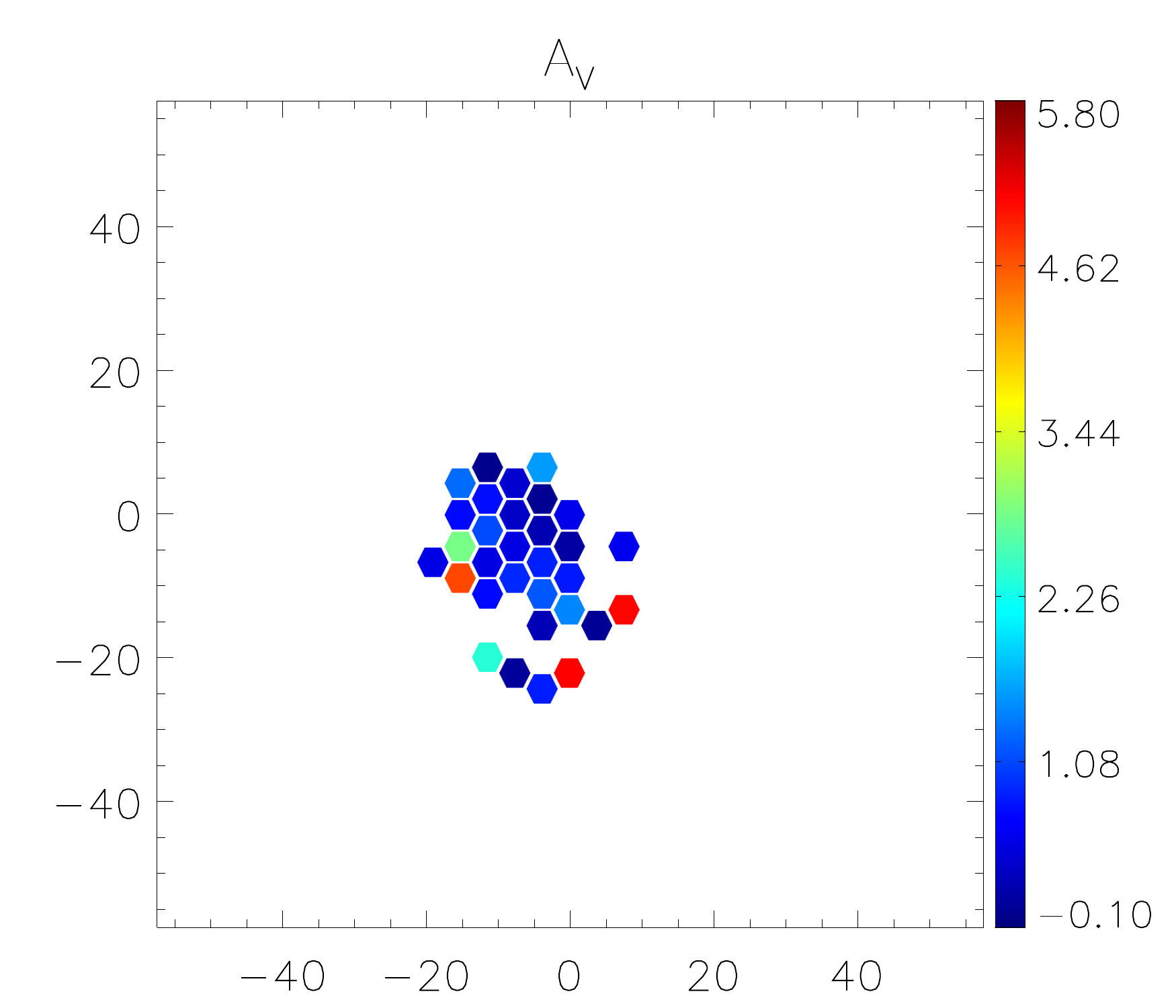}}
\hspace*{0.0cm}\subfigure{\includegraphics[width=0.20\textwidth]{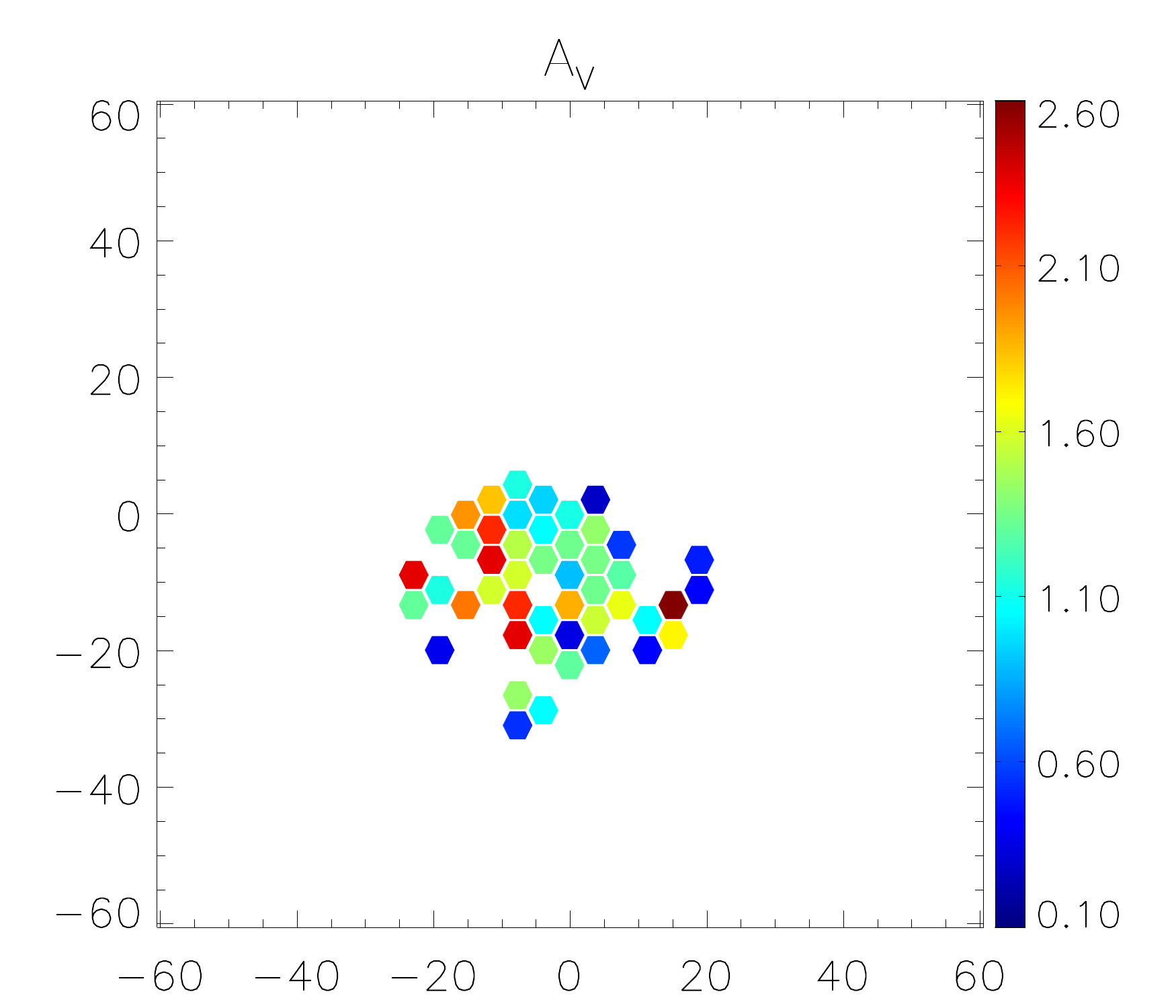}}
}}  
\mbox{
\centerline{
\hspace*{0.0cm}\subfigure{\includegraphics[width=0.20\textwidth]{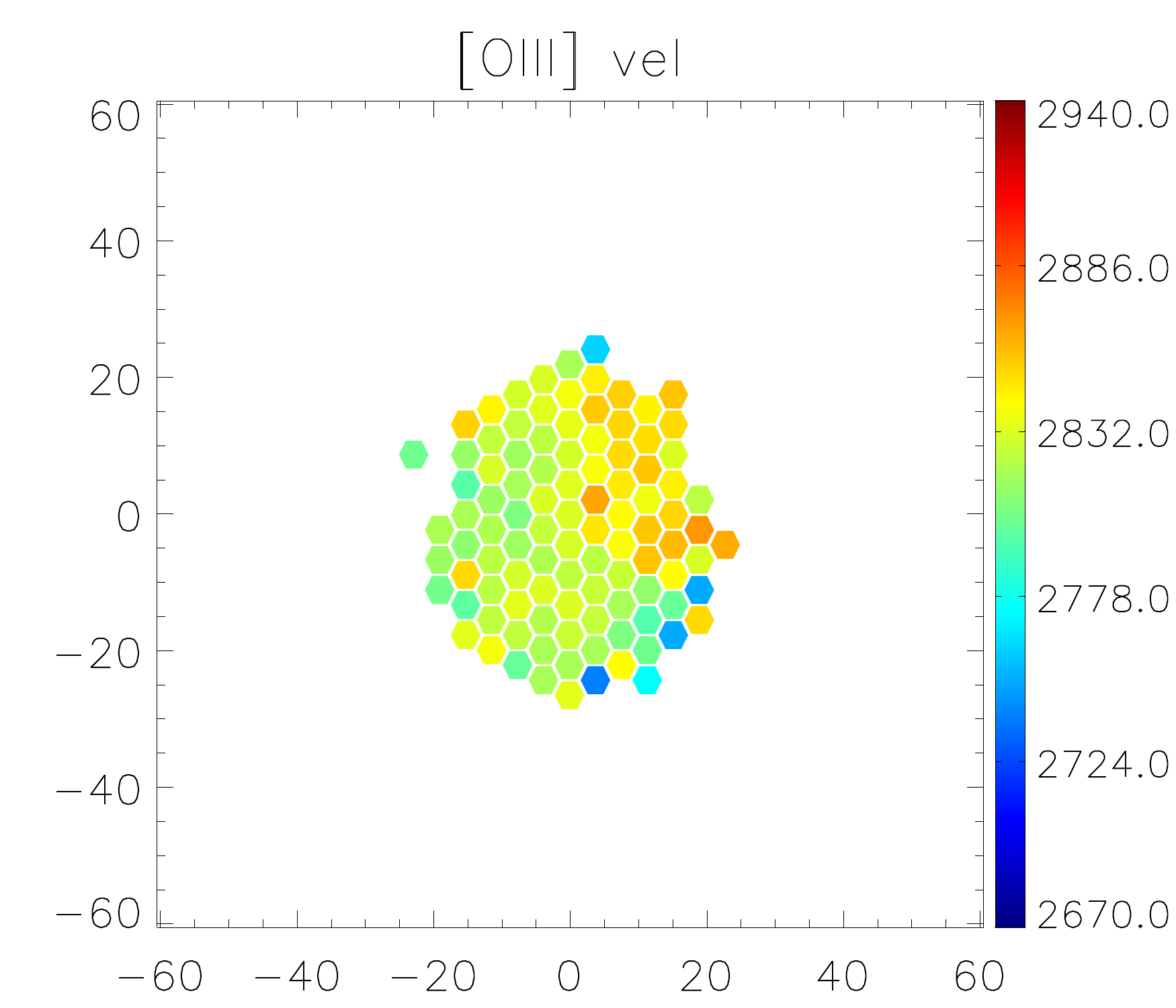}}
\hspace*{0.0cm}\subfigure{\includegraphics[width=0.20\textwidth]{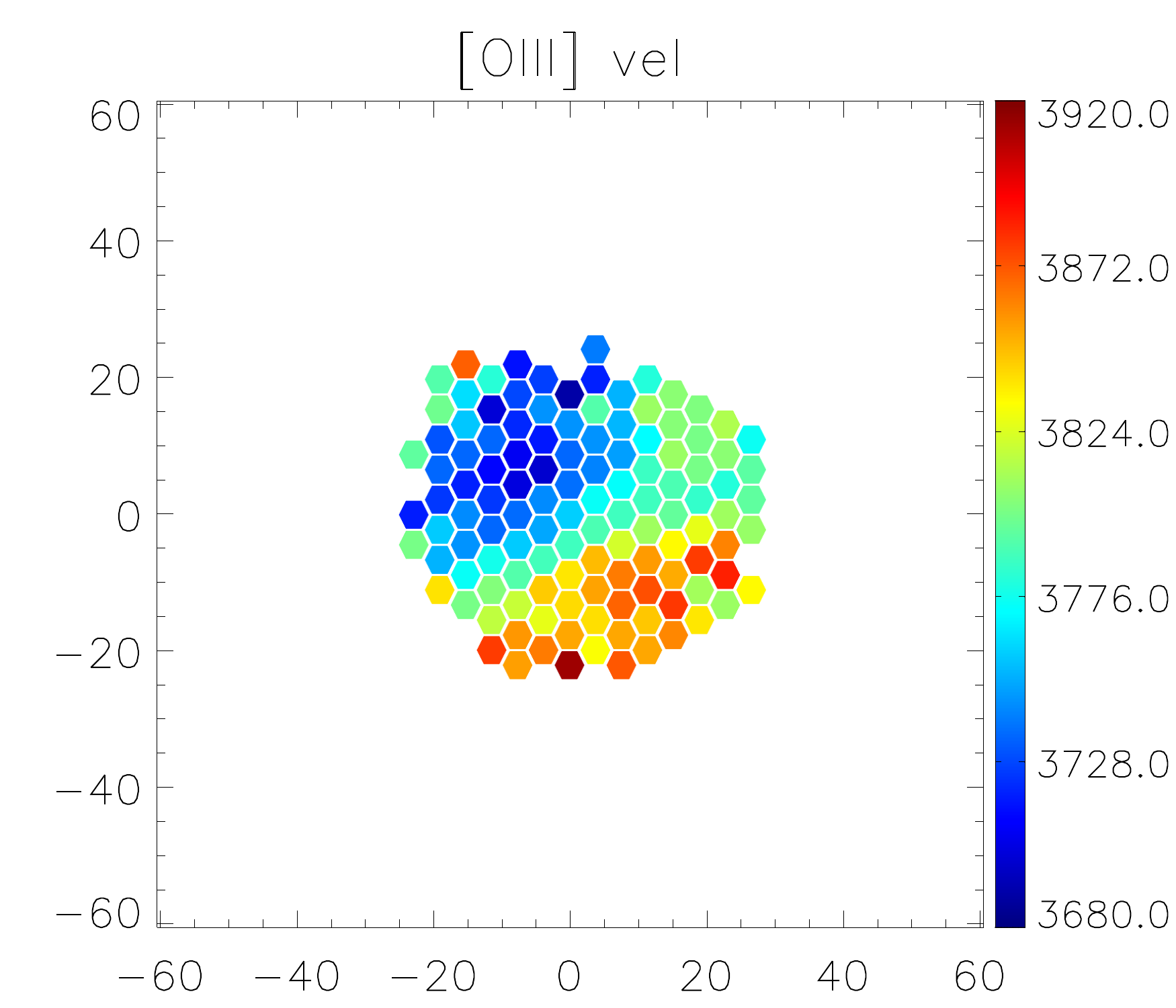}}
\hspace*{0.0cm}\subfigure{\includegraphics[width=0.20\textwidth]{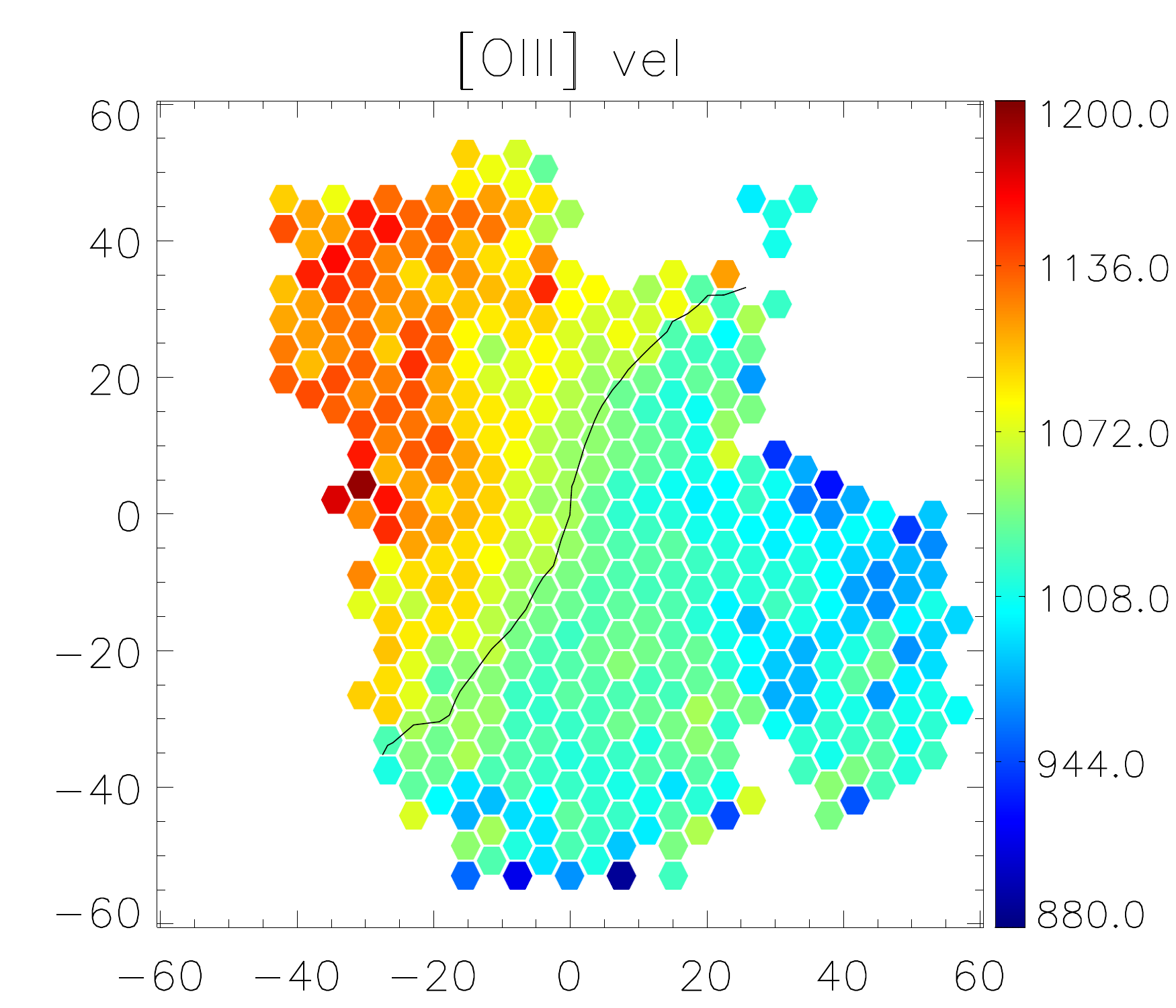}}
\hspace*{0.0cm}\subfigure{\includegraphics[width=0.20\textwidth]{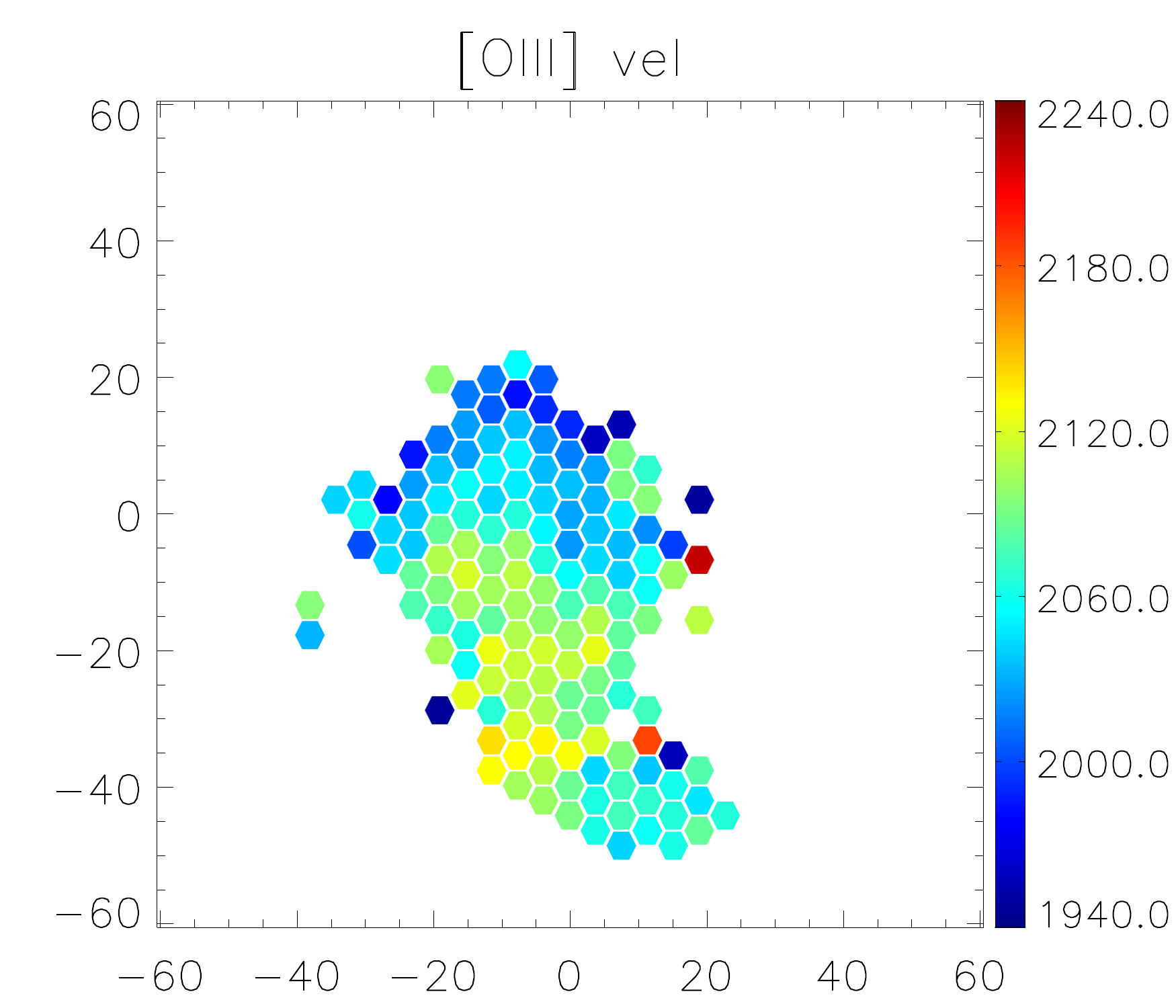}}
\hspace*{0.0cm}\subfigure{\includegraphics[width=0.20\textwidth]{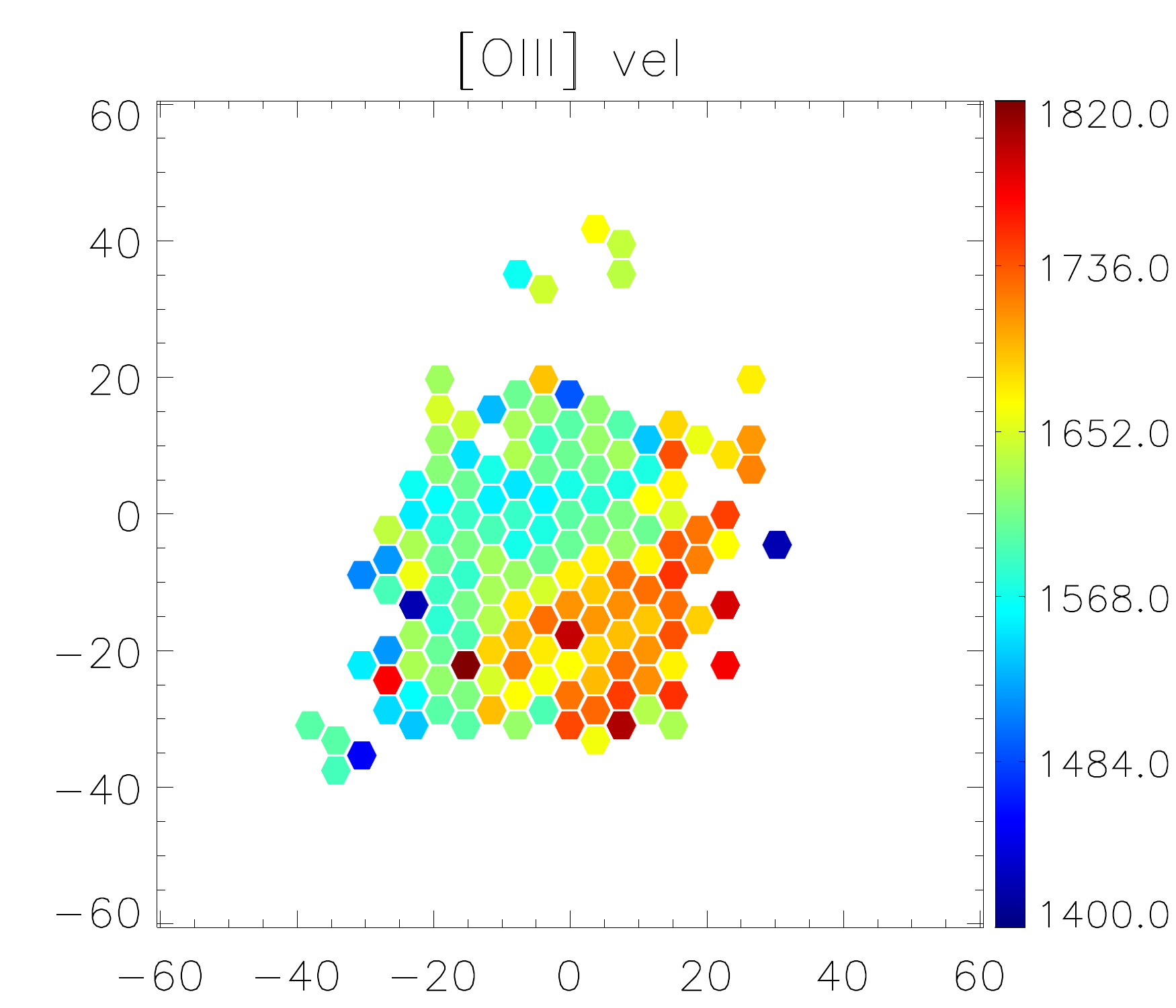}}
}}  
\mbox{
\centerline{
\hspace*{0.0cm}\subfigure{\includegraphics[width=0.20\textwidth]{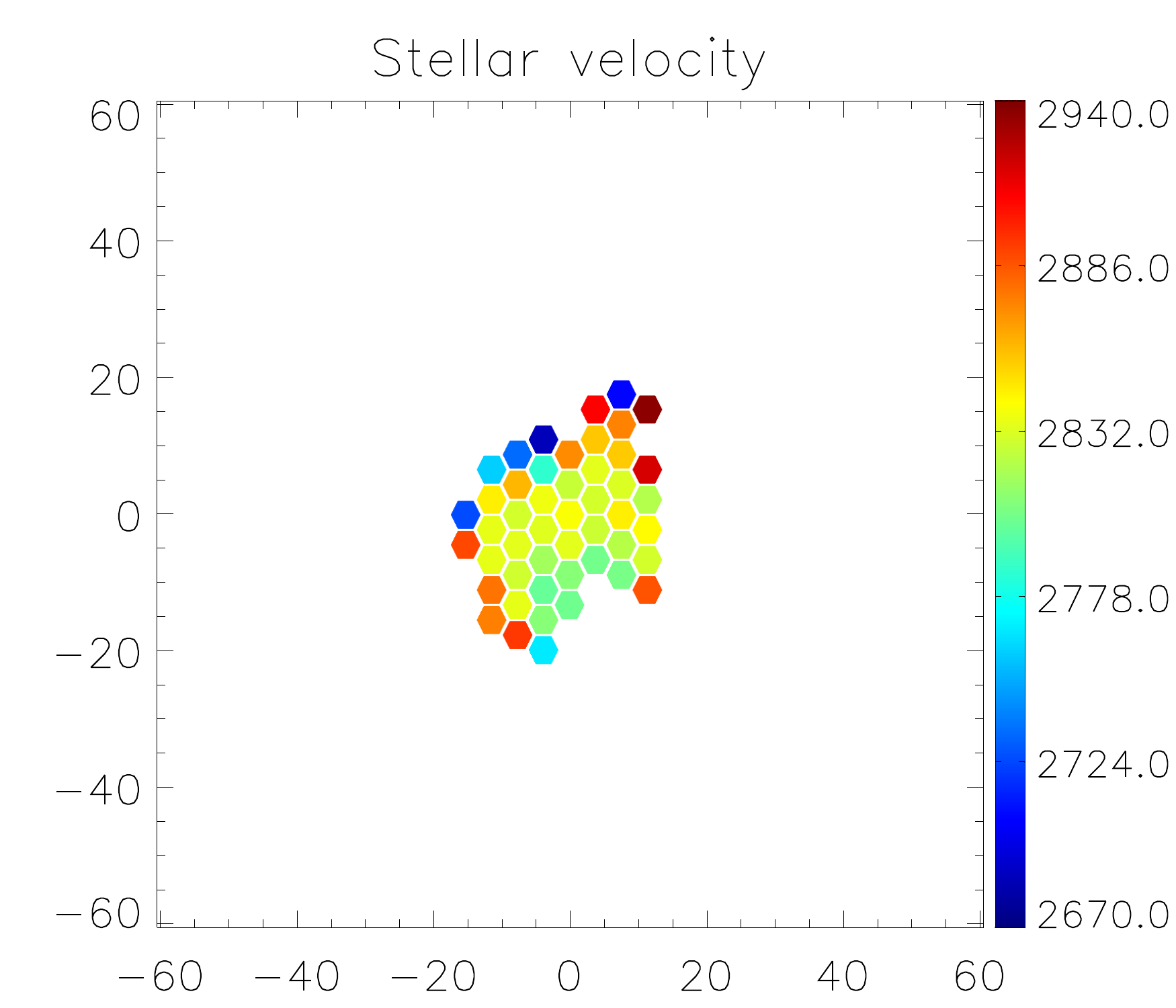}}
\hspace*{0.0cm}\subfigure{\includegraphics[width=0.20\textwidth]{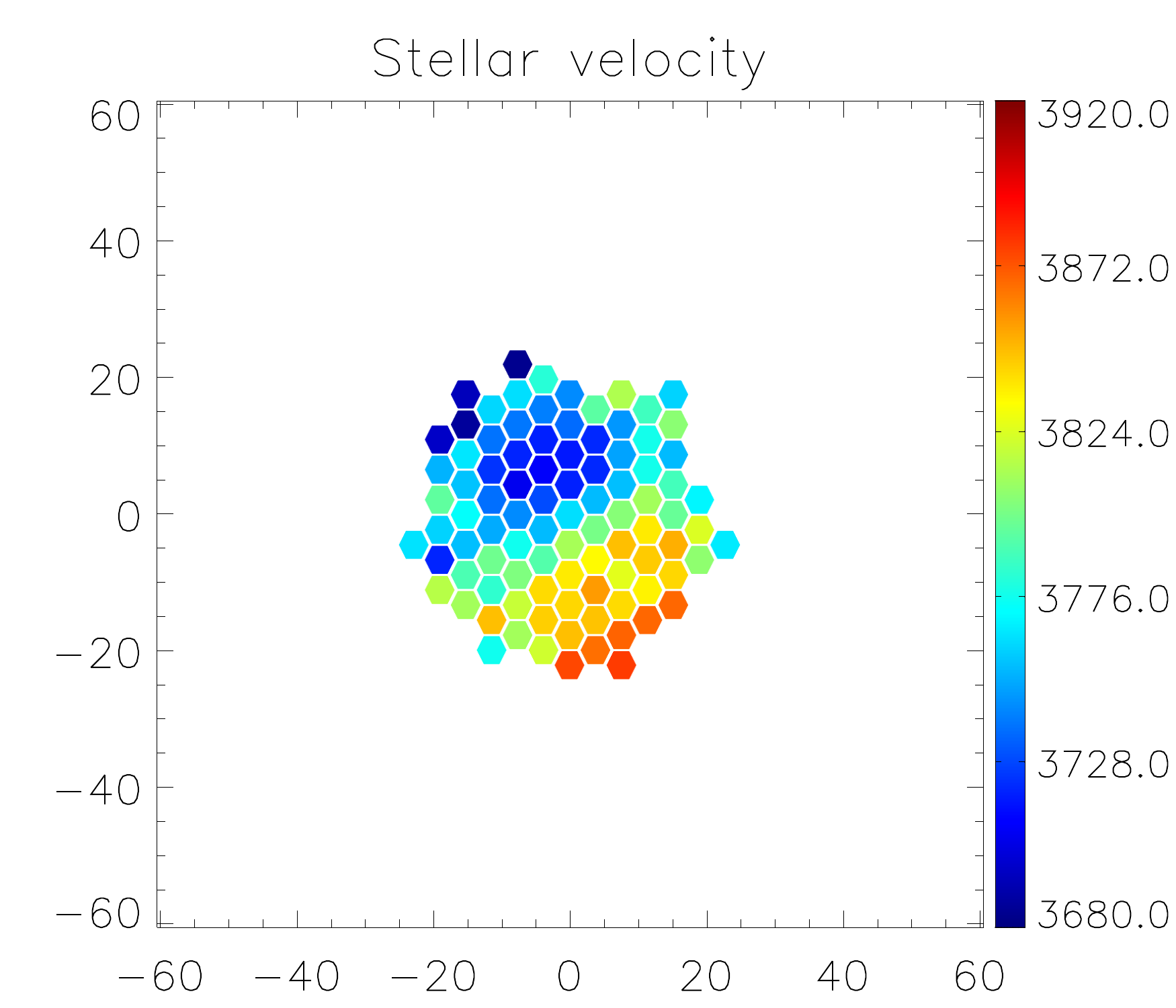}}
\hspace*{0.0cm}\subfigure{\includegraphics[width=0.20\textwidth]{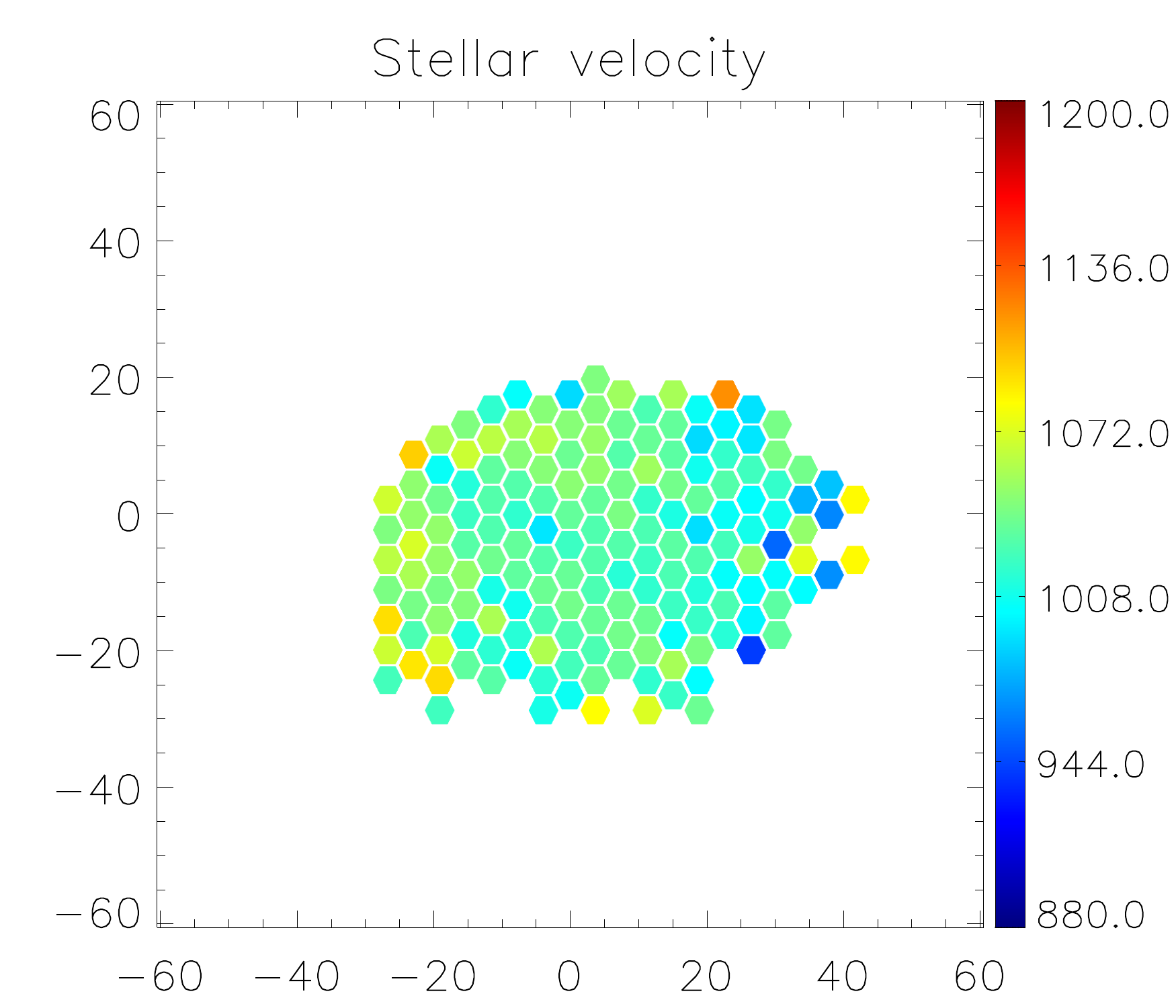}}
\hspace*{0.0cm}\subfigure{\includegraphics[width=0.20\textwidth]{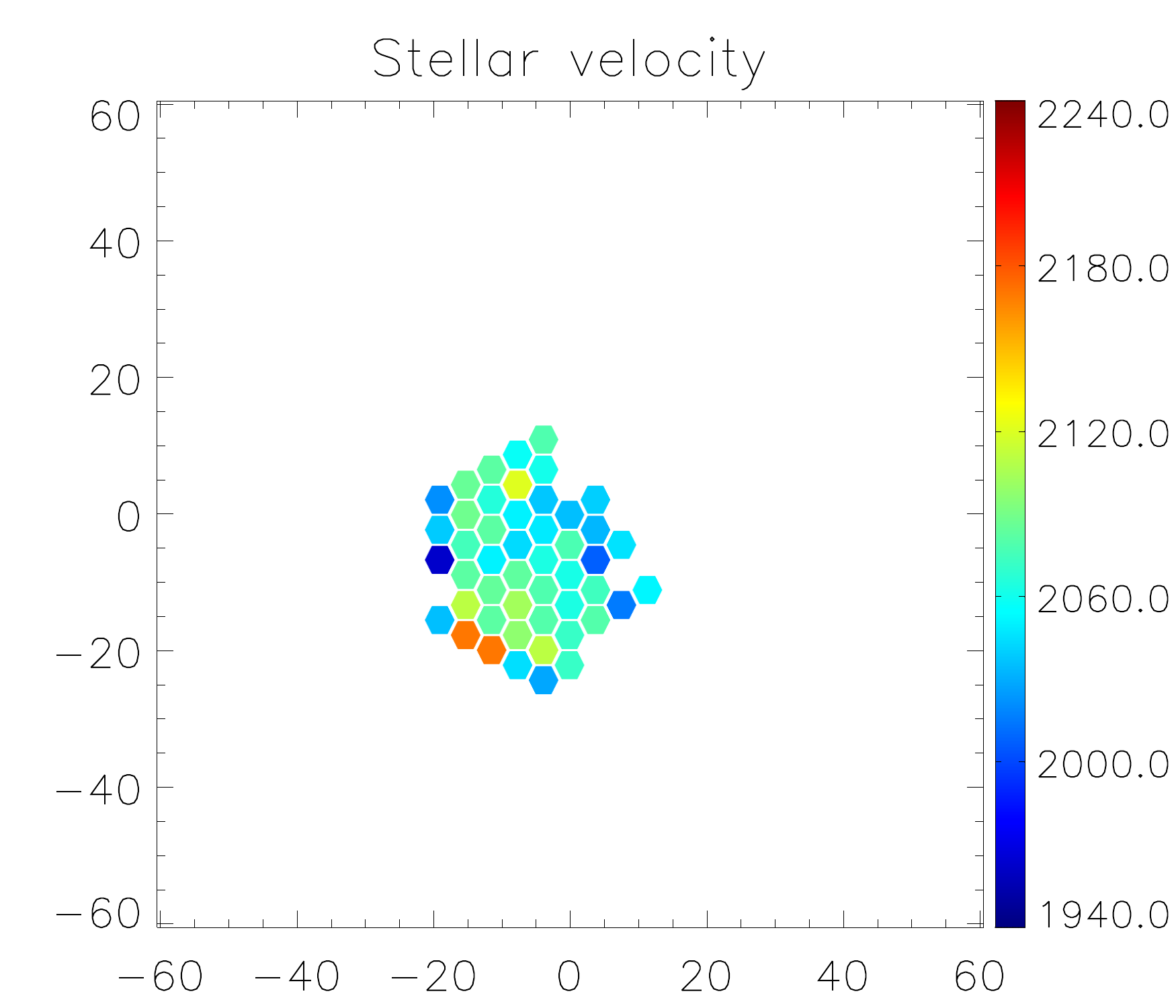}}
\hspace*{0.0cm}\subfigure{\includegraphics[width=0.20\textwidth]{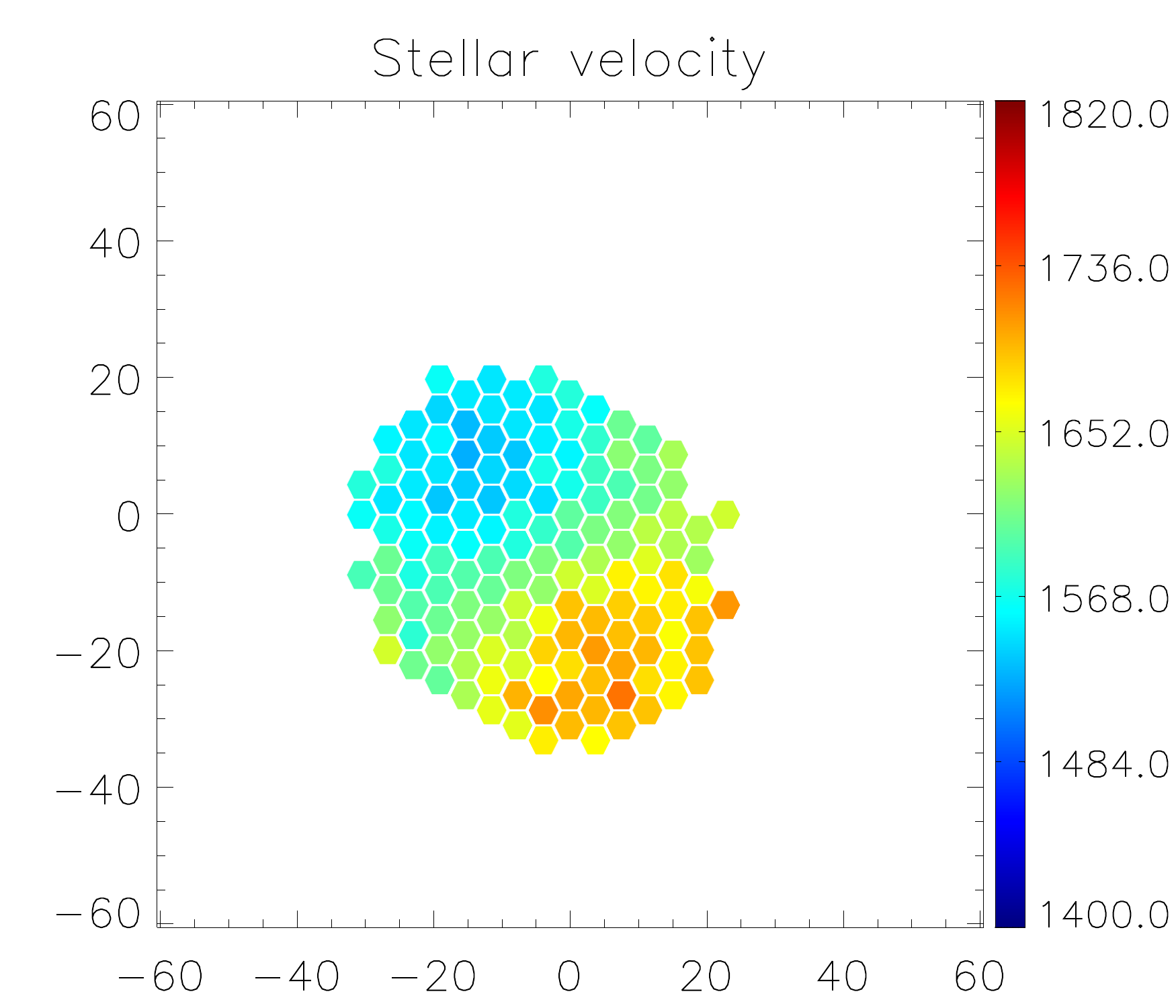}}
}}  
\caption{
First row: Observed (not reddening-corrected)
ionization ratio [\ion{O}{iii}]~$\lambda5007$/\Hb\ 
for the five galaxies.
Second row: Interstellar extinction \AV\ map, computed 
from the \Hb/\Hg\ ratio.
Row three: Velocity field (in km s$^{-1}$) of the ionized gas
[\ion{O}{iii}]~$\lambda5007$ emission line, except for III~Zw~102
for which the \Hb\ velocity field is shown.
Row four: Stellar velocity field. Axis units are arcseconds.  
The solid line in the gas velocity map of NGC~4670 is the isovelocity 
contour at $V=1050$ km s$^{-1}$ (see text for details).
}
\label{Figure:velocitymaps}
\end{figure*}

The first row of Figure~\ref{Figure:velocitymaps} displays the
[\ion{O}{iii}]~$\lambda5007$/\Hb\ maps for the five studied galaxies. This
ratio essentially probes the hardness of the stellar radiation field within the
nebula, and so high values of [\ion{O}{iii}]~$\lambda5007$/\Hb\ are  predicted
when the ionization is produced predominantly by UV photons from OB stars 
\citep{McCall1985}. However it is also sensitive to other factors as e.g. the
metallicity \citep{Searle1971}, the distance from the ionizing source to the
[\ion{O}{iii}] emitting layer \citep{Stasinska2009}, 
or the presence of shocks \citep{Dopita1995}. 
Therefore the information provided by [\ion{O}{iii}]~$\lambda5007$/\Hb\ is 
limited unless additional data regarding these other
parameters are added.

For \object{II Zw 33} and \object{Mrk 314}, the excitation maps roughly mimic
the emission-line maps, the highest [\ion{O}{iii}]~$\lambda5007$/\Hb\ values
($\geq 3$), spatially coinciding with the highest peak in emission-lines, as
usual in systems photoionized by hot stars. 

Both \object{Haro 1} and \object{III Zw 102} show more complex 
[\ion{O}{iii}]~$\lambda5007$/\Hb\ patterns, with the highest values of the
excitation  localized in the galaxy periphery. This behavior, frequently found
in spiral galaxies, has been interpreted as an abundance gradient
\citep{Smith1975}. Interestingly, Haro~1 and III~Zw~102 are the two most
luminous objects in the sample and both relatively far from the dwarf galaxy
regime.  

\object{NGC 4670} shows a somewhat complex excitation morphology, with several
[\ion{O}{iii}]~$\lambda5007$/\Hb\ peaks, located in the central giant
\ion{H}{ii} region and at the outer edge of the two larger filaments. For all
five galaxies, the excitation levels are consistent with \ion{H}{ii}-like
ionization.

Interstellar extinction can be probed by comparing the observed ratios of
hydrogen recombination lines with their theoretical values
\citep{Osterbrock2006}. In the optical domain, extinction maps are usually
derived from the ratio \Ha/\Hb\ as it involves the strongest lines, and
therefore, the easiest to measure. In our case, lacking \Ha, we have derived
the extinction from the \Hb/\Hg\ line ratio; the extent of the extinction maps
is dictated by the extent of the \Hg\ maps (as said in
Sect.~\ref{SubSection:CreatingMaps} we measured \Hg\ only in the
continuum-subtracted spectra.

The extinction maps are shown in the second row of
Fig.~\ref{Figure:velocitymaps} using the \AV\ parameter.
We adopted the Case B, low density limit, 
$T=10000$ K approximation, where the theoretical \Hb/\Hg\ ratio 
is 2.15 \citep{Osterbrock2006}. 
The Cardelli extinction law was used to derive \AV\ from the observed
\Hb/\Hg\ value.

\subsection{Ionized gas and stellar kinematics}
\label{SubSection:Kinematics}

We study the kinematics of the ionized gas using the  
[\ion{O}{iii}]~$\lambda5007$ emission-line, except for III~Zw~102 for which we
use \Hb, much stronger than [\ion{O}{iii}]. Due to the low spectral 
resolution of our data (about 5 \AA\ FWHM), no reliable velocity dispersion 
measurements could be obtained.

The velocity fields are shown in Figure~\ref{Figure:velocitymaps} (row three);
in these maps red and blue colors represent, respectively, redshifted and
blueshifted regions with respect to the recessional velocity of each BCG. 
To estimate the errors on the velocity data, we made a combined plot 
(all galaxies together) of the uncertainty on the [\ion{O}{iii}]~$\lambda5007$ 
emission-line centroid versus the logarithm of its flux.
For $\log(\mathrm{flux}) >4$ the error is $\lesssim 5$ km s$^{-1}$, in agreement 
with the mean uncertainty estimated from the scatter in the
velocity maps built on the 5577 \AA\ skyline. The error increases to about
10--15 km s$^{-1}$ at $\log(\mathrm{flux}) =3$, to 25--30 km s$^{-1}$ 
at $\log(\mathrm{flux}) =2$, and reaches or exceeds 40 km s$^{-1}$ at
fainter flux levels.

The pPXF program provides, besides the best fitting stellar template, the
radial velocity and associated uncertainty for each fitted spaxel, allowing us
to produce the galaxy's stellar velocity field (given in
Figure~\ref{Figure:velocitymaps}, row four).

The uncertainty on the stellar radial velocity varies between $\simeq 10$ km
s$^{-1}$ in the nuclear regions and $\simeq 40$ km s$^{-1}$ in the outer
spaxels.

\subsection{Integrated spectroscopy}
\label{SubSection:IntegratedSpectroscopy}

For every galaxy we have also produced its integrated and nuclear spectrum. In
order to  generate the integrated spectrum we included only those spaxels in
which the stellar contribution was successfully modeled as a sum of templates
(see  Section~\ref{SubSection:LineFit}); in this way we ensure a better
correction for the  contribution of the underlying stellar population and,
therefore, a more accurate determination of the emission-lines fluxes.  The
integrated spectrum was produced by summing together the individual 
\emph{pure} emission-line spectra of the selected spaxels. A corresponding
continuum spectrum was obtained by summing the best-fit continuum spectra for
the same spaxels. 

On the same line, we also obtained line-emission and best-fit continuum
spectra for the central region by summing over 6--8 spaxels around the
continuum peak (for short, we shall refer to them as ``nuclear spectra'').

\begin{figure}
\centering
\includegraphics[angle=0, width=0.50\textwidth]{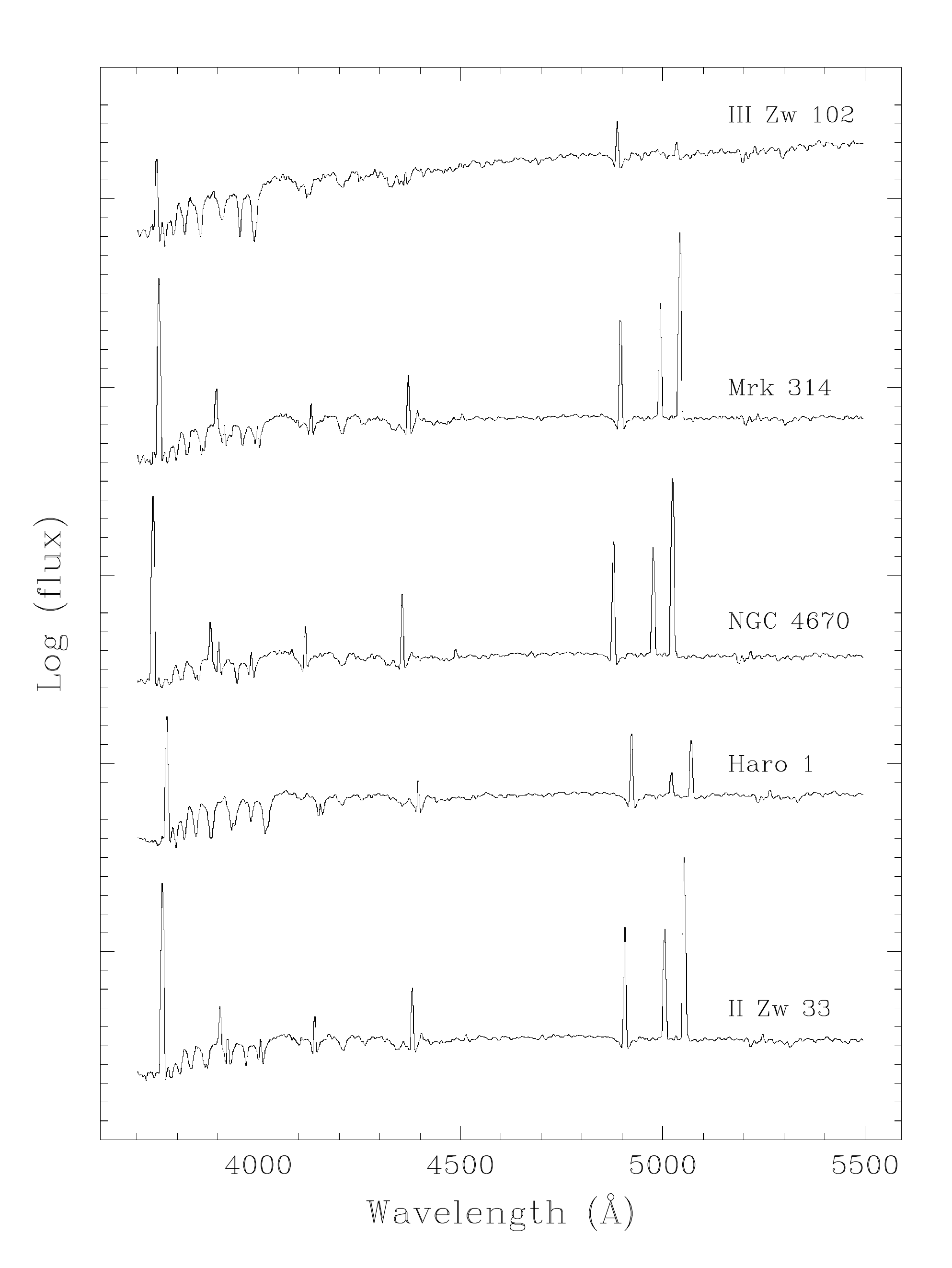}
\caption{Integrated spectra of the sample galaxies, obtained by summing
together the spectra of all those spaxels in which the stellar continuum could be
fitted by a template (see \S~\ref{SubSection:LineFit}). Spectra are plotted 
in logarithmic flux scale, and have been shifted vertically. Major tick-marks 
on the $y$ axis are spaced 1 dex apart. Nuclear spectra do not exhibit 
significantly distinct features from the integrated spectra and are not shown.} 
\label{Figure:SpectraPanel} 
\end{figure}

Figure~\ref{Figure:SpectraPanel} displays the integrated spectra for the
sample objects.  All objects but III~Zw~102 display a typical BCG spectrum, in
which strong optical emission-lines are superposed on an almost flat
continuum. On the contrary, III~Zw~102's spectral shape, with a high and red
continuum, resembles more those typical of an early type galaxy, rather than
those of BCGs. All objects display prominent absorption features, including
the higher-order Balmer lines, \Hd, \Hg, \Hb, \ion{Ca}{ii} H and K lines,  and
the G-band at 4304 \AA. 

For both the integrated and nuclear spectra, the flux of the visible
emission-lines was measured on the \emph{pure} emission-line spectrum using the
Gaussian profile  fitting option in the IRAF task \emph{splot}. In order to
compute equivalent widths, the continuum level was measured on the best-fit
continuum spectrum. For Balmer lines, the continuum level was determined by
interpolating over the absorption feature using two continuum windows on
either side. The interstellar extinction coefficient, \CHbeta, was computed
through a least-square fit to the Balmer decrement, using the \Hb, \Hd\ and
\Hg\ lines.

Reddening-corrected intensity ratios and equivalent widths of the integrated
and nuclear spectra are listed in Table~\ref{Table:LineFlux}. 
The listed uncertainties on the line ratios and the \Hb\ fluxes are derived 
by summing in quadrature measurement errors, uncertainties on the flux
calibration (Sect.~\ref{SubSection:FluxCalibration})
and the uncertainty on the computed \CHbeta.


\begin{table*}
\begin{scriptsize}
\caption{Reddening corrected line ratios, normalized to \Hb, for the
sample of galaxies}
\label{Table:LineFlux}
\begin{center}
\begin{tabular}{|ll|cccc|cccc|}       
\hline\hline                 
$\lambda$  & Ion  &  \multicolumn{4}{|c|}{II~Zw~33}   & 
                     \multicolumn{4}{|c|}{Haro~1}     \\
           &      &  \multicolumn{2}{|c}{Integrated}  & 
                     \multicolumn{2}{c|}{Nucleus}     &
                     \multicolumn{2}{|c}{Integrated}  & 
                     \multicolumn{2}{c|}{Nucleus}     \\
           &      &  $F_{\lambda}$  & $-W_{\lambda}$  & 
                     $F_{\lambda}$  & $-W_{\lambda}$  &
                     $F_{\lambda}$  & $-W_{\lambda}$  & 
                     $F_{\lambda}$  & $-W_{\lambda}$ \\
\hline  

3727 &    [\ion{O}{ii}] &    $3.312 \pm 0.284$  &  $58.94 \pm 1.43$  & 
                             $3.125 \pm 0.254$  &  $48.26 \pm 0.70$  & 
                             $2.419 \pm 0.217$  &  $24.25 \pm 0.32$  & 
                             $2.106 \pm 0.190$  &  $19.55 \pm 0.15$ \\ 

3835 &             H9   &    $0.055 \pm 0.006$  &   $0.80 \pm 0.05$  & 
                             $0.041 \pm 0.004$  &   $0.50 \pm 0.04$  & 
                              ---             &  ---               & 
                              ---             &  ---              \\ 

3868 & [\ion{Ne}{III}]  &    $0.259 \pm 0.018$  &   $3.53 \pm 0.06$  & 
                             $0.236 \pm 0.015$  &   $2.66 \pm 0.04$  & 
                             $0.062 \pm 0.006$  &   $0.44 \pm 0.02$  & 
                             $0.040 \pm 0.004$  &   $0.27 \pm 0.01$ \\ 

3889 &          HeI+H8  &    $0.166 \pm 0.012$  &   $2.39 \pm 0.04$  & 
                             $0.168 \pm 0.011$  &   $1.95 \pm 0.03$  & 
                             $0.132 \pm 0.009$  &   $0.94 \pm 0.02$  & 
                             $0.126 \pm 0.009$  &   $0.85 \pm 0.01$ \\ 

3968 & [\ion{Ne}{III}]  &    $0.228 \pm 0.014$  &   $3.21 \pm 0.04$  & 
                             $0.209 \pm 0.012$  &   $2.47 \pm 0.04$  & 
                              ---             &  ---               & 
                              ---             &  ---              \\ 

4101 &             \Hd  &    $0.256 \pm 0.012$  &   $3.68 \pm 0.03$  & 
                             $0.256 \pm 0.012$  &   $3.02 \pm 0.06$  & 
                             $0.244 \pm 0.013$  &   $1.84 \pm 0.02$  & 
                             $0.239 \pm 0.012$  &   $1.73 \pm 0.01$ \\ 

4340 &             \Hg  &    $0.479 \pm 0.017$  &   $7.82 \pm 0.07$  & 
                             $0.476 \pm 0.015$  &   $6.26 \pm 0.06$  & 
                             $0.469 \pm 0.018$  &   $3.96 \pm 0.01$  & 
                             $0.469 \pm 0.018$  &   $3.88 \pm 0.01$ \\ 

4363 &  [\ion{O}{III}]  &    $0.032 \pm 0.003$  &   $0.51 \pm 0.03$  & 
                             $0.024 \pm 0.002$  &   $0.31 \pm 0.03$  & 
                              ---             &  ---               & 
                              ---             &  ---              \\ 

4861 &             \Hb  &    $1.000 $            &  $19.57 \pm 0.13$  & 
                             $1.000 $            &  $15.57 \pm 0.09$  & 
                             $1.000 $            &  $10.49 \pm 0.05$  & 
                             $1.000 $            &  $10.64 \pm 0.02$ \\ 
 
4959 &  [\ion{O}{III}]  &    $0.903 \pm 0.023$  &  $18.66 \pm 0.10$  & 
                             $0.873 \pm 0.022$  &  $14.00 \pm 0.06$  & 
                             $0.222 \pm 0.006$  &  $ 2.45 \pm 0.02$  & 
                             $0.187 \pm 0.006$  &  $ 2.05 \pm 0.01$ \\ 
 
5007 &  [\ion{O}{III}]  &    $2.653 \pm 0.067$  &  $57.47 \pm 0.30$  & 
                             $2.554 \pm 0.064$  &  $42.93 \pm 0.19$  & 
                             $0.656 \pm 0.017$  &  $ 7.70 \pm 0.03$  & 
                             $0.532 \pm 0.013$  &  $ 6.22 \pm 0.02$ \\[2pt] 

&         &                  \multicolumn{2}{c}{ $I(\Hb)=  1061.6 \pm   18.5 $}   &  
                             \multicolumn{2}{c|}{$I(\Hb)=   489.6 \pm    8.5 $}   &  
                             \multicolumn{2}{c}{ $I(\Hb)=  4185.9 \pm   72.8 $}   &  
                             \multicolumn{2}{c|}{$I(\Hb)=  1532.0 \pm   26.5 $}  \\  

&         &                  \multicolumn{2}{c}{ $F(\Hb)=  2090.3 \pm  283.3 $}   &  
                             \multicolumn{2}{c|}{$F(\Hb)=   760.0 \pm   80.4 $}   &  
                             \multicolumn{2}{c}{ $F(\Hb)=  9800.7 \pm 1628.1 $}   &  
                             \multicolumn{2}{c|}{$F(\Hb)=  3600.5 \pm  597.4 $}  \\  

&         &                  \multicolumn{2}{c}{ $\CHbeta=0.294 \pm 0.058$}     &  
                             \multicolumn{2}{c|}{$\CHbeta=0.191 \pm 0.045$}     &  
                             \multicolumn{2}{c}{ $\CHbeta=0.369 \pm 0.072$}     &  
                             \multicolumn{2}{c|}{$\CHbeta=0.371 \pm 0.072$}    \\  

&         &                  \multicolumn{2}{c}{ $\AV=0.630 \pm 0.125$}     &  
                             \multicolumn{2}{c|}{$\AV=0.409 \pm 0.097$}     &  
                             \multicolumn{2}{c}{ $\AV=0.791 \pm 0.154$}     &  
                             \multicolumn{2}{c|}{$\AV=0.794 \pm 0.154$}    \\  

&         &                  \multicolumn{2}{c}{ Area: $739\; \square\arcsec = 24.3$ kpc$^{2}$}     &  
                             \multicolumn{2}{c|}{Area: $138\; \square\arcsec = 4.5$ kpc$^{2}$}      &  
                             \multicolumn{2}{c}{ Area: $1432\; \square\arcsec = 94.2$ kpc$^{2}$}    &  
                             \multicolumn{2}{c|}{Area: $108\; \square\arcsec = 7.1$ kpc$^{2}$}     \\  

&         &                  \multicolumn{2}{c}{ $V_\mathrm{stars}=2828.1$}     &  
                             \multicolumn{2}{c|}{$V_\mathrm{stars}=2829.8$}     &  
                             \multicolumn{2}{c}{ $V_\mathrm{stars}=3772.0$}     &  
                             \multicolumn{2}{c|}{$V_\mathrm{stars}=3758.0$}     \\ 

&         &                  \multicolumn{2}{c}{ $V_\mathrm{[OIII]}=2827.4$}     &  
                             \multicolumn{2}{c|}{$V_\mathrm{[OIII]}=2828.4$}     &  
                             \multicolumn{2}{c}{ $V_\mathrm{[OIII]}=3781.5$}     &  
                             \multicolumn{2}{c|}{$V_\mathrm{[OIII]}=3759.6$}    \\[5pt]  

\hline\hline                 
$\lambda$  & Ion  &  \multicolumn{4}{|c|}{NGC~4670}    & 
                     \multicolumn{4}{|c|}{Mrk~314}    \\
           &      &  \multicolumn{2}{|c}{Integrated}   & 
                     \multicolumn{2}{c|}{Nucleus}      &
                     \multicolumn{2}{|c}{Integrated}   & 
                     \multicolumn{2}{c|}{Nucleus}     \\
           &      &  $F_{\lambda}$  & $-W_{\lambda}$   & 
                     $F_{\lambda}$  & $-W_{\lambda}$   &
                     $F_{\lambda}$  & $-W_{\lambda}$   & 
                     $F_{\lambda}$  & $-W_{\lambda}$  \\
\hline     

3727 &   [\ion{O}{ii}]  &    $3.051 \pm 0.280$  &  $60.52 \pm 0.97$   & 
                             $2.487 \pm 0.200$  &  $44.16 \pm 0.30$   & 
                             $3.255 \pm 0.270$  &  $50.66 \pm 0.92$   & 
                             $2.859 \pm 0.264$  &  $54.09 \pm 0.60$  \\ 

3835 &              H9  &    $0.048 \pm 0.004$  &   $0.89 \pm 0.03$   & 
                             $0.044 \pm 0.004$  &   $0.71 \pm 0.01$   & 
                             $0.053 \pm 0.005$  &   $0.97 \pm 0.07$   & 
                             $0.037 \pm 0.003$  &   $0.78 \pm 0.02$  \\ 

3868 & [\ion{Ne}{III}]  &    $0.202 \pm 0.016$  &   $3.32 \pm 0.03$   & 
                             $0.208 \pm 0.012$  &   $3.37 \pm 0.01$   & 
                             $0.296 \pm 0.019$  &   $3.58 \pm 0.06$   & 
                             $0.297 \pm 0.022$  &   $4.24 \pm 0.02$  \\ 

3889 &          HeI+H8  &    $0.187 \pm 0.014$  &   $3.18 \pm 0.02$   & 
                             $0.182 \pm 0.011$  &   $2.84 \pm 0.01$   & 
                             $0.165 \pm 0.010$  &   $2.91 \pm 0.08$   & 
                             $0.148 \pm 0.010$  &   $2.12 \pm 0.02$  \\ 

3968 & [\ion{Ne}{III}]  &    $0.217 \pm 0.013$  &   $3.79 \pm 0.02$   & 
                             $0.215 \pm 0.010$  &   $3.55 \pm 0.01$   & 
                             $0.245 \pm 0.013$  &   $3.34 \pm 0.03$   & 
                             $0.214 \pm 0.014$  &   $3.07 \pm 0.01$  \\ 

4101 &             \Hd  &    $0.263 \pm 0.014$  &   $4.52 \pm 0.04$   & 
                             $0.259 \pm 0.013$  &   $4.53 \pm 0.02$   & 
                             $0.256 \pm 0.011$  &   $3.29 \pm 0.03$   & 
                             $0.241 \pm 0.012$  &   $3.40 \pm 0.01$  \\ 

4340 &             \Hg  &    $0.458 \pm 0.016$  &   $8.99 \pm 0.04$   & 
                             $0.468 \pm 0.013$  &   $9.71 \pm 0.04$   & 
                             $0.477 \pm 0.015$  &   $6.85 \pm 0.05$   & 
                             $0.469 \pm 0.017$  &   $7.88 \pm 0.04$  \\ 

4363 &  [\ion{O}{III}]  &     ---               &  ---                & 
                              ---               &  ---                & 
                             $0.037 \pm 0.002$  &   $0.53 \pm 0.02$   & 
                             $0.038 \pm 0.002$  &   $0.65 \pm 0.02$  \\ 
 
4861 &             \Hb  &    $1.000$            &  $21.93 \pm 0.09$   & 
                             $1.000$            &  $26.03 \pm 0.14$   & 
                             $1.000$            &  $17.29 \pm 0.16$   & 
                             $1.000$            &  $20.41 \pm 0.11$  \\ 

4959 &  [\ion{O}{III}]  &    $0.765 \pm 0.019$  &  $17.89 \pm 0.08$   & 
                             $0.868 \pm 0.021$  &  $24.11 \pm 0.08$   & 
                             $1.070 \pm 0.026$  &  $19.51 \pm 0.18$   & 
                             $1.175 \pm 0.029$  &  $25.34 \pm 0.20$  \\ 

5007 &  [\ion{O}{III}]  &    $2.223 \pm 0.056$  &  $54.01 \pm 0.23$   & 
                             $2.500 \pm 0.061$  &  $72.26 \pm 0.23$   & 
                             $3.053 \pm 0.075$  &  $59.08 \pm 0.55$   & 
                             $3.354 \pm 0.085$  &  $77.07 \pm 0.63$  \\[2pt] 

&         &                  \multicolumn{2}{c}{ $I(\Hb)=  6636.0 \pm  116.1 $}   &  
                             \multicolumn{2}{c|}{$I(\Hb)=  3101.8 \pm   54.3 $}   &  
                             \multicolumn{2}{c}{ $I(\Hb)=  1491.8 \pm   26.2 $}   &  
                             \multicolumn{2}{c|}{$I(\Hb)=   992.0 \pm   17.3 $}  \\  

&         &                  \multicolumn{2}{c}{ $F(\Hb)= 10489.3 \pm 1598.2 $}   &  
                             \multicolumn{2}{c|}{$F(\Hb)=  5565.2 \pm  382.6 $}   &  
                             \multicolumn{2}{c}{ $F(\Hb)=  3614.7 \pm  394.0 $}   &  
                             \multicolumn{2}{c|}{$F(\Hb)=  1976.5 \pm  330.3 $}  \\  

&         &                  \multicolumn{2}{c}{ $\CHbeta=0.199 \pm 0.066$}     &  
                             \multicolumn{2}{c|}{$\CHbeta=0.254 \pm 0.029$}     &  
                             \multicolumn{2}{c}{ $\CHbeta=0.384 \pm 0.047$}     &  
                             \multicolumn{2}{c|}{$\CHbeta=0.299 \pm 0.072$}    \\  

&         &                  \multicolumn{2}{c}{ $\AV=0.426 \pm 0.141$}     &  
                             \multicolumn{2}{c|}{$\AV=0.543 \pm 0.062$}     &  
                             \multicolumn{2}{c}{ $\AV=0.823 \pm 0.100$}     &  
                             \multicolumn{2}{c|}{$\AV=0.641 \pm 0.155$}    \\  

&         &                  \multicolumn{2}{c}{ Area: $2571\; \square\arcsec = 30.3$ kpc$^{2}$}     &  
                             \multicolumn{2}{c|}{Area: $92\;   \square\arcsec = 1.1$ kpc$^{2}$}      &  
                             \multicolumn{2}{c}{ Area: $739\;  \square\arcsec = 14.6$ kpc$^{2}$}     &  
                             \multicolumn{2}{c|}{Area: $108\;  \square\arcsec = 2.1$ kpc$^{2}$}      \\  

&         &                  \multicolumn{2}{c}{ $V_\mathrm{stars}=1026.2$}     &  
                             \multicolumn{2}{c|}{$V_\mathrm{stars}=1019.9$}     &  
                             \multicolumn{2}{c}{ $V_\mathrm{stars}=2067.6$}     &  
                             \multicolumn{2}{c|}{$V_\mathrm{stars}=2064.6$}    \\  

&         &                  \multicolumn{2}{c}{ $V_\mathrm{[OIII]}=1049.7$}     &  
                             \multicolumn{2}{c|}{$V_\mathrm{[OIII]}=1051.1$}     &  
                             \multicolumn{2}{c}{ $V_\mathrm{[OIII]}=2076.3$}     &  
                             \multicolumn{2}{c|}{$V_\mathrm{[OIII]}=2080.3$}    \\  


\hline
\hline                 
$\lambda$  & Ion  &  \multicolumn{4}{|c|}{III~Zw~102}    &
                     \multicolumn{4}{|c|}{ }            \\
           &      &  \multicolumn{2}{|c}{Integrated}     & 
                     \multicolumn{2}{c|}{Nucleus}        &
                     \multicolumn{2}{|c}{ }              & 
                     \multicolumn{2}{c|}{ }             \\
           &      &  $F_{\lambda}$  & $-W_{\lambda}$   & 
                     $F_{\lambda}$  & $-W_{\lambda}$   & 
                                    &                  & 
                                    &                  \\
\hline     

3727 &   [\ion{O}{ii}]  &    $1.618 \pm 0.155$  &  $ 9.81 \pm 0.20$   &  
                             $1.269 \pm 0.120$  &  $11.38 \pm 0.11$   &  
                                                &                     &
                                                &                    \\

3835 &              H9  &     ---               &  ---                &  
                              ---               &  ---                &  
                                                &                     &
                                                &                    \\

3868 & [\ion{Ne}{III}]  &     ---              &  ---                &  
                              ---              &  ---                &  
                                               &                     &                        
                                               &                     \\

3889 &          HeI+H8  &     ---              &  ---                &  
                              ---              &  ---                &  
                                               &                     &
                                               &                    \\

3968 & [\ion{Ne}{III}]  &     ---              &  ---                &  
                              ---              &  ---                &  
                                               &                     &
                                               &                    \\

4101 &             \Hd  &    $0.237 \pm 0.015$  &   $1.01 \pm 0.01$   &  
                             $0.233 \pm 0.012$  &   $1.52 \pm 0.02$   &  
                                                &                     &
                                                &                    \\ 

4340 &             \Hg  &    $0.469 \pm 0.019$  &   $2.20 \pm 0.02$   &  
                             $0.469 \pm 0.018$  &   $3.45 \pm 0.03$   &  
                                                &                     &
                                                &                    \\

4363 &  [\ion{O}{III}]  &     ---               &  ---                &  
                              ---               &  ---                &  
                                                &                     &
                                                &                    \\

4861 &             \Hb  &    $1.000$            &  $ 4.93 \pm 0.02$   &  
                             $1.000$            &  $ 8.53 \pm 0.04$   &  
                                                &                     &
                                                &                    \\
 
4959 &  [\ion{O}{III}]  &    $0.112 \pm 0.004$  &  $ 0.58 \pm 0.01$   &  
                             $0.086 \pm 0.003$  &  $ 0.76 \pm 0.01$   &  
                                                &                     &
                                                &                    \\

5007 &  [\ion{O}{III}]  &    $0.319 \pm 0.008$  &  $ 1.77 \pm 0.01$   &  
                             $0.230 \pm 0.006$  &  $ 2.18 \pm 0.01$   &  
                                                &                     &
                                                &                    \\[2pt]

&         &                  \multicolumn{2}{c}{ $I(\Hb)= 1652.9 \pm   29.2 $}  &  
                             \multicolumn{2}{c|}{$I(\Hb)=  809.2 \pm   14.2 $}  &  
                             \multicolumn{2}{c}{}                               &
                             \multicolumn{2}{c|}{}                             \\

&         &                  \multicolumn{2}{c}{ $F(\Hb)= 6031.4 \pm 1075.0 $}  &  
                             \multicolumn{2}{c|}{$F(\Hb)= 3647.1 \pm  628.9 $}  &  
                             \multicolumn{2}{c}{}                               &
                             \multicolumn{2}{c|}{}                             \\

&         &                  \multicolumn{2}{c}{ $\CHbeta=0.562 \pm 0.077$}     &  
                             \multicolumn{2}{c|}{$\CHbeta=0.654 \pm 0.075$}     &  
                             \multicolumn{2}{c}{}                               &
                             \multicolumn{2}{c|}{}                             \\

&         &                  \multicolumn{2}{c}{ $\AV=1.203 \pm 0.165$}         &  
                             \multicolumn{2}{c|}{$\AV=1.399 \pm 0.160$}         &  
                             \multicolumn{2}{c}{}                               &
                             \multicolumn{2}{c|}{}                             \\

&         &                  \multicolumn{2}{c}{ Area: $2171\; \square\arcsec = 27.2$ kpc$^{2}$}      &  
                             \multicolumn{2}{c|}{Area: $123\; \square\arcsec = 1.5$ kpc$^{2}$}        &  
                             \multicolumn{2}{c}{}                               &
                             \multicolumn{2}{c|}{}                             \\

&         &                  \multicolumn{2}{c}{ $V_\mathrm{stars}=1611.7$}      &  
                             \multicolumn{2}{c|}{$V_\mathrm{stars}=1621.9$}      &  
                             \multicolumn{2}{c}{}                               &
                             \multicolumn{2}{c|}{}                             \\

&         &                  \multicolumn{2}{c}{ $V_\mathrm{[OIII]}=1615.6$}    &  
                             \multicolumn{2}{c|}{$V_\mathrm{[OIII]}=1608.7$}    &  
                             \multicolumn{2}{c}{}                               &
                             \multicolumn{2}{c|}{}                             \\
\hline       

\hline       
\end{tabular}
\end{center}
\tablefoot{
Reddening-corrected line fluxes, normalized to $F(\Hb)=1$. 
The observed \Hb\ flux, $I(\Hb)$, the reddening-corrected \Hb\ flux, $F(\Hb)$
(both $\times 10^{-16}$ ergs cm$^{-2}$ s$^{-1}$), 
the reddening coefficient, $\CHbeta$, and \AV\  (derived as $2.14\times\CHbeta$) 
are listed for each region. Also shown are the area covered by the integrated spectra,
and the corresponding (luminosity weighted) average stellar and gas velocities.
The quoted uncertainties account for measurement, flux-calibration and 
reddening coefficient errors.
}
\end{scriptsize}
\end{table*}

\subsubsection{Heavy element abundances}
\label{SubSection:Abundances}

In the two galaxies Mrk~314 and II~Zw~33, where the
[\ion{O}{iii}]~$\lambda4363$ auroral line could be measured, electron
temperatures and oxygen and neon abundances were derived using the "direct"
method. 

First, we computed the electron temperature $T_\mathrm{e}$(\ion{O}{III}) from
the [\ion{O}{iii}]~$\lambda4363$/[\ion{O}{iii}]~$\lambda4959+\lambda5007$
ratio using the equations provided by \cite{Aller1984}. As in our spectra
there are no useful lines to obtain the electron density, $N_\mathrm{e}$, we
assumed $N_\mathrm{e}=100$ cm$^{-3}$.

The metal abundances were derived using the revised expressions in
\cite{IzotovStasinskaetal2006}, which take into account the latest atomic data
and incorporate an appropriate grid of photoionization models with
state-of-the-art model atmospheres. We used their equations (3) and (5) to
compute the O$^{+}$/H,  O$^{++}$/H and Ne$^{2+}$ abundances, after computing
$T_\mathrm{e}$(\ion{O}{II}) using equation (14). The total oxygen abundance
was obtained by simply summing the abundances of O$^{+}$ and O$^{++}$, as the
fraction of the O$^{+++}$ ion is generally negligible. The correction factor
for Ne, $ICF(\mathrm{Ne}^{2+})$, was computed using equation (19).
(Using the $\mathrm{Ne}^{++}/\mathrm{O}^{++}$ ratio instead of the ICF
we would get a Ne abundance about 0.2 dex larger.)

For the remaining three galaxies, where [\ion{O}{iii}]~$\lambda4363$ could not
be measured, we had to rely on the so-called "strong-line methods", in order
to derive the oxygen abundance. We adopted the revised calibration of the
P-method \citep{Pilyugin2000,Pilyugin2001}, provided in \cite{Pilyugin2005}.

Electron temperatures and abundances are listed in
Table~\ref{Table:Abundances}. For II~Zw~33 and Mrk~314, we provide only the
abundances derived from the direct method; since both galaxies lie in the
transition zone $8.0 \leq 12 + \log\,(\mathrm{O/H}) \leq 8.3$,  the
\cite{Pilyugin2005} calibration does not give reliable abundance values.

\begin{table*}
\begin{scriptsize}
\caption{Electron temperatures and heavy metal abundances.}
\label{Table:Abundances}
\begin{center}
\begin{tabular}{|l|cc|cc|cc|cc|cc|}       
\hline\hline                 
                                     & \multicolumn{2}{|c|}{II~Zw~33}               &   \multicolumn{2}{|c|}{Haro~1}  &   \multicolumn{2}{|c|}{NGC~4670}   & \multicolumn{2}{|c|}{Mrk~314}    &   \multicolumn{2}{|c|}{III~Zw~102}   \\
Value                                &  Integrated        & Nuclear                 &  Integrated        & Nuclear    &     Integrated        & Nuclear    &  Integrated        & Nuclear     &  Integrated        & Nuclear         \\       
\hline
$T_\mathrm{e}$(\ion{O}{III}) (K)     &  $ 12290 \pm  330$ & $ 11270 \pm  360$       & ---                & ---        & ---                   & ---        & $ 12330 \pm  250$  & $ 12090 \pm  210$ & ---          & ---             \\
$T_\mathrm{e}$(\ion{O}{II}) (K)      &  $ 12080 \pm  290$ & $ 11160 \pm  350$       & ---                & ---        & ---                   & ---        & $ 12110 \pm  210$  & $ 11910 \pm  190$ & ---          & ---             \\
$12 + \log\,(\mathrm{O/H})$\tablefootmark{1}      &  $ 8.06  \pm 0.03$ & $ 8.16  \pm 0.04$       & ---                & ---        &                       &            & $ 8.08  \pm 0.03$  & $ 8.10  \pm 0.03$ &              &                 \\
$12 + \log\,(\mathrm{O/H})$\tablefootmark{2}      &  ---               & ---                     & $ 8.36 $           & $ 8.40$    & $ 8.29$               & $ 8.37$    & ---                & ---               & 8.44         & 8.49            \\
$\log(\mathrm{Ne/O})$                &  $-0.85  \pm 0.05$ & $-0.85  \pm 0.06$       &  ---               & ---        & ---                   & ---        & $-0.83  \pm 0.04$  & $-0.82  \pm 0.04$ & ---          & ---             \\
\hline 
%
%
%
%

\end{tabular}
\end{center}
\tablefoot{\tablefoottext{1} Oxygen abundances computed using the $T_\mathrm{e}$ method.
\tablefoottext{2} Oxygen abundances derived from \cite{Pilyugin2005}; uncertainties are 
about 0.1 dex.}
\end{scriptsize}
\end{table*}

\subsubsection{Comparison with literature data}
\label{SubSection:ComparisonLiterature}

To assess the quality of our measurements, we compared our integrated fluxes
and equivalent widths with the integrated data published by 
\cite{MoustakasKennicutt2006}. Four of our objects are in common with their
sample, the exception being II~Zw~33. 

\begin{figure}
\centering
\includegraphics[angle=0, width=0.50\textwidth, bb=45 40 570 560]{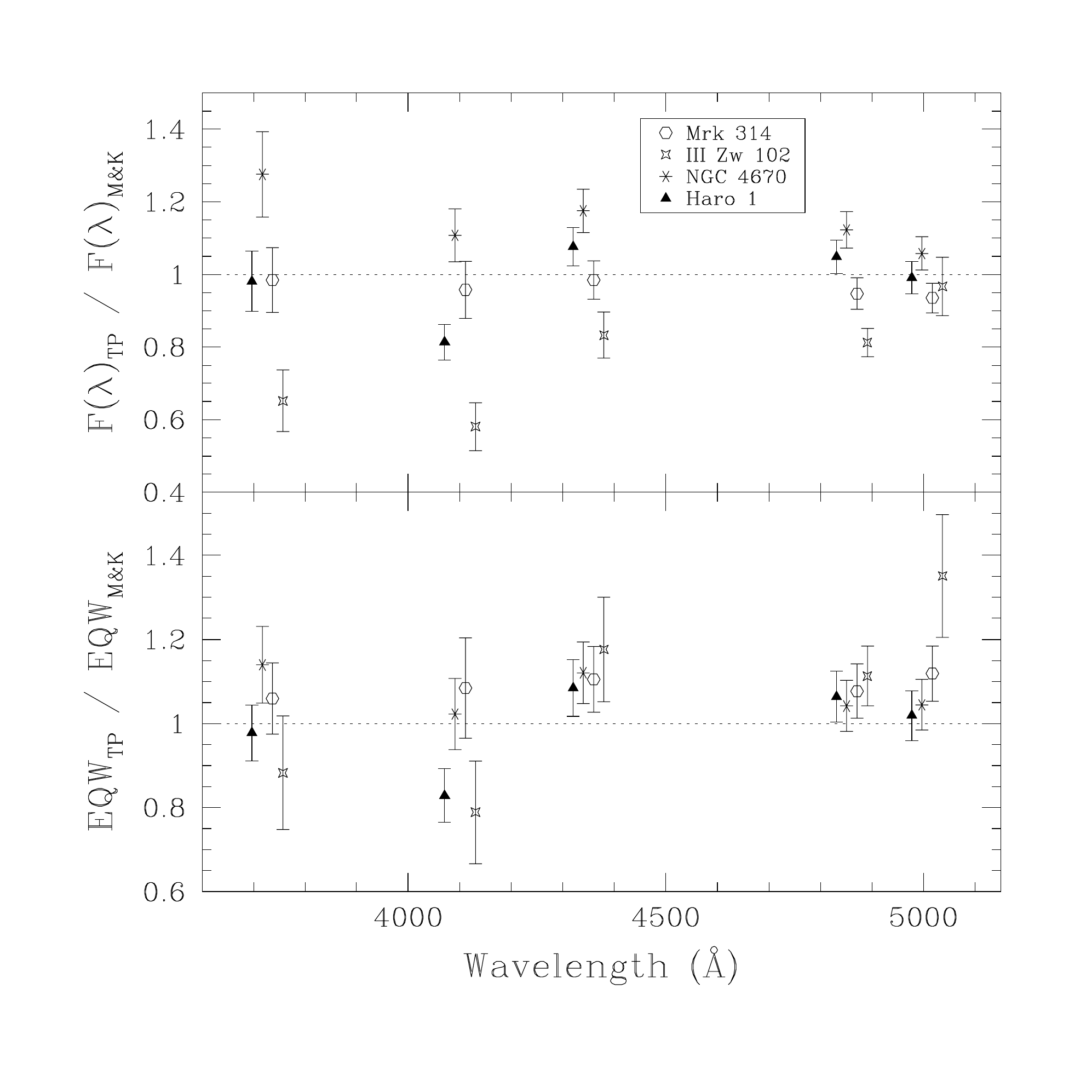}
\caption{Upper panel: ratio between the fluxes computed by direct summation
of our integrated spectra, after correction for foreground galactic extinction, 
and the (galactic extinction corrected) integrated fluxes published by 
\cite{MoustakasKennicutt2006}. Lower panel: ratio between equivalent
width measurements. To avoid overlapping of error bars (which include
the uncertainties on both datasets), datapoints have been slightly shifted 
from their nominal position on the $X$ axis.}
\label{Figure:Moustakas}
\end{figure}

For this comparison, our measurements have been corrected for foreground 
galactic extinction only. The results of the comparison are displayed  in
Figure~\ref{Figure:Moustakas}. Taking into account the different integration
area used in the \cite{MoustakasKennicutt2006} spectra (a
$30\arcsec\times50\arcsec$ aperture) and our own spectra, the comparison looks
fair.

\section{Discussion}
\label{Section:Discussion}

\subsection{II~Zw~33 (=Mrk 1094)}
\label{Section:IIZw33}

\object{II Zw 33} is a well-studied object, cataloged as BCG in
\cite{ThuanMartin1981}. Optical imaging and surface photometry has been
presented in several works \citep{Loose1987,Cairos2001-II,Cairos2001-I,
GildePaz2003, GildePaz2005}. The presence of a central SF bar and the striking
twisting of the isophotes have been interpreted as interaction signs
\citep{Mendez1999}. II~Zw~33 is also a well known Wolf-Rayet (WR) galaxy, included
in the WR galaxy lists by \cite{Vacca1992} and \cite{Schaerer1999}.

Maps of II~Zw~33 are displayed in Figures
~\ref{Figure:emissionmaps} and \ref{Figure:velocitymaps}. At the
II~Zw~33 distance, the $4\farcs2$ fiber diameter provides a spatial sampling of
760~pc per spaxel. 

The line emission is concentrated along a central bar-like
structure, with no clear peak, while the stellar emission extends further
out from the central SF regions, evidencing the presence of a much more
extended, more evolved stellar population. The continuum map displays a well
defined peak, located roughly at the center of the galaxy.

The excitation map ([\ion{O}{iii}]/\Hb) traces the regions of star-formation,
as expected for objects photoionized by stars. The peak in [\ion{O}{iii}]/\Hb\
is located at the southern tip of the bar (knot 'A' as labeled in
\citealp{Mendez1999}); their value of 5.5 agrees well with the peak 
value we find, $\simeq4.5$ (the small discrepancy is likely due to the
difference in spatial resolution). 

The extinction is relatively uniform in the central region, with \AV\
values around 0.4; these values are significantly smaller than those published
by \cite{Mendez1999}.

The [\ion{O}{iii}] velocity map does not show a clear rotation pattern,
although the northwest side seems to have a slightly higher radial velocity as
a whole. The ionized gas kinematics is in good qualitative agreement with the
\ion{H}{I} kinematical map published by \cite[][see their Fig.~6]{Walter1997}.
The stellar velocity field is not as extended as required to make a sensible
comparison, and is also much noisier. 

We derived an oxygen abundance of $12+\log\,(\mathrm{O/H})=8.06$,
in good agreement with the value of $12+\log\,(\mathrm{O/H})=7.99$ reported by
\cite{Shi2005}, who also used the $T_\mathrm{e}$ method.

The Wolf-Rayet bump around 4600--4680 \AA\ is visible in two spaxels, marked
by a star in the [\ion{O}{iii}] flux maps in Figure~\ref{Figure:emissionmaps}.

\subsection{Haro~1 (=NGC 2415)}
\label{Section:Haro1}

\object{Haro 1} is classified as BCG by \cite{Gordon1981} and  \cite{Klein1984}, and is
included in the \ion{H}{ii} galaxies sample studied by \cite{Deeg1997}. It has
been defined as a paired galaxy with little (if any) tidal disturbance
\citep{SchneiderSalpeter1992}. Its companion is UGC~3937, with a redshift
difference of only 210 km s$^{-1}$ and a projected distance of 350 kpc (23
arcmin at Haro~1's distance). Broad-band surface brightness photometry was
presented in \cite{Cairos2001-II,Cairos2001-I}.

Maps are shown in Figures ~\ref{Figure:emissionmaps} and
\ref{Figure:velocitymaps} (column~2). At the Haro~1 distance, the spatial 
resolution is about 1.1~kpc per spaxel. 

Haro~1 displays outer regular isophotes in the continuum maps. (The jagged
appearance is the result of masking several foreground stars; for the sake of
clarity, a smaller star, located at $\approx 30 \arcsec$ east from the center,
has been left, as well as the background galaxy on the west-side). The
continuum emission  peak is placed roughly at the center of the isophotes. The
emission-line maps display a more irregular pattern: a considerably large
filament extends out to the north-west, and the emission peak seems to be
slightly offset to the west with respect to the continuum peak.

The [\ion{O}{iii}]/\Hb\ ratio map has a shallow minimum, 0.40, in the galaxy's
central region (where the emission peaks), and increases slightly outwards, 
reaching its maximum at the position of the extension detected in emission-line
maps. Radial changes in the [\ion{O}{iii}]/\Hb\ ratio have been found in
spiral galaxies, and have been interpreted as variation in the heavy element
abundances \citep{Searle1971,Smith1975}. Interestingly, Haro~1 is the most
luminous object in our sample, and it has been also classified as a Sm galaxy 
(HyperLeda\footnote{http://leda.univ-lyon1.fr}; \citealp{Paturel2003}).

The extinction map shows little or no spatial variations, with
an average extinction $\AV \sim 0.90$. 

Haro~1's kinematics show an overall regular rotation about an axis oriented
southeast-northwest, with pretty much the same pattern and amplitude (about
160 km s$^{-1}$) both in the gas and in the stellar velocity maps. 

The value derived for the metallicity, $12+\log\,(\mathrm{O/H})=8.36$, is
slightly lower that the value of $12+\log\,(\mathrm{O/H})=8.52$ derived in
\cite{Shi2005}, which used the P-method (as calibrated in
\citealp{Pilyugin2001}).

\subsection{NGC~4670 (=Haro~9, Arp~163)}
\label{Section:NGC4670}

\object{NGC 4670} has been classified as a blue amorphous galaxy \citep{Hunter1994,
Marlowe1997}, as a peculiar, barred S0/a galaxy \citep{deVaucouleurs1991} and
as a BCG \citep{GildePaz2003}. It is listed as a peculiar galaxy with "diffuse
counter tails"  in Arp's catalogue \citep{Arp1966}, and as a WR galaxy  in
\cite{MasHesse1999} and \cite{Schaerer1999}. 
Broad-band photometry and \Ha\ imaging have been provided in
\cite{GildePaz2003}, \cite{GildePaz2005} and \cite{Marlowe1997}. It is also
cataloged as a WR galaxy .  

Maps are shown in Figures~\ref{Figure:emissionmaps} and
\ref{Figure:velocitymaps} (column~3). At the NGC~4670 distance, the VIRUS-P
IFU $4\farcs2$ fiber diameter corresponds to about 460~pc. 

The continuum map shows an outer regular morphology, with elliptical
isophotes; the bulge, extremely bright, is clearly elongated on an east-west
axis. The continuum peak appears to be a little displaced to the east, in
agreement with the $B$-band image (Figure~\ref{Figure:Bbandimages}). In
emission-lines, the galaxy displays a rather complex pattern, with a central,
roughly circular and very bright \ion{H}{ii} region, surrounded by an extended
filamentary envelope, which has several bubbles, loops and extensions.
Remarkable are the two biggest filaments, which expand to the southwest and
northeast, extending up to about 7.5 and 6.5 kpc respectively. 

The ionization map displays an intriguing structure. The [\ion{O}{iii}]/\Hb\
ratio clearly shows higher values in the central region, with a peak value
$\sim 3$, and at the outer edge of the two larger extensions. Highest values
of [\ion{O}{iii}]/\Hb\ in the central knot, where the line-emission intensity 
also peaks, are consistent with photoionization produced by hot stars; it is not
clear however what the ionizing mechanism acting on the filaments is. Although the
morphology suggests a shock-like excitation mechanism, no conclusions can be
drawn without the measurement of additional diagnostic lines. The excitation
values we found agree very well with those reported in
\cite{Hunter1994}, [\ion{O}{iii}]/\Hb\ = 2.97, derived for the spectra
positioned on the center of the galaxy. 

The extinction map shows a roughly flat behavior in the inner region, with 
\AV\ around 0.40, and seems to be dominated by noise in the outer parts.

\cite{Hunter1994}, analyzing VLA data, report an ``ordered rotation east-west,
and the motion appears to be that of a disk at the same position angle as the
optical galaxy ($\sim -100\deg$)''. Our velocity maps show a more complex
picture. At first sight a rotation pattern, though somewhat irregular, is
visible in the gaseous component, with an amplitude of about 150 \kms, around
an axis oriented roughly northwest-southeast; the velocity gradient seems to
be a little steeper in the northeast side of the galaxy compared to the
southwest part. The stellar velocity field is overall flat, and broadly agrees
with the gas kinematics in the same area, except for the slight velocity rise
toward its east edge, visible in the [\ion{O}{iii}] radial velocity map but
not in the stellar one.

\begin{figure}
\centering
\includegraphics[angle=0, width=0.50\textwidth, bb=22 68 560 750]{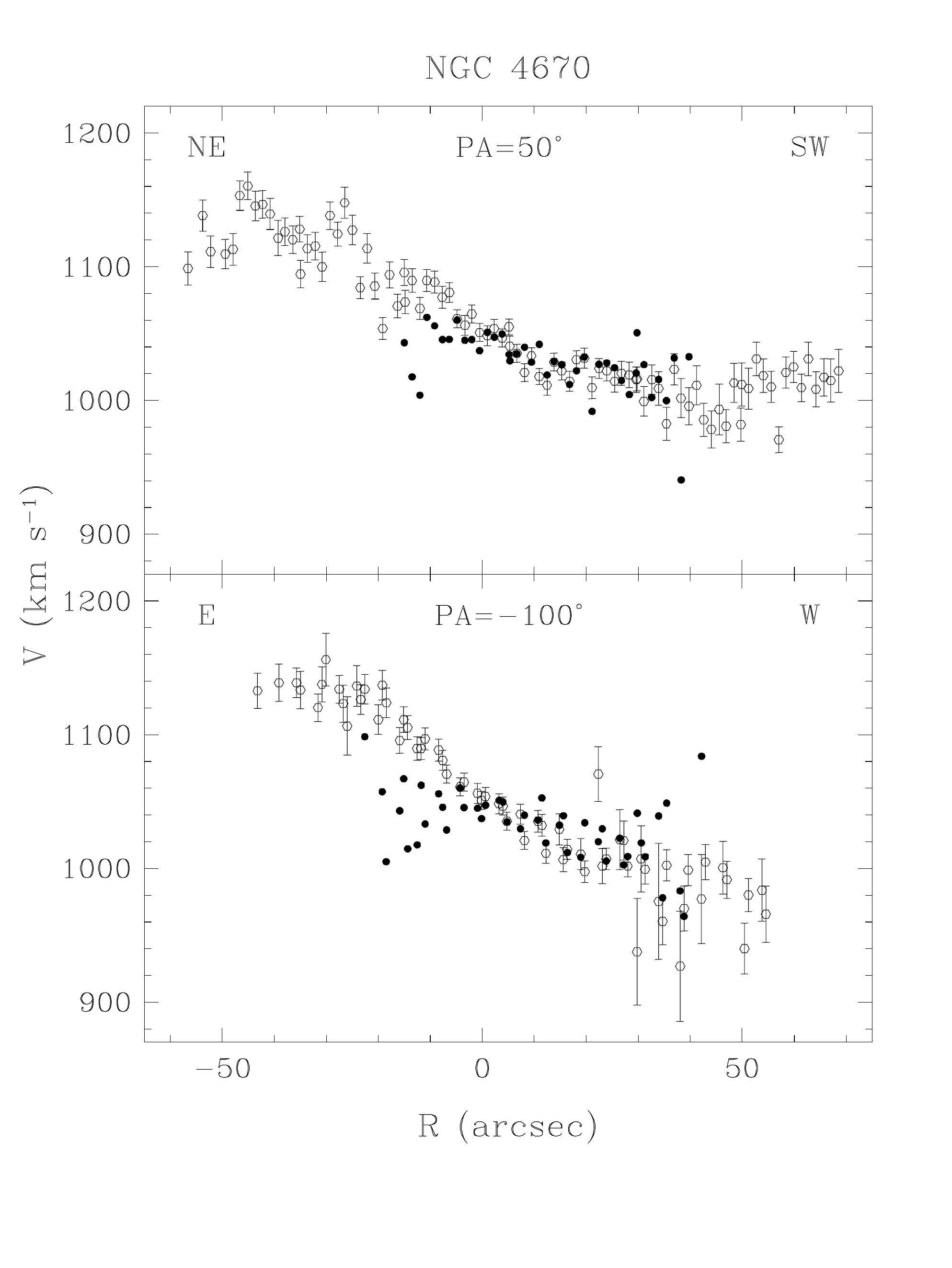}
\caption{Upper panel: Kinematical profile of NGC~4670 along a spatial cut 
through the reference point (0.0,8.0) and $\mathrm{PA}=50\deg$. 
The $x$ axis is the distance of the projection of each spaxel on the above line 
to the reference point; only spaxels whose distance to the line is less than 
$6\arcsec$ have been selected and plotted. Lower panel: Kinematical profile
through the same point as above and $\mathrm{PA}=-100\deg$. Open points are 
[\ion{O}{iii}] velocities, solid points stellar velocities.}
\label{Figure:KinematicsNGC4670}
\end{figure}

The behavior of the gas and stellar kinematics can be better perceived in
Fig~\ref{Figure:KinematicsNGC4670}, where we plot a velocity profile along a
line with $\mathrm{PA}=50\deg$, where the gas distribution extends the most from the
galaxy center, and a velocity profile at $\mathrm{PA}=-100\deg$, the value adopted by
\cite{Hunter1994}.
Both graphs show a somewhat irregular rise of the gas velocity
in a central region of about $40\arcsec$ in size, and are flat, though
noisier, outside it.
The stellar velocity profile shows that the stars co-rotate with the gas,
but with a lower velocity gradient, and seems to flatten out at somewhat 
smaller radial distance than the gas.
Furthermore, the isovelocity contour at $V=1050$ \kms\ in the
[\ion{O}{iii}] velocity field is clearly S-shaped, a characteristic
typical of barred galaxies \citep{Peterson1978}.

Thus, the kinematics of NGC~4670 is unlike the one of a rotating disk, and
resembles more the kinematics of a barred galaxy, in line with its
morphological classification, SB(s)0/aP?, listed in \cite{deVaucouleurs1991}.

The derived metallicity, $12+\log\,(\mathrm{O/H})=8.29$, is slightly 
lower but consistent with the value of $12+\log\,(\mathrm{O/H})=8.4$ 
reported in \cite{MasHesse1999}.

The Wolf-Rayet bump around 4600--4680 \AA\ is visible in three spaxels, marked
by a star in the [\ion{O}{iii}] flux maps in Figure~\ref{Figure:emissionmaps}.

\subsection{Mrk~314 (=NGC 7468)}
\label{Section:Mrk314}


\object{Mrk 314} has been cataloged as a polar-ring galaxy candidate
\citep{Whitmore1990,vanDriel2000} and is included in the \cite{MazBor1993}
catalogue of multiple nuclei galaxies. Surface photometry in the optical and in
the NIR was presented in \cite{Cairos2001-II,Cairos2001-I}, \cite{Caon2005} and
\cite{Noeske2005}. 

The maps of Mrk~314 are shown in
Figures~\ref{Figure:emissionmaps} and \ref{Figure:velocitymaps} (column~4).
The spatial sampling is about 590~pc per spaxel. 

The galaxy displays a similar morphology in both emission-line and continuum
maps, with the SF knots aligned along the northeast southwest direction. In
the continuum map, a more extended stellar component, with elliptical
isophotes, surrounds the SF region. The brighter knot is located
roughly at the center of the outer isophotes, and a tail-like feature departs
in the southwest direction; a second filament extends out to the northwest. 
In the emission-line maps, the emission is concentrated in a central SF bar, the
emission peak being slightly displaced from the continuum peak.

The peak of the [\ion{O}{iii}]/\Hb\ ratio is spatially co-located with the
line emission maximum, as expected for objects ionized by OB hot stars; the
excitation peak value of $\simeq 4.0$ agrees well with the value
found by \cite{GarciaLorenzo2008}.

The extinction map does not show any definite pattern, but a more or
less homogeneous distribution with somewhat higher values at the south, and
a \AV\ value around 0.7--0.8 in the central part; this result is
consistent with the \Ha/\Hb\ map presented in \cite{GarciaLorenzo2008},
measured in a smaller FOV ($16\arcsec\times12\arcsec$).

The ionized gas velocity field displays a complicated pattern. Around the main
body of the galaxy, the velocity field is similar to a rotating disk, with the
rotation axis oriented east-west, the northern side approaching and the
southern one receding. This rotation is however perturbed in the central
galaxy regions (within $10\arcsec$), where the velocity field shows an S-shape
distortion, indicative of the presence of a second rotating system.  South of
the main body (this area coincides with the tail-like feature), the ionized
gas is not rotating with the main-body galaxy disk. 

An extensively kinematic analysis of this galaxy has been done by
\cite{Shalyapina2004}, who found that Mrk~314 harbors two kinematics
subsystems of ionized gas: the main gaseous disk, and a smaller inner disk
oriented orthogonally to it. Our coarse spatial resolution does not allow us
to either confirm or disprove the existence this inner disk, but the
kinematics of the main disk agree well with the data shown by
\cite{Shalyapina2004}: see for instance the peak in the velocity profile at
$\simeq 30\arcsec$ south of the nucleus, and then the sharp decrease (their
Fig. 2a).

The stellar velocity field looks more irregular, though there seems to be some
rotation along an axis at about $\mathrm{PA}=\sim 45\deg$, with a total
amplitude of about 70--80 km s$^{-1}$.

The oxygen abundance we found in the integrated spectrum, $12+\log\,(\mathrm{O/H})=8.08$,  
is somewhat lower than the one reported by \cite{Shi2005},
$12+\log\,(\mathrm{O/H}) = 8.26$ (derived using also the $T_\mathrm{e}$ method), and the ones
reported by \cite{GarciaLorenzo2008}, which range spatially from 8.14 to 8.35 
(using the empirical calibrator N2, \citealp{Denicolo2002}).

\subsection{III~Zw~102 (=NGC~7625, Arp~212)}
\label{Section:IIIZw102}

\object{III Zw 102} is a galaxy well studied in a large range of wavelengths,  but
still a puzzling system. It is included in the Vorontsov-Velyaminov
catalogue of interacting galaxies \citep{Vorontsov1959} and in Arp's {\em
Atlas of Peculiar Galaxies} \citep{Arp1966}.  It has been classified as a
SA(rs)~a~pec in RC2 \citep{deVaucouleurs1976}, as a S0~pec \citep{Lynds1973},
as an elliptical with dust lanes \citep{Hawarden1981,Bertola1987,Yasuda1992},
and as a BCG  \citep{ThuanMartin1981}. \cite{Whitmore1990} considered
III~Zw~102 to be "related to polar-ring galaxies". Broad-band photometry was
presented in \cite{Cairos2001-II,Cairos2001-I}. III~Zw~102 kinematics was
first studied by \cite{Demoulin1969}, which concludes, from optical
spectroscopy, that the galaxy rotates as a nearly solid body within the
central $30\arcsec$. \ion{H}{i} gas displays a more complicated behavior, with
the outer \ion{H}{i} showing apparent counter-rotation with respect to the
stellar and molecular gas components \citep{Li1993}. 

Results from 3D investigations of III~Zw~102 have been published in
\cite{GarciaLorenzo2008}, where the central $33.6\arcsec\times29.4\arcsec$ of
the galaxy were mapped with a spatial resolution of $2.7\arcsec$
spaxel$^{-1}$, using the IFS \textsc{integral}, and in \cite{Moiseev2008},
where the distribution and kinematics of the ionized gas was studied by means
of Fabry-Perot scanning interferometry. 

Maps of III~Zw~102 are shown in Figures ~\ref{Figure:emissionmaps} and
\ref{Figure:velocitymaps} (column~5). The VIRUS-P
IFU $4\farcs2$ fiber diameter corresponds to about 470~pc.

In the continuum the galaxy displays outer circular isophotes, indicative
of the presence of an underlying population of older stars, extending up to a
radius of about 7 kpc. The central regions are more irregular, and have
elliptical isophotes, with the major axis oriented in the northeast-southwest 
direction. The presence of dust is clear seen on the continuum
map: a dust lane crosses the galaxy in a southwest-northeast direction,
almost parallel to the ellipses' major axis, and large dust patches are also
evident at the south, almost parallel to the galaxy minor axis. In the
narrow-band maps the morphology is more irregular, with several filaments,
bubbles and holes. The extended filaments, departing to the southeast and
to the northwest, are very well traced in the [\ion{O}{ii}] and \Hb\ maps.

The [\ion{O}{iii}]/\Hb\ ratio (see Figure~\ref{Figure:velocitymaps})  
is lower in the central part of the galaxy, with a minimum value of
$\simeq0.22$, in good agreement with the values published by
\cite{GarciaLorenzo2008}. 
The excitation, like in Haro~1, 
increases towards the galaxy outskirts, a behavior often seen in spiral 
galaxies \citep{Smith1975}. 
Just like Haro~1, III~Zw~102 is a luminous object, 
which has been classified as a spiral by \cite{deVaucouleurs1976}. 

The extinction map is highly inhomogeneous, and traces well the dust lanes
detected in optical frames and color maps. The values are indicative of a
considerable amount of dust.

The stellar kinematics display an overall rotation around an axis
oriented approximately northwest-southeast, with the northeast side
approaching and the southwest receding, and a velocity amplitude of about 160
\kms. The gas kinematics, which in this galaxy is better mapped by the \Hb\
line (stronger than [\ion{O}{iii}]) broadly agrees with the stellar kinematics
in the inner region. Notable are the two features protruding from either side
of the galaxy, and which, together with the knot $40\arcsec$ northeast of the
galaxy center, seem to form a structurally and kinematically independent
system. These findings are in line with the results of the detailed analysis
of this galaxy  published by \cite{Moiseev2008}: III~Zw~102 is made up of two
distinct systems: an inner disk, where gas and stars co-rotate, and a polar
ring orthogonal to it.

The oxygen abundances, $12+\log\,(\mathrm{O/H})=8.44$ in the integrated
spectrum,  and $12+\log\,(\mathrm{O/H}=8.49$ in the nuclear spectrum, are much
lower than  those derived in \cite{GarciaLorenzo2008} using the empirical
calibrator N2 \citep{Denicolo2002}, ranging from 8.81 to 8.84 in the
circumnuclear regions and about 8.85 for the nucleus. While the disagreement
can be due partly to the different corrections for the underlying absorption
at \Hb, discrepancies of up to  0.4 dex between values obtaining using the
P-method and N2 have already been reported \citep{Shi2005}.

\section{Summary}
\label{Section:Summary}

We present results of an IFS analysis of five luminous BCG galaxies, based on 
VIRUS-P data covering the wavelength interval 3550--5850 \AA\ with a
resolution of about 5 \AA\ FWHM. The mapped area is $1\farcm8$ by $1\farcm8$,
each fiber corresponding to a circle of $4\farcs2$ on the sky.

For each galaxy, spaxel by spaxel, we modeled and subtracted the stellar
continuum before fitting the most important emission-lines with a Gaussian. We
produced an atlas of maps of the most relevant emission-lines
in the spectra: [\ion{O}{ii}]~$\lambda3727$, \Hg, \Hb, and 
[\ion{O}{iii}]~$\lambda5007$, of the continuum, of the extinction (\AV)
and the excitation ([\ion{O}{iii}]~$\lambda5007$/\Hb) ratios, as well as of
the gas and the stellar kinematics. Integrated and nuclear spectroscopic
properties (line fluxes, extinction coefficient, and metal abundances) were
also derived. From this study, we highlight the following results:

\begin{itemize}

\item All galaxies show a regular morphology in the continuum, with outer 
elliptical or circular isophotes; the ionized gas displays a more complex
pattern, with filamentary emission and the presence of bubbles and holes. 
Remarkable are NGC~4670, where the gas emission shows two large extensions
north and southwest of the center, and III~Zw~102, where the two narrow
features protruding almost symmetrically from the main body, and a knot north
of the galaxy center, form part of a polar ring.

\item The galaxies display different ionization patterns: in II~Zw~33 and
Mrk~314 the excitation ratio, [\ion{O}{iii}]~$\lambda5007$/\Hb\, traces the SF
regions, as expected in objects ionized by stars. NGC~4670 has a somewhat
complex pattern, with [\ion{O}{iii}]~$\lambda5007$/\Hb\ peaking in the central
[\ion{H}{ii}] region and at the outer edge of the two large filaments. These
three galaxies have excitation ratios larger than one. In Haro~1 and
III~Zw~102, \Hb\ is stronger than [\ion{O}{iii}], and the excitation does not
trace the SF regions, but has a minimum in the central galaxy regions (where
the emission peaks) and increases towards the galaxy outskirts.

\item All galaxies possess significant amounts of dust, with interstellar
reddening values up to $\AV \sim 1.0$. 

\item A variety of kinematical behaviors are present in the sample galaxies.
II~Zw~33 shows little or no rotation, while in Haro~1 gas and stars both
display the same overall regular rotation pattern. The kinematics of
NGC~4670 is typical of a barred galaxy, with S-shaped isovelocity contours in
the gaseous component; stars co-rotate with the gas, albeit with a smaller
gradient and amplitude.
Mrk~314 has an irregular rotation pattern, with a drop in velocity in its
southern lobe. The most complex object is III~Zw~102, with two kinematically
independent systems: an inner disk, where gas and stars co-rotate, and a
polar ring orthogonal to it.

\end{itemize}

\begin{acknowledgements} 

L.~M.~Cair{\'o}s acknowledges the Alexander von Humboldt Foundation. VIRUS-P
has been made possible by a generous donation from the Cynthia and George
Mitchell Foundation. This paper includes data taken at The McDonald
Observatory of the University of Texas at Austin. We thank
J.~N.~Gonz{\'a}lez-P{\'e}rez and J.~Falc{\'o}n Barroso for stimulating
discussions and tips on the data processing and analysis. N.~Caon is grateful
for the hospitality of the Leibniz-Institut f{\"u}r Astrophysik. This research
has made use of the NASA/IPAC Extragalactic Database (NED), which is operated
by the Jet Propulsion Laboratory, Caltech, under contract with the National
Aeronautics and Space Administration. This work has been funded by
the Spanish Ministry of Science and Innovation (MCINN) under the collaboration
"Estallidos" (grants HA2006-0032, AYA2007-67965-C03-01 and AYA2010-21887-C04-04). 
P.~Papaderos is supported by Ciencia 2008 Contract, funded by FCT/MCTES (Portugal) 
and POPH/FSE (EC).

\end{acknowledgements}

\bibliographystyle{aa}
\bibliography{virus}

\end{document}